\documentclass[aps,amsmath, amssymb, prd, twocolumn, superscriptaddress,floatfix,showpacs,bibliography,10pt]{revtex4-1}
\usepackage{graphicx}  
\graphicspath{{figures/}}

\usepackage{hyperref}			
\usepackage[all]{hypcap}
\usepackage{sistyle}
\usepackage{enumitem}
\usepackage{color}
\usepackage{cancel}

\usepackage{tikz}
\tikzset{dot/.style={shape=circle,fill=black,scale=0.3}}

\usepackage{mathtools}
\newcommand{\eqcolon}{\mathrel{\resizebox{\widthof{$\mathord{=}$}}{\height}{ $\!\!=\!\!\resizebox{1.2\width}{0.8\height}{\raisebox{0.23ex}{$\mathop{:}$}}\!\!$ }}}
\newcommand{\coloneq}{\mathrel{\resizebox{\widthof{$\mathord{=}$}}{\height}{ $\!\!\resizebox{1.2\width}{0.8\height}{\raisebox{0.23ex}{$\mathop{:}$}}\!\!=\!\!$ }}}

\begin{document}

\title{Analytic series expansion of the overlap reduction function for gravitational wave search with pulsar timing arrays}

\author{Adrian \surname{Bo\^itier}}
\email[]{boitier@physik.uzh.ch}
\affiliation{Physik-Institut, Universit\"at Z\"urich, Winterthurerstrasse 190, 8057 Z\"urich, Switzerland}

\author{Shubhanshu \surname{Tiwari}}
\email[]{stiwari@physik.uzh.ch}
\affiliation{Physik-Institut, Universit\"at Z\"urich, Winterthurerstrasse 190, 8057 Z\"urich, Switzerland}

\author{Philippe \surname{Jetzer}}
\email[]{jetzer@physik.uzh.ch}
\affiliation{Physik-Institut, Universit\"at Z\"urich, Winterthurerstrasse 190, 8057 Z\"urich, Switzerland}

\date{\today} 
\newcommand{\shubh}[1]{\textcolor{red}{Shubhanshu: #1}}
\newcommand{\philippe}[1]{\textcolor{green!50!black}{Philippe: #1}}
\begin{abstract}
In our previous paper \cite{PTA1} we derived a generic expression for the pulse redshift, the main observable for the Pulsar Timing Array (PTA) experiment for detection of gravitational waves for all possible polarizations induced by modifications of general relativity (GR). In this work, we provide a generic expression of the overlap reduction function for PTA without using the short wavelength approximation for tensorial polarization.
We find that when the overlap reduction function does not have the exponential terms, which is the case when using short wavelength approximation, it leads to discontinuities and poles.
In this work, we provide a series expansion to calculate the integral exactly and investigate the behavior of the series for short wavelength values via numerical evaluation of the analytical series. We find a disagreement for the limit of colocated pulsars with the Hellings \& Downs curve. Otherwise our formalism agrees with the Hellings \& Downs curve for a broad range of pulsars.
\end{abstract}

\pacs{04.30.-w, 04.80.Nn}

\maketitle

\section{Introduction}

In our previous paper, we calculated the redshift of PTA pulses under the influence of gravitational waves with all six possible polarizations in modified GR  \cite{PTA1}. We use our previous result up to first order in the strain $h$ and zeroth order in gravitational wave frequency times pulsar period $\omega T_a$ to calculate the cross-correlated signal for the gravitational wave background (GWB). We collect all direction integrals over geometric terms in the overlap reduction function. 
This includes the interference terms which we exclude from the pattern functions as we argued in our previous work; thus, we arrive at the same integral as the present literature \cite{Chamberlin:2011ev} currently in use with the search of GW with PTA \cite{EPTA, IPTA, PPTA}. When calculating the integral however, we do not rely on the short wavelength approximation. In Appendix~\ref{sec: e^ikx}, we prove that the short wavelength can be safely used under the condition~\eqref{eq: sw-cond}. Since this condition is not met by the overlap reduction function due to the poles in the integrand, we cannot be sure that the approximation can be made. Instead, we express the integrand as a Laurent series and determine the integral via residue theorem. We find that our series agrees with the Hellings \& Downs (H\&D) curve in the entire parameter space except for a neighborhood around $\phi=0$ and $L=L'$, i.e. colocated pulsars like PSR J0737-3039 \cite{alma99116780715905508}, which should both be visible again in 2035. This puts the approximation on solid mathematical footing. Our formalism is also applicable to other polarizations and leads to a generically valid expression for any pulsar found and still holds for much higher frequencies and shorter distances.

A gravitational wave background would introduce a characteristic redshift imprint on a collection of pulsars, which depends on their location on the sky and distance from Earth. Knowing these quantities would allow to recognize (identify) such a signal (a signal with these properties) as a GWB. This can be achieved via matched filtering, where one uses the expected imprint as a filter function to extract a signal from the noisy data that match the filter. For this to work, the expected signal must be predicted correctly up to an error of roughly $\pm3\%$ \cite{Owens}.\\
The geometric part of the expected response of the cross-correlated signal of two pulsars is called the overlap reduction function and is an essential part of the filter function. It is an integral over the unit sphere, and since the integrand contains exponential terms which make an integration difficult, one usually refers to the short wavelength approximation to drop out these terms. This way, one arrives at the Hellings \& Downs curve \cite{H&D}. We would like to point out, that this function is a priori not defined at $0$ but can be continuously extended.\\
By plotting the pattern functions, which are usually used, we spotted that there always is a discontinuity for $+, \times$, and a pole for $x, y$, and $l$ polarization for each pulsar in the integration domain. This always occurs when the source lies directly behind the pulsar. This ill-defined point is forced to zero by the exponential terms. So, if the exponential terms are neglected due to short wavelength approximation, then the integrand is ill defined on its integration domain. Therefore, we calculate the integral without this approximation. Since it is not obvious to us how one could compute the limit of our result, we investigate its behavior for long wavelengths by numerically evaluating the resulting series for increasing values of $L\omega$.\\
We find that, except of one special case, the series converges to the H\&D curve as expected. However, for the special case where the two pulsar distances are almost the same and they are very close to each other on the sky (best example: double pulsar system), the series tends to the value $2$ instead of $1$.\\

We derive the signal-to-noise ratio (SNR) for a GWB measurement using PTA's in Sec.~\ref{sec: CorrPTA} and extract the expression for the overlap reduction function for the tensor mode from this calculation. We then outline the integration of this overlap reduction function in Sec.~\ref{sec: Gamma} and improve on the convergence of the resulting power series in Sec.~\ref{sec: Optimization}. We present and discuss the results of numerical evaluations of the series in Sec.~\ref{sec: Discussion}. Finally, we conclude our formalism which we developed in the previous \cite{PTA1} and this paper in Sec.~\ref{sec: Conclusions}.\\
The detailed calculations were moved to Supplemental Material to improve the readability of the main text.

\section{Correlations of PTA signals}\label{sec: CorrPTA}
We cross-correlate two PTA signals which have been measured during the same time period and define their mutual observation time as the length of the intersection of the time intervals $I$ and $I'$ in which the two pulsars $a$ and $a'$ have been observed:
\begin{equation}
T_{obs} \coloneq T(I \cap I').
\end{equation}
We will abbreviate $T_{obs}$ to $T$.\\

The data $d$ consist of the timing residuals $r$ and the noise $n$,
\begin{equation}
d(t) = r(t,\vec{\lambda}) + n(t),
\end{equation}
where $\vec{\lambda}$ is a parameter vector as, for example the polarization modes $\vec{\lambda} = (T,V,S)$ in the case of a gravitational wave background or the position of a source in the sky and the polarizations $\vec{\lambda} = (\theta,\phi,+,\times,x,y,b,l)$ in the case of point sources.\\
To use matched filtering, we multiply a filter $Q$ to the correlation:
\begin{equation}
Y \coloneq \int_0^T d(t)d'(t')Q(t-t')dt'dt.
\end{equation}

The cross-correlated signal is the expectation of the filtered correlation and only depends on the two residuals, since the noise $n_a$ is uncorrelated:
\begin{equation}
\mu \coloneq \mathbb{E}[Y] = \int_0^T \mathbb{E}\left[ r(t,\vec{\lambda})r'(t',\vec{\lambda'}) \right] Q(t-t') dt'dt.
\end{equation}
The variance of cross-correlated signals is given by \cite{GWPol}
\begin{widetext}
	\begin{align}
		\sigma^2 \coloneq& \mathbb{V}[Y]|_{h=0} = \int_0^T \mathbb{E}\left[ d(t)d(t')d'(\tau)d'(\tau') \right] Q(t-\tau)Q(t'-\tau') d\tau'd\tau dt'dt = \frac{T}{4}\int P(|f|)P'(|f|)|\tilde{Q}(f)|^2 df,
	\end{align}
\end{widetext}
where $P_a$ is the noise power spectrum of pulsar $a$.

\subsection{Gravitational wave background}
The GWB can be described by the power spectrum
\begin{equation}
\mathbb{E}[\tilde{h}_A^*(f,\hat{\Omega})\tilde{h}_{A'}(f',\hat{\Omega}')] = \delta(f-f') \frac{1}{4\pi}\delta(\hat{\Omega}-\hat{\Omega}') \delta_{AA'} \frac{1}{2}S_h^{A'}(|f'|).
\end{equation}

To reexpress the correlated signal in terms of the power spectral density of the GWB we first have to Fourier transform it:
\begin{align}
&\mu = \int_0^T \mathbb{E}\left[ r(t,\vec{\lambda})r'(t',\vec{\lambda'}) \right] \int\tilde{Q}(f)e^{-2\pi\mathrm{i}f(t-t')}df dt'dt \notag \\
&= \int \mathbb{E}\left[ \underset{\tilde{r}^*(f)}{\underbrace{\int_0^Tr(t,\vec{\lambda})e^{-2\pi\mathrm{i}ft}dt}} \underset{\tilde{r}'(f)}{\underbrace{\int_0^Tr'(t',\vec{\lambda}')e^{2\pi\mathrm{i}ft'}dt'}} \right] \tilde{Q}(f) df.
\end{align}

The residual is defined as the time integral of the measured redshift $z$,
\begin{equation}
r(t) = \int_0^t z(\tau) d\tau;
\end{equation}
therefore, its Fourier transform is given by
\begin{align}\label{eq: FT[r]}
	\tilde{r}(f) =& \int_0^T \underset{r(t)}{\underbrace{\int_0^t z(\tau) d\tau}} e^{2\pi\mathrm{i}ft} dt \notag \\
	=& -\frac{1}{2\pi\mathrm{i}}\int \frac{\tilde{z}(\nu)}{\nu} \int_0^Te^{-2\pi\mathrm{i}(\nu-f)t}dt d\nu \notag \\
	&+ \frac{1}{2\pi\mathrm{i}}\int\frac{\tilde{z}(\nu)}{\nu}d\nu
		\underset{\approx 0, \textit{ $ $ } fT \gg 1}{\underbrace{\int_0^Te^{2\pi\mathrm{i}ft}dt}} \notag \\
	=& -\frac{1}{2\pi\mathrm{i}}\int \frac{\tilde{z}(\nu)}{\nu}\delta_T(\nu-f)d\nu.
\end{align}
As seen in the proof of $\lim_{\omega\to\infty} \int_a^b f(x) e^{\mathrm{i}\omega x} dx = 0$ in~\eqref{eq.: sw approx}, only the two slices at the boundaries contribute to the integral, since the function $f(x)=1$ is constant in this case. If $fT \gg 2\pi$, then the integrand cancels on the largest part of the integration domain, and thus the integral is small compared to $T$, whereas the first integral is of order $T$ ($\delta_T(0)=T$). Therefore, the contribution of the second integral can be neglected.\\

To use the power spectrum, we need to express $\tilde{z}$ in terms of $\tilde{h}$. To simplify the calculation, we express the frequency in terms of the angular frequency of the gravitational waves, $2\pi f = \omega$,
\begin{align}
	\tilde{z}&(\omega) = \int_{-\frac{T}{2}}^{\frac{T}{2}} z_P(t) e^{\mathrm{i}\omega t} dt \notag \\
	=& \int_{-\frac{T}{2}}^{\frac{T}{2}} \left\lbrace \frac{F^A \Delta h_A(t)}{1+\gamma}
		- \frac{\omega T_a}{2}\frac{F^A \Delta \dot{h}_A(t)}{1+\gamma} + \frac{\delta\theta(t)}{2\pi}
		\right\rbrace e^{\mathrm{i}\omega t} dt \notag \\
	=& F^A \int_{-\frac{T}{2}}^{\frac{T}{2}} \frac{h_A(t)-h_A(t-\frac{L}{c}[1+\gamma])}{1+\gamma}
		e^{\mathrm{i}\omega t} dt \notag \\
	&- \frac{\omega T_a}{2} F^A\int_{-\frac{T}{2}}^{\frac{T}{2}} \frac{\Delta \dot{h}_A(t)}{1+\gamma}
		e^{\mathrm{i}\omega t} dt + \int_{-\frac{T}{2}}^{\frac{T}{2}} \frac{\delta\theta(t)}{2\pi} e^{\mathrm{i}\omega t} dt.
\end{align}
A gravitational plane wave can be described as
\begin{equation}
h_A(t,\vec{x}) = \frac{1}{2\pi}\int \tilde{h}_A(\omega)e^{-\mathrm{i}\omega\left(t-\frac{\hat{\Omega}\cdot\vec{x}}{c}\right)} d\omega.
\end{equation}
And as derived in the previous paper~\cite{PTA1}, the amplitude difference is given by
\begin{align}
	&\Delta h_A(t) = h_A(t) - h_A(t_a) \notag \\
	&= \frac{1}{2\pi}\int \tilde{h}_A(\omega)[1-e^{\mathrm{i}\omega\tau}]e^{-\mathrm{i}\omega t} d\omega,
\end{align}
where $t_a = t - \tau$ is the time, retarded by the retardation time $\tau = \frac{L}{c}[1+\gamma]$.\\

We calculate the Fourier transform of the redshift for the leading-order term:
\begin{align}
	\tilde{z}(\omega) &\approx F^A\int \tilde{h}_A(f')\frac{1-e^{2\pi\mathrm{i}f'\tau}}{1+\gamma}
		\underbrace{\int_{-\frac{T}{2}}^{\frac{T}{2}} e^{-2\pi\mathrm{i}(f'-f)} dt}_{\delta_T(f'-f)} df' \notag \\
	&\approx F^A\tilde{h}_A(\omega)\frac{1-e^{\mathrm{i}\omega\tau}}{1+\gamma}.
\end{align}
\begin{widetext}
	Plugging this expression back into the correlation signal, we get
	\begin{align}
		\mu &= \frac{1}{4\pi^2}\int \frac{1}{\nu\nu'}\mathbb{E}\left[ \tilde{z}^*(\nu)\tilde{z}'(\nu') \right] 
			\delta_T(\nu-f)\delta_T(\nu'-f) d\nu'd\nu \tilde{Q}(f) df \notag \\
		&= \frac{1}{4\pi^2}\sum_A\int \frac{1}{2\nu^2}S_h^A(\nu)\delta_T^2(\nu-f)
			\frac{1}{4\pi}\int_{\mathbb{S}^2} F^A(\hat{\Omega})F'^A(\hat{\Omega})
			\frac{1-e^{2\pi\mathrm{i}\nu\tau}}{1+\gamma}\frac{1-e^{-2\pi\mathrm{i}\nu\tau'}}{1+\gamma'} d\hat{\Omega}\,
			d\nu\, \tilde{Q}(f) df  \notag \\
		&= \frac{T}{24\pi^2}\int \frac{1}{f^2}\left[\sum_MS_h^M(f)\Gamma_M\right]\tilde{Q}(f) df,
	\end{align}
\end{widetext}
where the power spectrum of a polarization mode is defined as the sum of the power spectra of both polarizations of that mode, $S_h^M \coloneq S_h^{M_1} + S_h^{M_2}$, and the overlap reduction functions are given by:
\begin{align}
	\Gamma_M \coloneq& \beta\int_{\mathbb{S}} 
		\left(F^{M_1}(\hat{\Omega})F'^{M_1}(\hat{\Omega}) + F^{M_2}(\hat{\Omega})
		F'^{M_2}(\hat{\Omega}) \right) \notag \\
	&\qquad \cdot \frac{1-e^{\mathrm{i}\frac{L\omega}{c}[1+\gamma]}}{1+\gamma}\frac{1-e^{-\mathrm{i}\frac{L\omega}{c}[1+\gamma']}}{1+\gamma'} d\hat{\Omega},
\end{align}
where $\beta = \frac{3}{4\pi}$ is a normalization factor, which we chose to adopt from Ref. \cite{Anholm2009} for ease of comparison.\\

The overlap reduction function for the tensor mode is also called Hellings \& Downs curve. It is usually assumed, that the exponential terms can be neglected. We will calculate this overlap reduction function up to first order in $h$ and zeroth order in $h\omega T_a$ in the next section (Sec.~\ref{sec: Gamma}):
\begin{align}\label{eq.: GammaT}
	\Gamma_T & = \beta\int_{\mathbb{S}}
		\left(F^+(\hat{\Omega})F'^+(\hat{\Omega}) + F^\times(\hat{\Omega})F'^\times(\hat{\Omega}) \right)
		\notag \\
	&\qquad \cdot \frac{1-e^{\mathrm{i}\frac{L\omega}{c}[1+\gamma]}}{1+\gamma}
		\frac{1-e^{-\mathrm{i}\frac{L\omega}{c}[1+\gamma']}}{1+\gamma'} d\hat{\Omega}.
\end{align}

To maximize the scalar product
\begin{equation}
(A|B) \coloneq \int \tilde{A}^*(f)\tilde{B}(f)P(f)P'(f) df,
\end{equation}
we choose the filter function to be
\begin{equation}
\tilde{Q}(f) = \frac{\sum_M S_h^M(f)\Gamma^M}{f^2P(f)P'(f)}.
\end{equation}

Rewriting signal and variance in terms of the scalar product and filter function, we get
\begin{align}
\mu = \frac{T}{8\pi^2}(Q|Q),		&&	\sigma^2 = \frac{T}{4}(Q|Q).
\end{align}
So, the signal-to-noise ratio is given by
\begin{equation}
SNR = \frac{1}{4\pi^2}\sqrt{T(Q|Q)} = \frac{1}{4\pi^2}\sqrt{T\int \frac{|\sum_M S_h^M(f)\Gamma^M|^2}{f^4P(f)P'(f)} df}.
\end{equation}

For more information about the matched filtering technique, see Appendix~\ref{sec: MatchedFilter}.

\section{Overlap reduction function}\label{sec: Gamma}
As derived in the previous section, the overlap reduction function for the tensor mode for PTAs using natural units $c = 1$ is given by
\begin{widetext}
	\begin{equation}
		\Gamma_T = \beta\sum_{A\in \{+,\times\}}\int_{\mathbb{S}^2} F^A(\hat{\Omega}) F'^A(\hat{\Omega})
			\frac{1-e^{\mathrm{i}L\omega[1+\gamma]}}{1+\gamma} \frac{1-e^{-\mathrm{i}L'\omega[1+\gamma']}}{1+\gamma'} d\hat{\Omega}.
	\end{equation}
\end{widetext}

To calculate this integral, we use the residue theorem on the $\varphi$-integral:
\begin{widetext}
	\begin{align}
		\Gamma_T &= \sum_{A\in \{+,\times\}}\int_0^{2\pi}\int_0^\pi F^A(\hat{\Omega}) F'^A(\hat{\Omega}) \Delta h(\hat{\Omega}) \Delta h'(\hat{\Omega}) \sin\theta d\theta\,d\varphi
			= \int_0^\pi \oint_{C_1} f(z) dz\,d\theta = 2\pi\mathrm{i} \int_0^\pi \text{Res}[f(z),0] d\theta, \notag \\
		\vphantom{a}\notag\\
		f(z) &= \frac{\left(F^A {F'}^A\Delta h\Delta h'\right)(\theta,z)}{\mathrm{i}z}\sin\theta,
			\quad \Delta h = \frac{1-e^{\mathrm{i}L\omega[1+\gamma]}}{1+\gamma}, \quad z = e^{\mathrm{i}\varphi}.
	\end{align}
\end{widetext}

The poles from the denominators $\gamma = -1$ and $\gamma' = -1$ are canceled by the nominators. Thus, the only pole left is the one at zero. To find the residue, we write the Laurent series of $f$ around zero and read out the $a_{-1}$-term:
\begin{align}
f(z) = \sum_{n\in\mathbb{Z}}a_n z^n, \quad Res[f(z),0] = a_{-1}.
\end{align}

We chose our reference frame such that the pulsar $a$ is located at $\hat{x} = (1,0,0)$ and the second pulsar $a'$ at $\hat{x}' = (\cos\phi,\sin\phi,0)$. Then, complexified pattern functions form a "generalized" polynomial in z. We write the exponential terms as a power series and use the geometric series to calculate the Laurent series of the $\frac{1}{1+\gamma}$-terms. We collect the (generalized) polynomial part in $P(z) \eqcolon z\sum_{n=-4}^4 b_n z^n$; \quad $b_{-n} = \bar{b}_n$\\
The powers in $P$ are going to shift our $a_{-1}$ term up and down the remaining series, which is a series multiplication of the exponential series part $E$ and the geometric series part $G$. The exponential part contains positive and negative powers of $z$,
\begin{widetext}
	\begin{align}
		E(z) =& \sum_{n=0}^\infty \frac{1}{n!} \left(\frac{\mathrm{i}\omega}{2z}\sin\theta\right)^n
			\left\lbrace -e^{\mathrm{i}L\omega}L^n\left(1+z^2\right)^n - e^{-\mathrm{i}L'\omega}(-L')^n\left(e^{\mathrm{i}\phi}
			+ e^{-\mathrm{i}\phi}z^2\right)^n \right. \notag \\
		 &\left.\qquad\qquad\qquad\qquad\quad +\ e^{\mathrm{i}(L-L')\omega}\left( L-L'e^{\mathrm{i}\phi}
			 + \left[L-L'e^{-\mathrm{i}\phi}\right]z^2 \right)^n\right\rbrace,
	\end{align}
	while the geometric series part only has positive powers of $z$,
	\begin{align}
		G(z) &= \sum _{n=0}^\infty \sum _{k=0}^n (-1)^n e^{-\mathrm{i}k\phi} \left(\frac{2}{\sin\theta}
			+ e^{-\mathrm{i}\phi}z\right)^k \left(\frac{2}{\sin\theta} + z\right)^{n-k} z^n.
	\end{align}
\end{widetext}

With these definitions, we can write $f$ as
\begin{equation}
f(z) = -\frac{\mathrm{i}e^{-\mathrm{i}\phi}}{8\sin\theta} P(z) \left( 1 + E(z) \vphantom{\sqrt{2}}\right)G(z),
\end{equation}
use the Cauchy product to write $f$ as a series of nested sums, and sort the powers of $z$ to read out the coefficient $a_{-1}$ to calculate the $\varphi$-integral,
\begin{align}\label{eq.: phi-int}
	&\int_0^{2\pi} F^A(\hat{\Omega}) F'^A(\hat{\Omega}) \Delta h(\hat{\Omega}) \Delta h'(\hat{\Omega}) \sin\theta\, d\varphi
		= \oint_{C_1} f(z) dz \notag \\
	&= 2\pi\mathrm{i}\,a_{-1},
\end{align}
which results in a series of nested sums of Bessel functions with $\sin\theta$ as arguments. We use the linearity of integration and the fact that the resulting series converges absolutely to pull the $\theta$-integration in to the innermost terms, which are powers of sines and cosines,
\begin{align}
\int_0^{\pi } \cos ^m(\theta ) \sin ^n(\theta ) \, d\theta
   =\frac{\left(1+(-1)^m\right) \Gamma \left(\frac{1+m}{2}\right) \Gamma
   \left(\frac{1+n}{2}\right)}{2 \Gamma \left(\frac{1}{2} (2+m+n)\right)},
\end{align}
for $\Re(m)$, and $\Re(n) > -1$.\\

We finally obtain an analytic expression for the overlap reduction function for the tensor mode in form of an absolutely convergent series of nested sums:

We finally obtain an analytic expression in form of a series of nested sums for the overlap reduction function without short wavelength approximation. This series is absolutely convergent and thus the expression is well defined for all angles $\phi\in[0,\pi]$ (for the proof, refer to the Supplemental Material),
\begin{widetext}
	\begin{align}\label{eq: GT-first}
		\Gamma_T =& \frac{\pi}{3} (3+\cos\phi) - \frac{\pi}{8} e^{\mathrm{i}\phi} \left[ \left(1+e^{-2\mathrm{i}\phi}\right)
			h^+_{0,0,3,0,0}(\phi) + 2\mathrm{i}\left(1+e^{-\mathrm{i}\phi}\right) h^-_{0,0,2,0,1}(\phi) + h^-_{0,0,3,0,2}(\phi) \right] \notag \\
		 &+ \frac{\pi}{4} e^{-\mathrm{i}\phi} \left(\sum_{j=0}^1 (-2\mathrm{i})^j \sum_{k=0}^j \sum_{m=0}^k \sum_{l=0}^{j-k}
			\left(\frac{\mathrm{i}}{2}\right)^{m+l} \binom{k}{m} \binom{j-k}{l} e^{-\mathrm{i}(k+m)\phi} \right. \notag \\
		 &\cdot \left\lbrace -\frac{e^{-2\mathrm{i}\phi}}{2} h^+_{j,m+l,3,0,6}(\phi) + \left(1+e^{2\mathrm{i}\phi}\right) f_{j,m+l}(\phi)
			+ \left(1+e^{-2\mathrm{i}\phi}\right) \left(h^+_{j,m+l,-1,4,4}(\phi) - h^+_{j,m+l,-1,0,4}(\phi)\right) \right.\notag\\
		 &\left.\qquad - \cos(2\phi) \left( h^+_{j,m+l,-1,0,2}(\phi) + 6h^+_{j,m+l,-1,2,2}(\phi) + h^+_{j,m+l,-1,4,2}(\phi)\right)
			 - 2h^+_{j,m+l,3,0,2}(\phi) \vphantom{\frac{1}{2}}\right\rbrace \notag\\
		 &+ \sum_{j=2}^\infty (-2\mathrm{i})^j \sum_{k=0}^j \sum_{m=0}^k \sum_{l=0}^{j-k} \left(\frac{\mathrm{i}}{2}\right)^{m+l}
			 \binom{k}{m} \binom{j-k}{l} e^{-\mathrm{i}(k+m)\phi} \notag \\
		 &\cdot \left\lbrace -\frac{1}{2} \left( e^{2\mathrm{i}\phi} h^+_{j,m+l,3,0,-2}(\phi)
			 + e^{-2\mathrm{i}\phi} h^+_{j,m+l,3,0,6}(\phi) \right) + \left(1+e^{2\mathrm{i}\phi}\right) f_{j,m+l}(\phi) \right.\notag\\
		 &\qquad + \left(e^{-2\mathrm{i}\phi}+1\right) \left(h^+_{j,m+l,-1,4,4}(\phi) - h^+_{j,m+l,-1,0,4}(\phi)\right) \notag\\
		 &\left.\left.\qquad - \cos(2\phi) \left(h^+_{j,m+l,-1,0,2}(\phi) + 6h^+_{j,m+l,-1,2,2}(\phi) + h^+_{j,m+l,-1,4,2}(\phi)\right)
			 - 2h^+_{j,m+l,3,0,2}(\phi) \vphantom{\frac{1}{2}}\right\rbrace \vphantom{\sum_j^1}\right),
	\end{align}
where we used the following definitions:
	\begin{align}\label{eq.: def h}
		h^\pm&_{j,b,s,t,N}(\phi) \coloneq \int_0^\pi \sin^{-j+b+s}\theta \cos^t\theta g_{j+b+N}(\theta)\, d\theta \\
		=& \Gamma\left(\frac{t+1}{2}\right) \Gamma\left(k+\frac{s+N+1}{2}\right) \left[
			e^{\mathrm{i}(L-L')\omega} \left(\frac{1}{2}(L-e^{\pm\mathrm{i}\phi}L')\omega\right)^{j+b+N} \right. \notag \\
		&\qquad \times\, _1\tilde{F}_2\left(b+\frac{s+N+1}{2};j+b+N+1,b+\frac{s+t+N}{2}+1;-(L^2+L'^2-2LL'\cos\phi)\frac{\omega^2}{4}
			\right) \notag \\
		&- e^{\mathrm{i}L\omega} \left(\frac{L\omega}{2}\right)^{j+b+N}\, 
			_1\tilde{F}_2\left(b+\frac{s+N+1}{2};j+b+N+1,b+\frac{s+t+N}{2}+1;-\left(\frac{L\omega}{2}\right)^2\right) \notag \\
		&\left. - e^{\mathrm{i}(\pm(j+b+N)\phi-L'\omega)} \left(-\frac{L'\omega}{2}\right)^{j+b+N}\, 
			_1\tilde{F}_2\left(b+\frac{s+N+1}{2};j+b+N+1,b+\frac{s+t+N}{2}+1;-\left(\frac{L'\omega}{2}\right)^2\right) \right] \notag
	\end{align}
and
	\begin{align}
		f_{j,b}(\phi) \coloneq& \int_0^{\pi} \sin^{-j+b-1}\theta \left(\cos^4\theta-1\right)g^+_{j+b}(\theta)\, d\theta \\
		=& \sqrt{\pi}2^{-j-b}\Gamma(b+1)\Gamma(b+3) \left[ e^{\mathrm{i}L\omega}(L\omega)^{j+b}\, 
			_2\tilde{F}_3\left(b+1,b+3;j+b+1,b+2,b+\frac{5}{2};-\left(\frac{L\omega}{2}\right)^2\right) \right. \notag \\
		&+ e^{\mathrm{i}((j+b)\phi-L'\omega)}(-L'\omega)^{j+b} \,
			_2\tilde{F}_3\left(b+1,b+3;j+b+1,b+2,b+\frac{5}{2};-\left(\frac{L'\omega}{2}\right)^2\right) \notag \\
		&\left. - e^{\mathrm{i}(L-L')\omega} \left((L-e^{i\phi}L')\omega\right)^{j+b}\,
			_2\tilde{F}_3\left(b+1,b+3;j+b+1,b+2,b+\frac{5}{2};-(L^2+L'^2-2LL'\cos\phi)\frac{\omega^2}{4}\right)
			\vphantom{\left(\frac{L}{2}\right)^2}\right]. \notag
	\end{align}
\end{widetext}

\section{Optimization of the overlap reduction function}\label{sec: Optimization}
If we evaluate~\eqref{eq: GT-first} and add the terms in that sequence, we would not add the largest terms first, and thus in this way, the numerical convergence is computationally inefficient, since one would add a lot of small and potentially irrelevant terms, before one adds the next larger term.\\
Since the series is absolutely convergent, which we show in the Supplemental Material, we can reorder the series such that the largest terms are added first. This also allowed us to simplify some terms and in doing so get rid of two sums. With that, we arrive at the final expression,
\begin{widetext}
	\begin{align}
		\Gamma_T =& \frac{\pi}{3} (3+\cos\phi) - \frac{\pi}{8} e^{\mathrm{i}\phi} \left[ \left(1+e^{-2\mathrm{i}\phi}\right)
			h^+_{0,0,3,0,0}(\phi) + 2\mathrm{i}\left(1+e^{-\mathrm{i}\phi}\right) h^-_{0,0,2,0,1}(\phi) + h^-_{0,0,3,0,2}(\phi) \right] \notag \\
		 &+ \frac{\pi}{4} e^{-\mathrm{i}\phi} \left(\sum_{j=0}^1 (-2\mathrm{i})^j \sum_{k=0}^j \sum_{m=0}^k \sum_{l=0}^{j-k}
			\left(\frac{\mathrm{i}}{2}\right)^{m+l} \binom{k}{m} \binom{j-k}{l} e^{-\mathrm{i}(k+m)\phi} \right. \notag \\
		 &\cdot \left\lbrace -\frac{e^{-2\mathrm{i}\phi}}{2} h^+_{j,m+l,3,0,6}(\phi) + \left(1+e^{2\mathrm{i}\phi}\right) f_{j,m+l}(\phi)
			+ \left(1+e^{-2\mathrm{i}\phi}\right) \left(h^+_{j,m+l,-1,4,4}(\phi) - h^+_{j,m+l,-1,0,4}(\phi)\right) \right.\notag\\
		 &\left.\qquad - \cos(2\phi) \left( h^+_{j,m+l,-1,0,2}(\phi) + 6h^+_{j,m+l,-1,2,2}(\phi) + h^+_{j,m+l,-1,4,2}(\phi)\right)
			 - 2h^+_{j,m+l,3,0,2}(\phi) \vphantom{\frac{1}{2}}\right\rbrace \notag\\
		&+ \sum_{a=2}^\infty (-2\mathrm{i})^a \sum_{b=0}^{\left\lfloor\frac{a}{2}\right\rfloor} \left(-\frac{1}{4}\right)^b
    	\left\lbrace -\frac{1}{2}\left(e^{2\mathrm{i}\phi} h^+_{a,b,3,0,-2}(\phi) + e^{-2\mathrm{i}\phi} h^+_{a,b,3,0,6}(\phi)\right)
    	+ \left(1+e^{2\mathrm{i}\phi}\right) f_{a,b}(\phi) \right.\notag\\
    	&\qquad + \left(e^{-2\mathrm{i}\phi}+1\right) \left(h^+_{a,b,-1,4,4}(\phi) - h^+_{a,b,-1,0,4}(\phi)\right)
    		- \cos(2\phi)\left(h^+_{a,b,-1,0,2}(\phi) + 6h^+_{a,b,-1,2,2}(\phi) + h^+_{a,b,-1,4,2}(\phi)\right) \notag\\
    	&\left.\left.\qquad- 2 h^+_{a,b,3,0,2}(\phi) \vphantom{\frac{1}{2}}\right\rbrace \sum _{k=0}^{a-b}
    	\begin{cases}
    		e^{-\mathrm{i}k\phi} \binom{a-b-k}{b}\, _2F_1\left(-b,-k;1+a-2b-k;e^{-\mathrm{i}\phi}\right) & a \geq 2b + k \\
    		e^{\mathrm{i}(a-2(b+k))\phi} \binom{k}{a-2b}\, _2F_1\left(-a+2b,-a+b+k;1-a+2b+k;e^{-\mathrm{i}\phi}\right) & a < 2b + k
    	\end{cases} \vphantom{\sum_j^1}\right),
	\end{align}
\end{widetext}
where $h^{\pm}_{a,b,s,t,N}(\phi)$ and $f_{a,b}(\phi)$ were redefined by using the substitution
\begin{equation}
	j,\ k,\ m,\ l \quad \to \quad a = j+m+l,\ b = m+l,\ k,\ m.
\end{equation}

\section{Discussion of the Results}\label{sec: Discussion}
\begin{figure}[h!]
    \begin{minipage}{\linewidth}
        \centering\includegraphics[width=\linewidth]{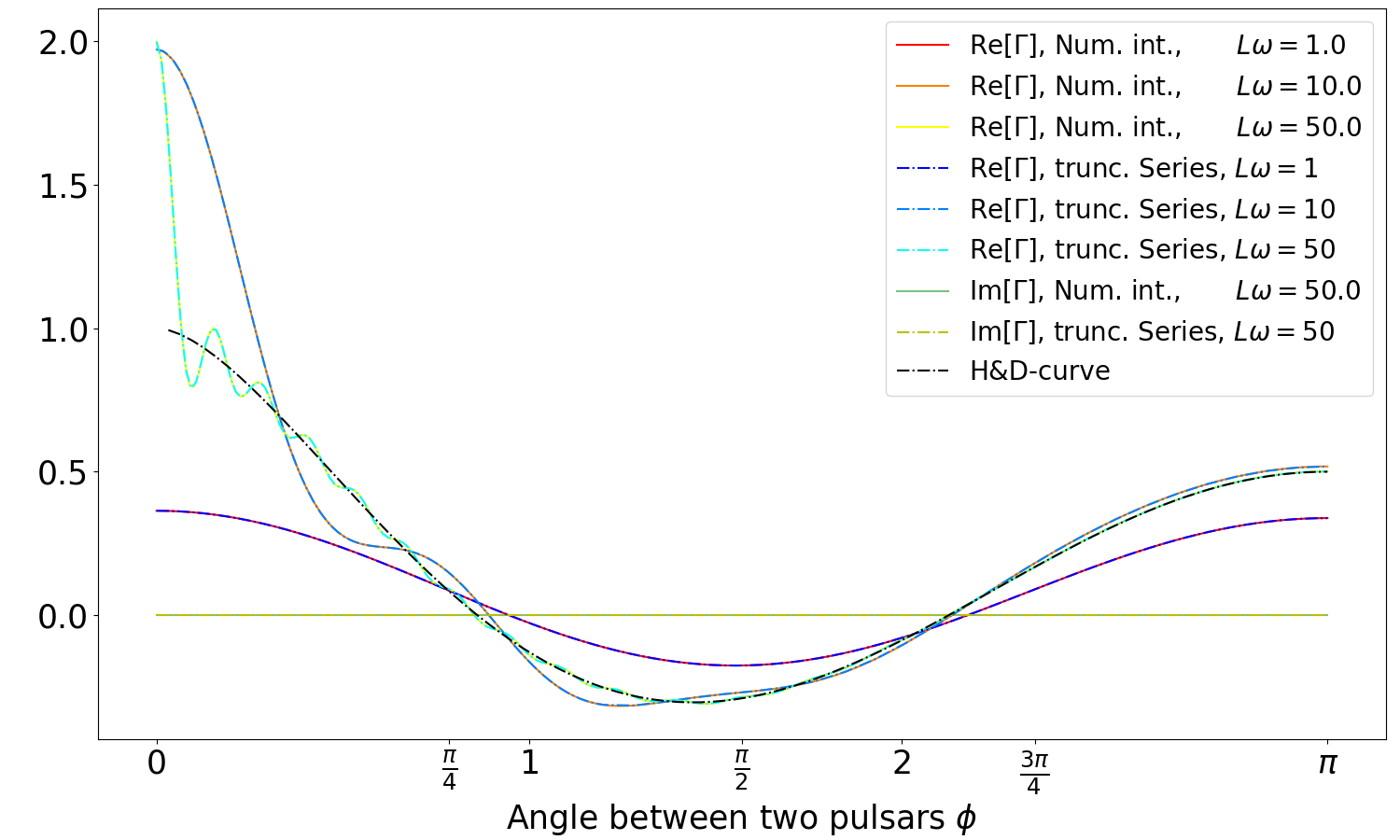}
    \end{minipage}
    \hfill
    \begin{minipage}{\linewidth}
        \centering\includegraphics[width=\linewidth]{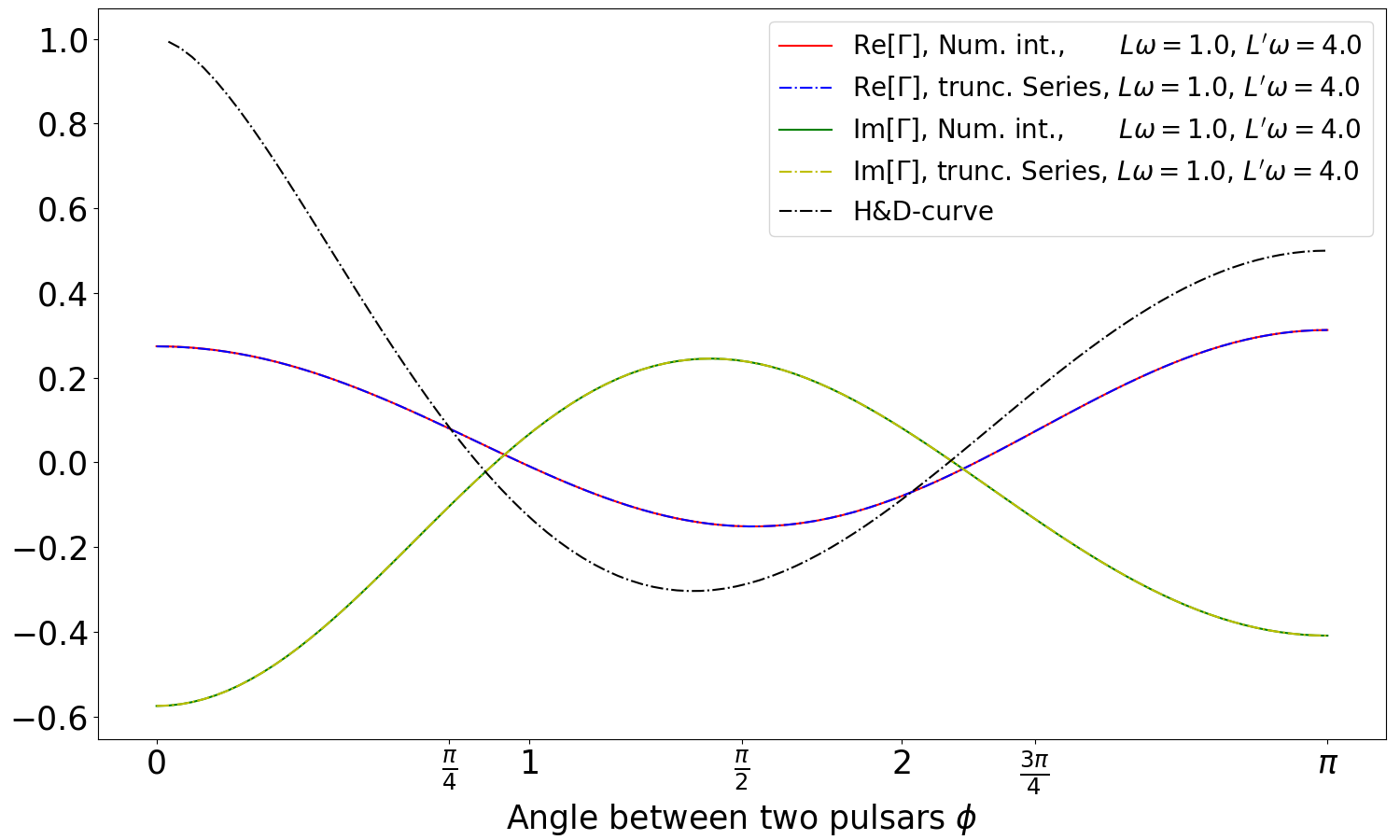}
    \end{minipage}
    \caption{A comparison between numerical integration and evaluation of the truncated analytical series for different values of $L\omega$ and $L'\omega$, in the cases where $L=L'$ (upper) and $L\neq L'$ (lower). The solid lines show the numerical integrations, the dot-dashed lines show the truncated series, and the black dashed line is the Hellings and Downs curve.}\label{fig.: AvsN}
\end{figure}

We assure ourselves, that the numerical evaluation of the truncated series (sum until a suitable cutoff) agrees with a straightforward numerical integration via the trapezoidal rule and compare the analytic series to the approximation obtained by Hellings and Downs.\\
To investigate the dependence on $L\omega$, we compare the results for different values of $L\omega$ for the special case $L= L’$, where the two correlated pulsars are at the same distance and for large distance ratios $\frac{L}{L'}$.\\
Some representative results are plotted in Fig.~\ref{fig.: AvsN}.
The imaginary part always vanishes in the case where $L=L'$ since the only complex part of the integrand is phase differences $(1-e^{\mathrm{i}L\omega[1+\gamma_a]})(1-e^{-\mathrm{i}L'\omega[1+\gamma_{a'}]})$ and in this special case they become complex conjugates of each other and thus the integrand is real. It is not a surprise then, that we observe bigger imaginary parts for larger distance ratios $\frac{L}{L'}$. It might sound a bit strange that an overlap reduction function is complex, but it appears in the SNR only with its absolute value, and thus this physical quantity remains real.\\

\begin{figure}[h!]
    \centering\includegraphics[width=\linewidth]{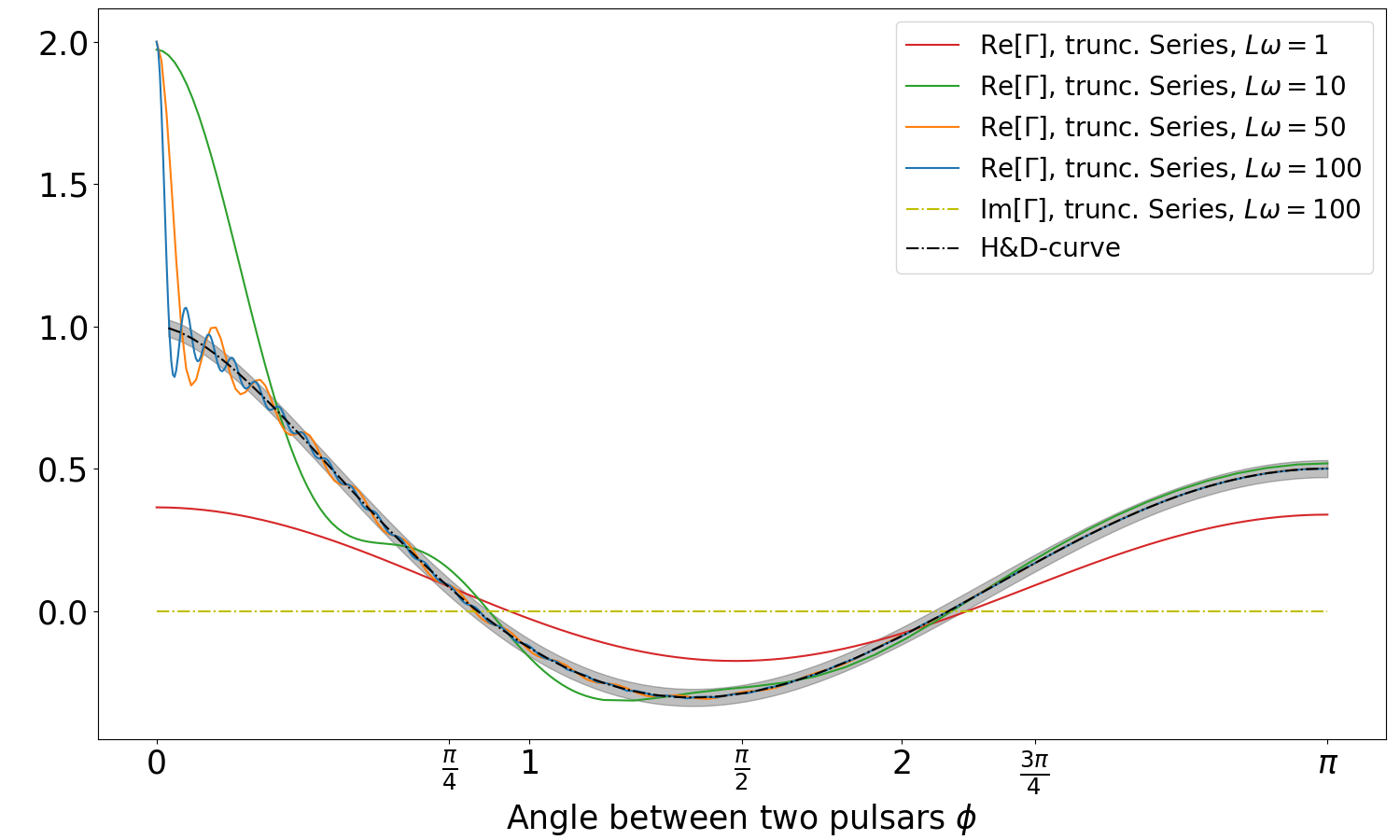}
    \caption{We plot the truncated series for $L\omega$ over 3 orders of magnitude in the special case of $L=L'$, to empirically investigate the convergence for large $L\omega$. We compare our results to the Hellings and Downs curve (black dashed line). The area in gray marks the $3\%$ deviation from the H\&D curve, which is unproblematic for matched filtering.}\label{fig.: Conv}
\end{figure}

\begin{figure}[h!]
\begin{minipage}{\linewidth}
    \centering\includegraphics[width=\linewidth]{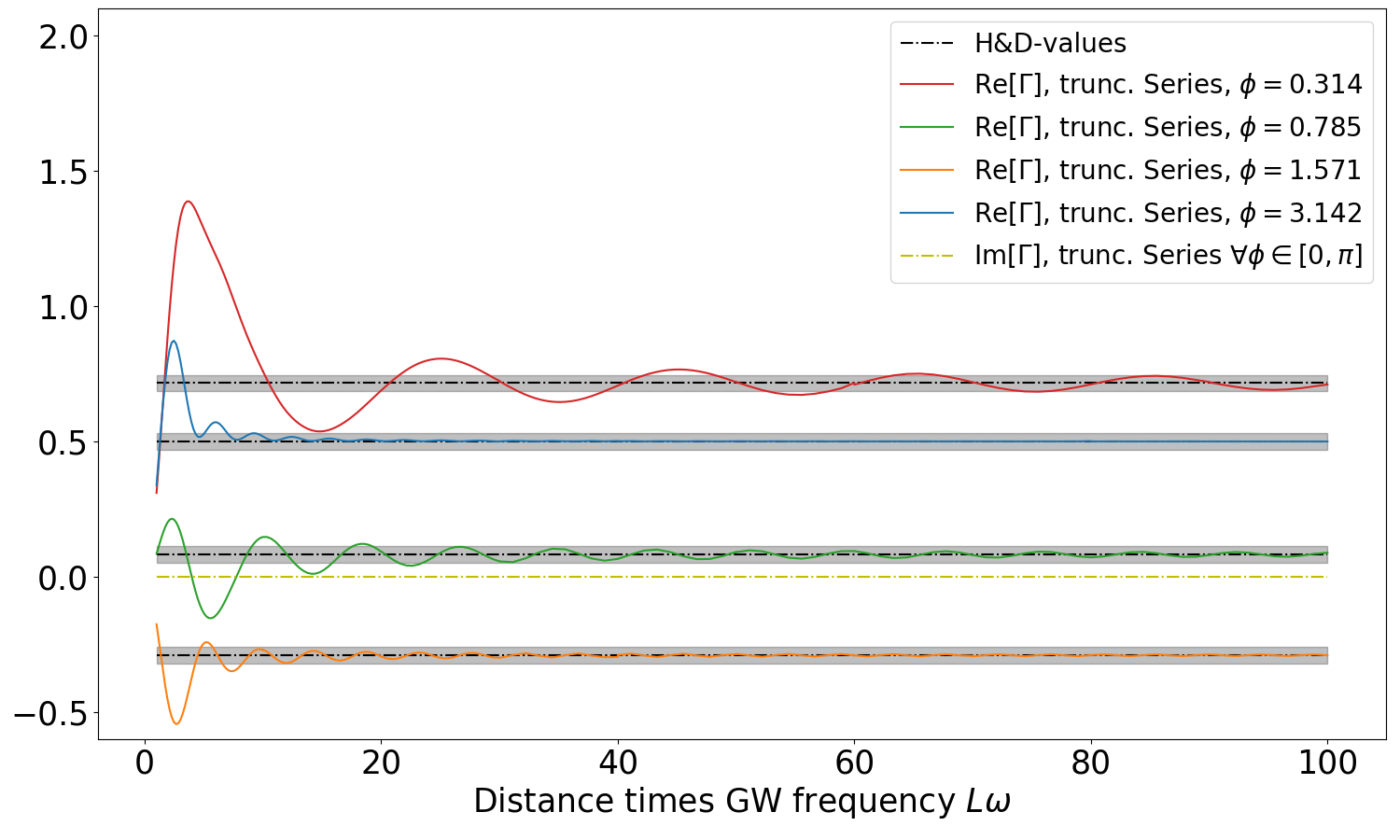}
\end{minipage}
\hfill
\begin{minipage}{\linewidth}
    \centering\includegraphics[width=\linewidth]{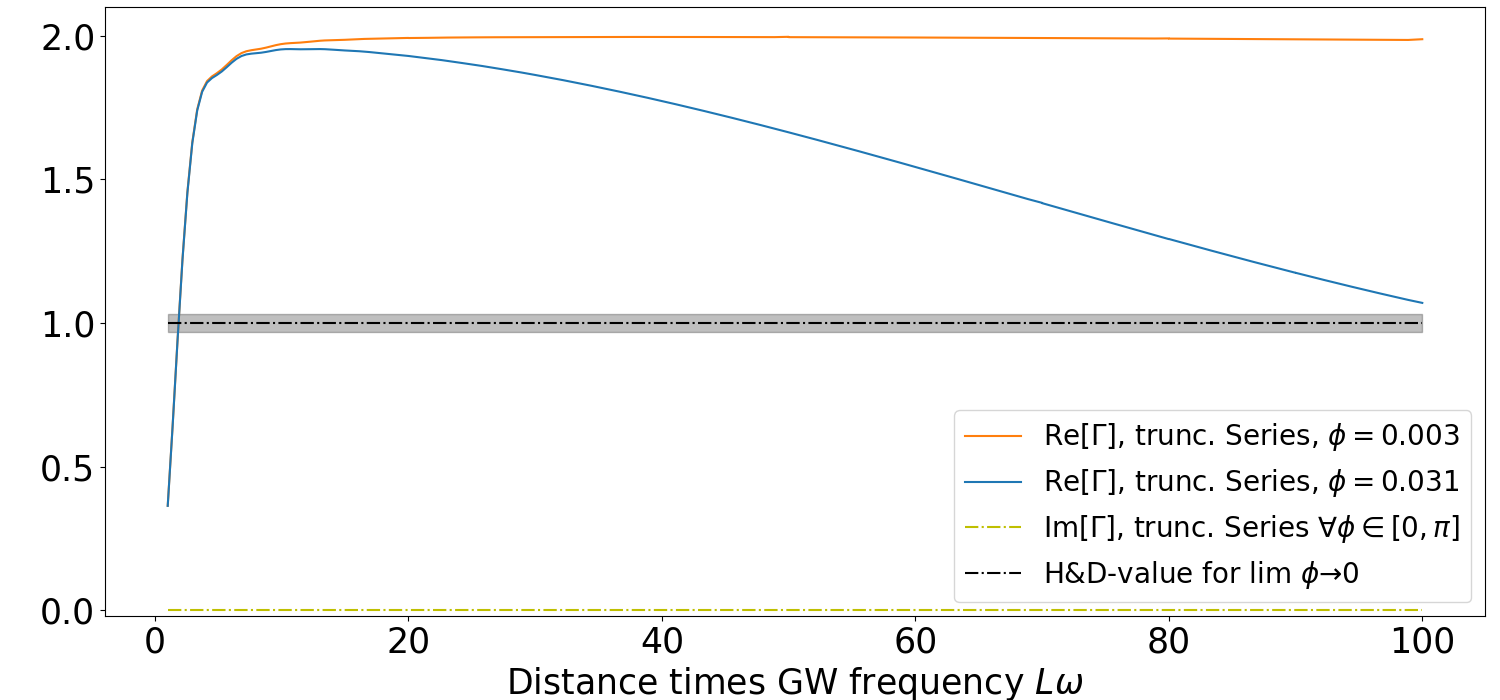}
\end{minipage}
    \caption{In the plot above, we sample the parameter space for the separation angle $\phi$ and plot the overlap reduction function as a function of distance times GW frequency $L\omega$ in the case of $L=L'$. Below, we show the same plot for $\phi$-values close to zero. We compare the two curves to the limit of the H\&D curve for $\phi$ tending to zero.}
    \label{fig.: Lodep_L=L'}
\end{figure}

We investigate the limit of large $L\omega$ in the case $L=L'$ by plotting the truncated series from $L\omega=1$ up to $L\omega=100$, where a evaluation on a laptop is still feasible. After that the numerical evaluation of the three nested sums becomes very expensive even for a single point. Since the inner sums are dependent on the outer ones, further parallelization is a nontrivial task, which we chose to not engage in. It can already be seen from Fig.~\ref{fig.: Conv} that the problematic region around $\phi=0$ becomes smaller for larger $L\omega$ and that the value at $\phi=0$ converges to $2$, contrary to the limit of $\phi\to0$ obtained from the H\&D curve. We also observe, that the function always oscillates in the region $[0,\frac{\pi}{2}]$. A deviation from the signal by $\pm3\%$ would still give good result when doing matched filtering. This behavior becomes even more obvious in Fig.~\ref{fig.: Lodep_L=L'}.\\

We conclude that for large $L\omega$ in the case where both pulsars are at the same distance from Earth $L=L'$ the series converges to the H\&D curve everywhere except of $\phi = 0$ where it converges to 2.\\

\begin{figure}[h!]
    \begin{minipage}{\linewidth}
        \centering\includegraphics[width=\linewidth]{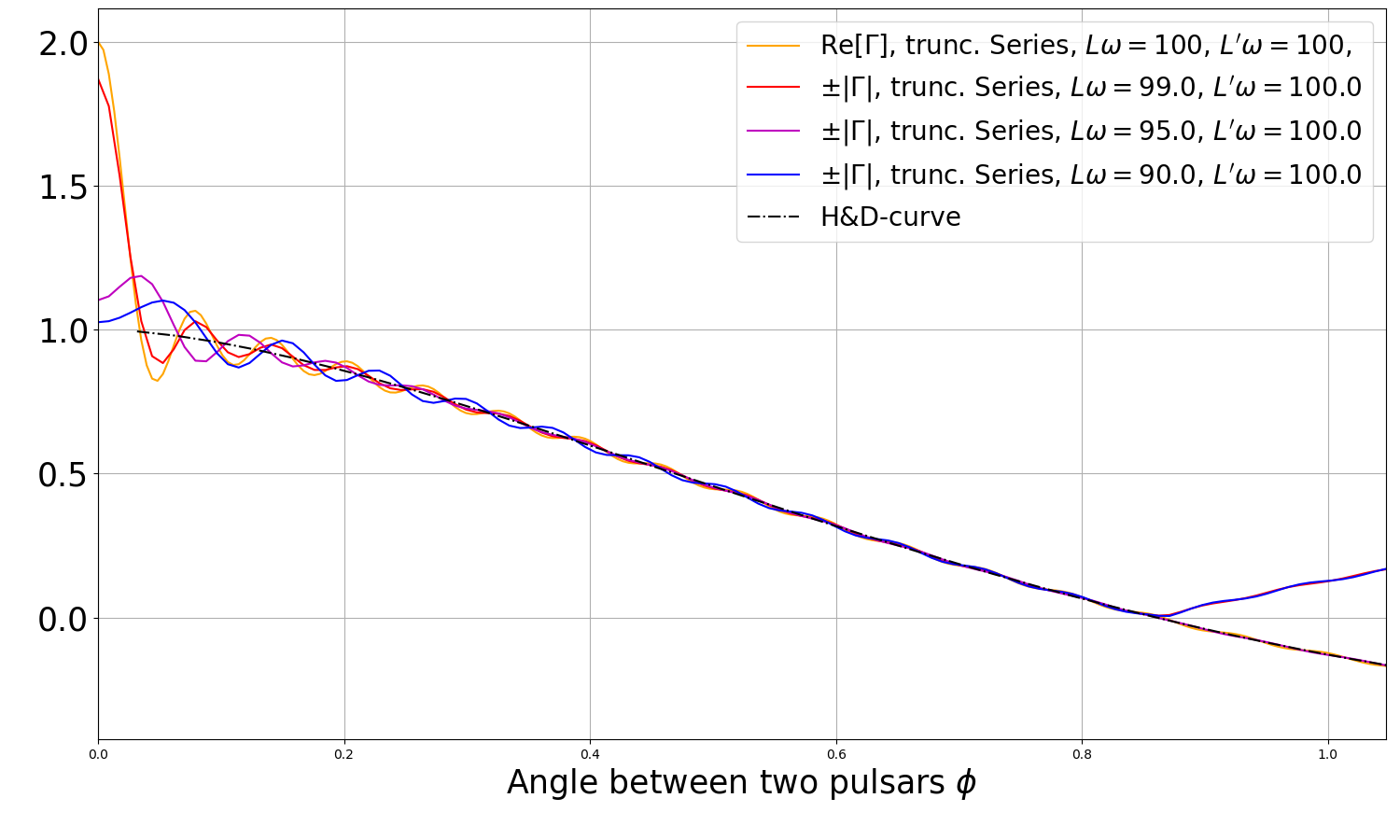}
    \end{minipage}
    \hfill
    \begin{minipage}{\linewidth}
        \centering\includegraphics[width=\linewidth]{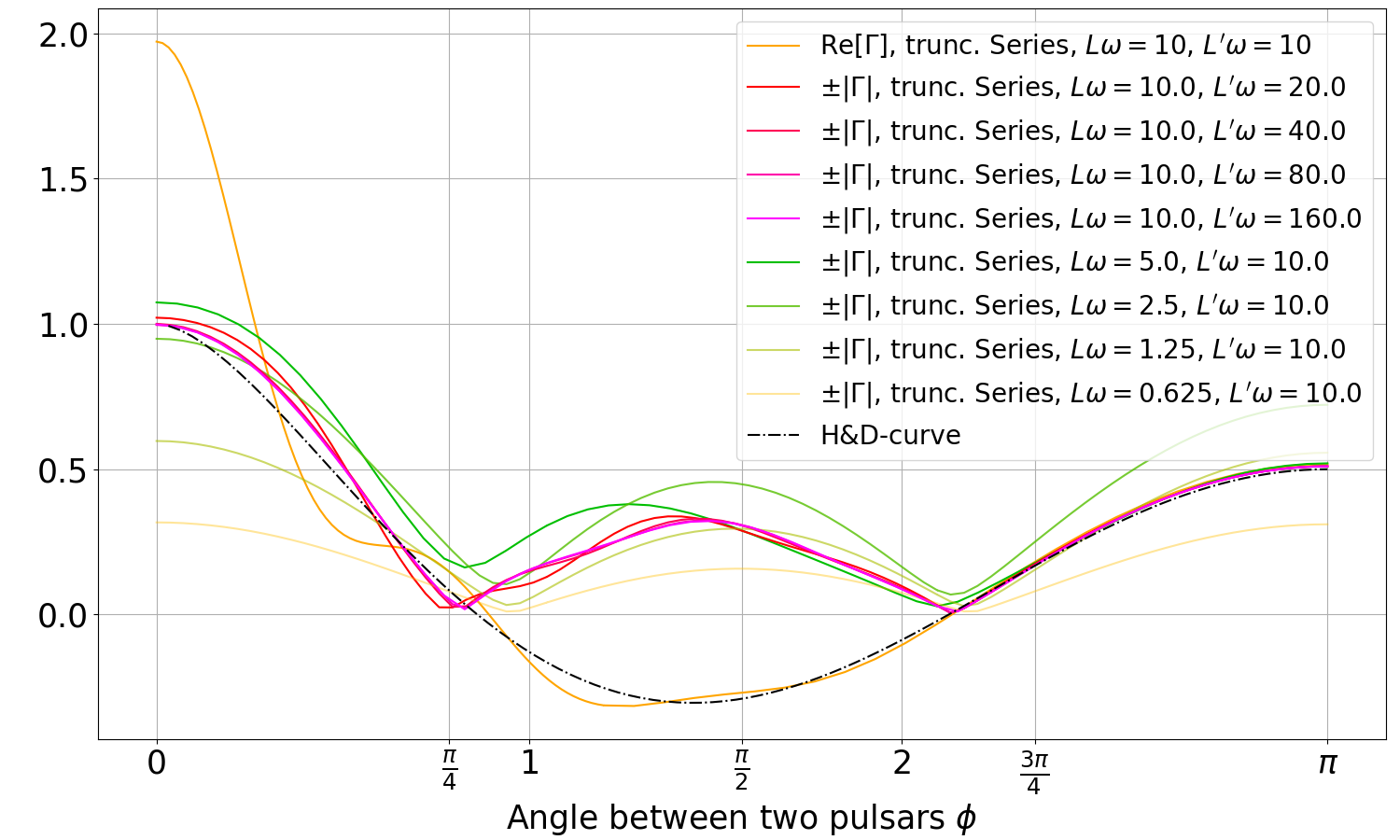}
    \end{minipage}
    \caption{We plot the series for small differences in the pulsar distance $\frac{L}{L'}\approx1$ around $L\omega=100$  in the upper panel and for large ratios in the lower panel. There, we use $L\omega=10$ as a reference value. The curves where $L'\omega$ is below $10$ are the purple ones and the ones, where $L'\omega$ is bigger than $10$ are red. We plot the special case $L=L'$ in orange and the H\&D curve in dashed black for comparison.}\label{fig.: DistRatio}
\end{figure}

Now that we know what happens at different orders of magnitudes of $L\omega$ for equal distances, we are interested in how different pulsar distances affect the result. We observe in Fig.~\ref{fig.: DistRatio} that the correlation at $\phi=0$ decays from $2$ to $1$ over a difference of only $5\%$ in $L\omega$, which then agrees with the limit obtained from H\&D, and that it converges more quickly to the H\&D curve than when $L=L'$. This means that it is only realistic to have a factor of 2 for a double pulsar system. The complex overlap reduction functions do not always cross zero. In the cases in which they do, we change the sign in front of the absolute value for a better comparison with the H\&D curve. Since only the absolute value is relevant for the SNR, the sign choices are inconsequential.
\begin{figure}[h!]
    \centering\includegraphics[width=\linewidth]{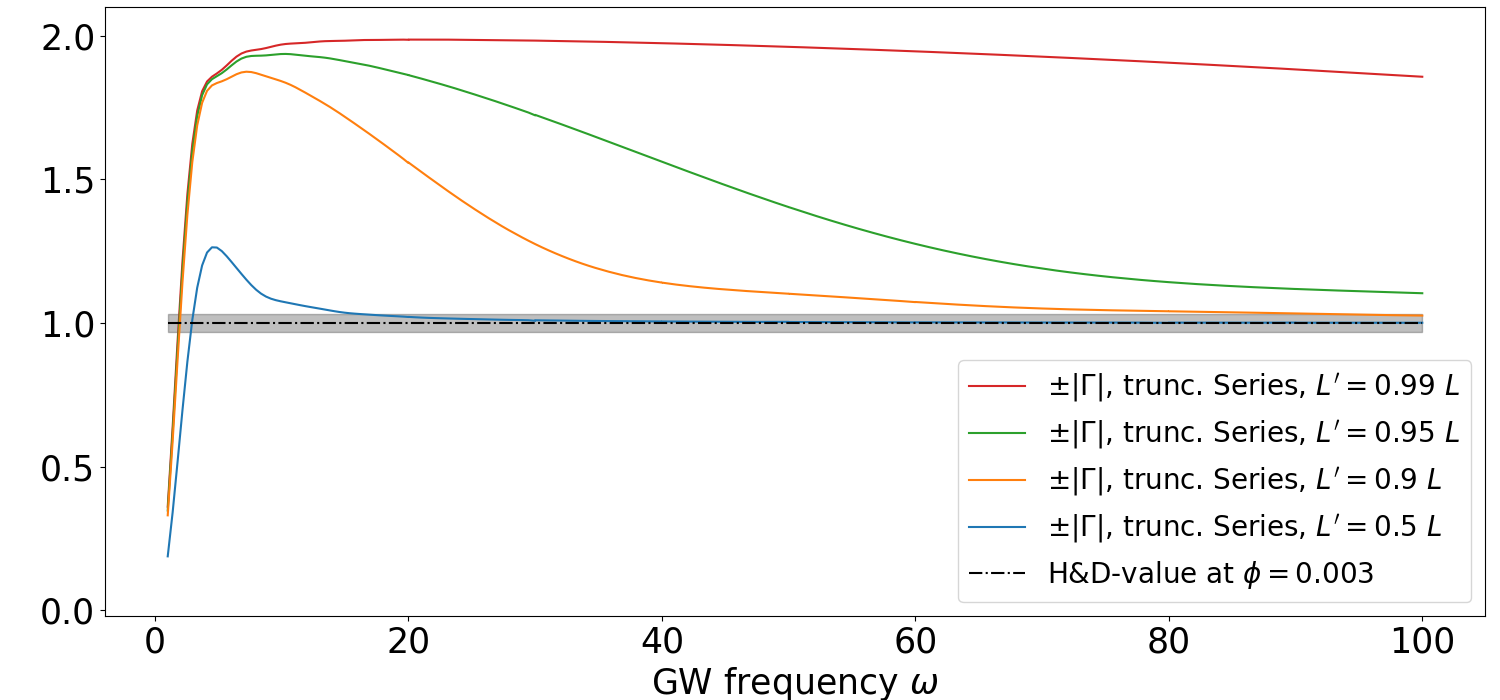}
    \caption{We plot the GW frequency $\omega$ dependence of the overlap reduction function, focusing on distance ratios close to 1, at the separation angle $\phi=\frac{\pi}{1000}$.}
    \label{fig.: odep_L!=L'}
\end{figure}

From the plots above Figs.~\ref{fig.: AvsN},~\ref{fig.: Conv} and ~\ref{fig.: DistRatio},~\ref{fig.: odep_L!=L'}, we see that for higher $L\omega$ the peak at $\phi=0$ and $L=L'$ becomes narrower and the neighborhood around this point where the short wavelength approximation does not apply becomes smaller. This is due to the pulsar term becoming relevant in this part of the parameter space. Thus, we expect that for all current observationally relevant $L\omega$ our result agrees with the Hellings and Downs curve, with the only exception being double pulsar systems for, e.g., Ref. \cite{alma99116780715905508}. Despite the fact that the overlap reduction function depends on the pulsar distances, one does not require precise distance measurements in this frequency regime. We either have a double pulsar system, in which the pulsars orbit each other and they are for all intents and purposes colocated; i.e., both the angular separation and difference in distance are negligible and there is a much larger redshift due to the pulsar orbits. In this case, the overlap reduction function would assume the value $2$. Or the pulsar systems do not form a double pulsar system and are separated by much larger distances, such that they do not affect each others orbit. In this case, for $L\omega$ around $1000$ or higher, the function would assume the value $1$ due to the rapid decay of the peak at such high $L\omega$ values.

\section{Conclusions}\label{sec: Conclusions}
Together with our previous paper \cite{PTA1}, we developed a methodology to calculate the redshift and overlap reduction functions, which is generally applicable for all polarizations and pulsar locations. 
We expect it to work also for higher-order terms in the strain $h$ and gravitational wave frequency times pulsar period $\omega T_a$ and even without any approximation in $\omega T_a$. We gave a semianalytical form of the redshift for a generic GW with all possible (including non GR) polarizations without expansion in $\omega T_a$ and a power series for the overlap reduction function to first order in $h$ and zeroth order in $\omega T_a$ for the tensor mode. (This can be extrapolated to other polarization modes.)\\

The approximation which was used by~\cite{H&D} and described as noise by Ref.~\cite{Anholm2009} is the one we derive in Appendix~\ref{sec: e^ikx}. We find, that this is only justified under the condition Ref.~\eqref{eq.: approx} which cannot be satisfied at $\phi = 0$ where the integrand has a pole but works fine on $\phi\in[\epsilon,2\pi]$ for a large enough $\epsilon$ dependent on the required precision and parameter $L\omega$ one is interested in. The Hellings and Downs curve
\begin{align}
    &\Gamma_0 = 3\left\{\frac{1}{3} + \frac{1-\cos\phi}{2}\left[\ln\frac{1-\cos\phi}{2} - \frac{1}{6}\right]\right\}, \notag \\
    &\lim_{\phi\to0} \Gamma_0 = 1
\end{align}
is not well defined at $\phi = 0$ since the argument of the logarithm goes to zero. The limit of $\phi\to0$ exists, however, but disagrees with the exact result, which is always $2$ at that point since the two signals in the correlation optimally stack up. It is basically putting two detectors on top of each other and thus gaining twice the signal since the only difference in the two measurements is their independent noise. If the distances $L$ and $L'$ to the two pulsars are not the same, however, the two signals differ more with increasing $\Delta L = |L-L'|$, and thus the correlation decays, and $\Gamma_T$ tends to 1.

It turns out that as $L\omega$ increases the absolute values of the first summands in the series become larger and larger. In fact, their absolute values lie orders of magnitudes above the value of the final result, which is between 0 and 2. And even though the peak goes farther away from the start $a=2$, the first terms cannot be neglected since they are always of order 1 and the way larger terms will cancel precisely until the result becomes of order 1. So, the most successful method has been to find out at which index $a$ the terms become smaller than $10^{-5}$ and sum until there, and then one only has to add a handful of additional terms until one cannot see the curve change anymore (additional term was $\lesssim 10^{-3}$ everywhere). Thus, the variable $aStart$ in the code can actually be deleted; it always has to be 5 anyways.\\
Due to the cancelling in extreme cases from $10^{90}$ down to 1, the precision was a big issue. We used \textit{Mathematica} \cite{Mathematica} for this reason to do the numerical calculations since we could just choose higher initial precisions to make sure that at the end of the calculation enough meaningful digits were left.\\
Another surprise was that for larger $\phi$ one had to choose larger cutoffs, although the function very quickly converged to the Hellings and Downs curve and was not oscillating as strong as for $\phi$ around zero.\\
It should be noted that the formalism and the results presented in this paper do not change the value of the overlap reduction function obtained using H\&D for most of the pulsar systems currently used for GW searches with PTA. Our formalism changes the value of the overlap reduction function only for a sub-population of co-located pulsars which can be used for GW searches. These systems can be potentially rare as the pulsars used for GW searches are traditionally millisecond pulsars. Pulsars in double-neutron-star binaries are unlikely to get fully recycled and hence will not be fast spinning with sufficiently low spin down rates \cite{form-mili}. However, in the future with SKA one will discover a huge population of pulsars with chances of finding rare systems. If a suitable double-neutron-star binary is discovered, it would be the most sensitive system for GW searches and hence applying the formalism presented in this paper would be necessary to make use of them.  

\begin{acknowledgements}
We thank the anonymous referee for suggesting numerous improvements to the paper, especially for suggesting the inclusion of Fig. \ref{fig.: odep_L!=L'}.  A.B. is supported by the Forschungskredit of the University of Zurich Grant No. FK-20-083 and by the Tomalla Foundation. S.T. is supported by Swiss National Science Foundation Grant No. 200020 182047.\\
\end{acknowledgements}

\appendix

\section{Matched filtering}\label{sec: MatchedFilter}
The idea is, that if one multiplies the signal $h$ to the strain $s = h + n$ in the integral then the signal is squared and thus positive on the entire domain, while the signal times the noise $n\,h$ can have both signs and since they are not correlated this contribution becomes small,
\begin{equation}
\int_U s(t)h(t) dt = \int_U h^2(t) + n(t)h(t) dt \approx \int_U h^2(t) dt,
\end{equation}
where we integrate over the domain $U \subset \mathbb{R}$.\\

To use this effect, we multiply our cross-correlated strain from detectors $a$ and $b$ with a filter function $Q$. The goal will then be, to find the best possible filter function to do this. Since we want to overlay signals that arrived at different times at the two detectors and thus get rid of the choice of $t_0 = 0$, we chose the filter to depend on the time difference:
\begin{equation}
Y \coloneq \int s(t)s'(t')Q(t-t') dt'dt.
\end{equation}

However, now our expression $Y$ is dependent on the filter function, which we choose, and does not represent the correlated signal anymore. To get rid of this dependence, we have to divide it out. Therefore, we take the signal-to-noise ratio and expand it with $Q$ by multiplying it into the integrals of the nominator and denominator:
\begin{widetext}
	\begin{align}
		SNR &= \frac{\mu}{\sigma} = \frac{\int h(t)h'(t') dt'dt}{\sqrt{\int n(t)n(t')n'(\tau)n'(\tau') d\tau' d\tau  dt'dt}} \\
		\mapsto \quad SNR[Q] &= \frac{\int s(t)s'(t')Q(t-t') dt'dt}{\sqrt{\int n(t)n(t')n'(\tau)n'(\tau')Q(t-\tau)Q(t'-\tau') d\tau' d\tau dt'dt}}
	\end{align}
\end{widetext}
This is not a strict expansion of the fraction, however, and the effect of the filter function does not divide out, which will allow us to maximize the SNR by choosing an appropriate filter.

To maximize the SNR with the filter function, we define a scalar product and reexpress the signal-to-noise ratio in terms of the scalar product:
\begin{align}
&(A|B) \coloneq \int \tilde{A}^*(f)\tilde{B}(f)P(f)P'(f) df \notag \\
&\rightarrow \quad \mu[Q] = \left(\left.\frac{h h'}{P P'}\right|Q\right), \quad \sigma[Q] = \sqrt{\frac{T}{4}(Q|Q)}.
\end{align}
The SNR is maximal if the scalar product $\mu[Q]$ is maximal, and thus the optimal filter function $Q_{opt}$ has to be chosen parallel to $\frac{h h'}{P P'}$.

With this choice, the filtered signal can be rewritten in terms of the optimal filter function, and the best signal-to-noise ratio one can achieve this way is given by
\begin{equation}
SNR[Q_{opt}] = \frac{\mu[Q_{opt}]}{\sigma[Q_{opt}]} = \frac{(Q_{opt}|Q_{opt})}{\sqrt{\frac{T}{4}(Q_{opt}|Q_{opt})}} = 2\sqrt{\frac{(Q_{opt}|Q_{opt})}{T}}.
\end{equation}

To get to the signal we get by using matched filtering, we cannot just plug $Q_{opt}$ into $\mu[Q]$ since this expression does not represent the signal due to its dependency on $Q$. The SNR, however, does represent the signal-to-noise ratio we get by using this method because we divide the filter function out. So, what we can do to get to our signal is multiplying the SNR with the noise that remains after matched filtering.\\
To get an expression for our filter independent noise, we use the same trick as above and expand the noise with $Q$,
\begin{equation}
\sigma^2 = \mathbb{V}[\mu] \quad \mapsto \quad \sigma_{MF}^2 = \frac{\mathbb{V}[\mu[Q]]}{\int Q^2 dt'dt},
\end{equation}
where $\sigma_{MF}$ denotes the remaining noise after the signal has been filtered with $Q$.\\

We can again express everything in terms of the scalar product, for $Q = Q_{opt}$:
\begin{align}
	&\mathbb{V}[Q_{opt}] = \frac{T}{4}(Q_{opt}|Q_{opt}), \quad \int |Q_{opt}(f)|^2 df = \frac{(Q_{opt}|Q_{opt})}{P(f_0)P'(f_0)} \notag \\
	&\Rightarrow \quad \sigma_{MF}^2 = \frac{T}{4}P(f_0)P'(f_0).
\end{align}

We now finally get an expression for the matched filtered signal:
\begin{equation}
\mu_{MF} = SNR[Q_{opt}]\cdot\sigma_{MF} = \sqrt{(Q_{opt}|Q_{opt})P(f_0)P'(f_0)}.
\end{equation}

\section{Short wavelength approximation}\label{sec: e^ikx}
\begin{figure}[h!]
	\centering\includegraphics[width=0.9\linewidth]{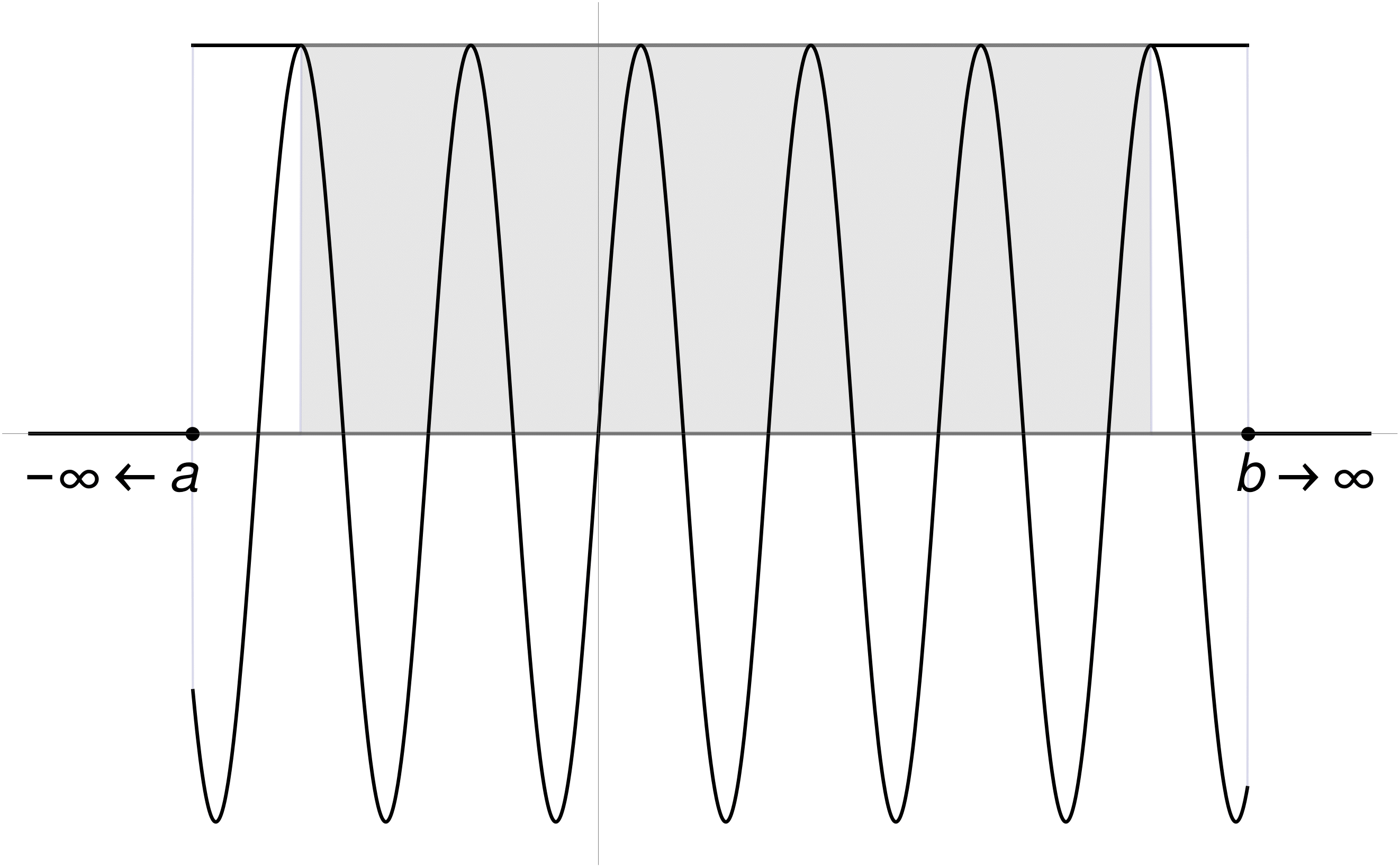}
	\caption{The part of the function, which lies in the grey area does not contribute to the integral, since it’s oscillations cancel. Only a segment on each side contributes, which is always smaller than one period.}\label{fig: OscCancel}
\end{figure}

In many cases in physics, rapidly oscillating terms are being neglected (short wavelength approximation) and considered as small with respect to other terms in the integral if the frequency is very high. This comes from the fact that a constant function times $e^{i\omega x}$ is arbitrarily small compared to the integral over the constant function as one lets the integration domain become infinite. The idea is sketched in Fig. \ref{fig: OscCancel}:
\begin{widetext}
	\begin{equation}
		\lim_{a\to-\infty}\lim_{b\to\infty} \frac{1}{b-a} \int_a^b e^{\mathrm{i}\omega x} dx
			\leqslant \lim_{a\to-\infty}\lim_{b\to\infty} \frac{1}{b-a} \left( \int_a^{a+\frac{2\pi}{\omega}} \left|e^{\mathrm{i}\omega x}\right| dx
			+ \int_{b-\frac{2\pi}{\omega}}^b \left|e^{\mathrm{i}\omega x}\right| dx \right) = 0.
	\end{equation}
\end{widetext}

We give a proof that in the following case, which is essentially the limit for $\omega\to\infty$ of an inverse Fourier transform restricted on the interval $[a,b]\subset\mathbb{R}$, this holds:
\begin{equation}\label{eq.: sw approx}
\lim_{\omega\to\infty} \int_a^b f(x) e^{\mathrm{i}\omega x} dx = 0, \qquad \text{if }\ f'(x) < \infty \quad \forall x\in[a,b].
\end{equation}

\textbf{Proof:}\\
Let $\varepsilon = \frac{2\pi}{\omega}$, $N = \left\lfloor\frac{b-a}{\varepsilon}\right\rfloor$, $x_n = a + n\varepsilon$, and $f_m' = \max\{f'(x)|x\in[a,b]\}$.\\

In the limit of $\epsilon\to 0$, the approximation of a smooth function $f$ with a piecewise linear one becomes precise. We do this on intervals which comprise exactly one period of the oscillation. At the end, there will be a small piece of the domain left due to rounding with a length smaller than $\varepsilon$:
\begin{widetext}
	\begin{align}
		\lim_{\omega\to\infty} \int_a^b f(x) e^{i\omega x} dx = \lim_{\varepsilon\to 0}\left\lbrace \sum_{n=0}^{N-1} \int_0^\varepsilon
			\left(f'(x_n)x+f(x)\right) e^{\mathrm{i}\omega x} dx + \int_{a+N\varepsilon}^b f(x) e^{\mathrm{i}\omega x} dx \right\rbrace.
	\end{align}

	The first integral can be solved by partial integration:
	\begin{align}
	&\left| \lim_{\varepsilon\to 0} \sum_{n=0}^{N-1} \int_0^\varepsilon \left(f'(x_n)x+f(x)\right) e^{\mathrm{i}\omega x} dx \right|
		= \left| \lim_{\varepsilon\to 0} \sum_{n=0}^{N-1} f'(x_n)\left[ 
		e^{\mathrm{i}\omega\varepsilon}\left(\frac{1}{\omega^2}-\frac{\mathrm{i}\varepsilon}{\omega}\right) - \frac{1}{\omega^2} \right]
		+ \lim_{\varepsilon\to 0} \sum_{n=0}^{N-1} \frac{f(x_n)}{\mathrm{i}\omega}\left[ e^{\mathrm{i}\omega\varepsilon} - 1 \right] \right| \notag \\
	&\quad \overset{\omega\varepsilon=2\pi}{=} \left| \frac{1}{2\pi\mathrm{i}} \sum_{n=0}^{N-1} f'(x_n)\varepsilon^2 \right|
	\leqslant \frac{b-a}{2\pi} \lim_{\varepsilon\to 0} |f_m'|\varepsilon = 0.
	\end{align}
\end{widetext}

Since the second integral is only over a fraction of the domain, smaller than $\varepsilon$, we can construct an upper bound:
\begin{align}\label{eq.: simple sw approx}
	\lim_{\varepsilon\to 0}& \left| \int_{a+N\varepsilon}^b f(x) e^{\mathrm{i}\omega x} dx \right|
	\lim_{\varepsilon\to 0} \leqslant \varepsilon \left|\max\{f(x)|x\in[b-\varepsilon,b]\}\right| \notag \\
	=& \lim_{\varepsilon\to 0} \varepsilon f(b) = 0. \qquad \square
\end{align}

One could use $\frac{b-a}{2\pi} |f_m'|\varepsilon + \epsilon|f(b)|$ as an error estimate for large but finite $\omega$. It is not proven to be a strict upper bound, however, since we approximated the function with a piecewise linear function and we did not give an upper bound for this error in the finite case.

In the case of the overlap reduction function for PTA's, the exponential term is of the form $e^{\mathrm{i}\omega g(x)}$, and thus we have to generalize the result above to use it:
\begin{equation}\label{eq.: approx}
\lim_{\omega\to\infty} \int_a^b f(x) e^{i\omega g(x)} dx = 0, \ \text{ if }\ \frac{f'g'-fg''}{(g')^3} < \infty \ \forall\,x\in[a,b].
\end{equation}

\textbf{Proof:}\\
\begin{align}
\lim_{\omega\to\infty} \int_a^b f(x) e^{\mathrm{i}\omega g(x)} dx
	= \lim_{\omega\to\infty} \int_{g(a)}^{g(b)} \frac{f}{g'}(g^{-1}(y)) e^{\mathrm{i}\omega y} dy.
\end{align}
With the substitution $y = g(x)$, \ $dy = g'(x)dx$, we brought the integral into the simpler form and can now apply the result from above:
\begin{widetext}
	\begin{align}\label{eq: sw-cond}
		&\left| \lim_{\omega\to\infty} \int_{g(a)}^{g(b)} \frac{f}{g'}(g^{-1}(y)) e^{\mathrm{i}\omega y} dy \right|
			\leqslant \frac{|g(b)-g(a)|}{2\pi} \lim_{\varepsilon\to 0} \max_y\left\{ \frac{d}{dy} \left(\frac{f}{g'}\circ g^{-1}\right)(y) \right\} \varepsilon \notag\\
		 &\qquad = \frac{|g(b)-g(a)|}{2\pi} \lim_{\varepsilon\to 0} \max_x\left\{ \frac{d}{dx} \left(\frac{f}{g'}\right)(x)\frac{dy}{dx} \right\} \varepsilon
			 = \frac{|g(b)-g(a)|}{2\pi} \lim_{\varepsilon\to 0} \max_x\left\{ \frac{f'g'-fg''}{(g')^3}(x) \right\} \varepsilon = 0. \\
		& &\square \notag
	\end{align}
\end{widetext}

\bibliographystyle{apsrev4-1}
\bibliography{PTARef}

\end{document}


\title{Supplemental to Analytic series expansion of the overlap reduction function for gravitational wave search with pulsar timing arrays}

\author{Adrian \surname{Bo\^itier}}
\email[]{boitier@physik.uzh.ch}
\affiliation{Physik-Institut, Universit\"at Z\"urich, Winterthurerstrasse 190, 8057 Z\"urich}

\author{Shubhanshu \surname{Tiwari}}
\email[]{stiwari@physik.uzh.ch}
\affiliation{Physik-Institut, Universit\"at Z\"urich, Winterthurerstrasse 190, 8057 Z\"urich}

\author{Philippe \surname{Jetzer}}
\email[]{jetzer@physik.uzh.ch}
\affiliation{Physik-Institut, Universit\"at Z\"urich, Winterthurerstrasse 190, 8057 Z\"urich}



\maketitle

\section{Overlap reduction function}\label{sec: Gamma}
As derived in the previous section the overlap reduction function for the tensor mode for PTA's using natural units $c = 1$ is given by:
\begin{widetext}
	\begin{equation}
		\Gamma_T = \beta\sum_{A\in \{+,\times\}}\int_{\mathbb{S}^2} F^A(\hat{\Omega}) {F^A}'(\hat{\Omega})
			\frac{1-e^{\mathrm{i}L\omega[1+\gamma]}}{1+\gamma} \frac{1-e^{-\mathrm{i}L'\omega[1+\gamma']}}{1+\gamma'} d\hat{\Omega}.
	\end{equation}
\end{widetext}

To calculate this integral we use the residue theorem on the $\varphi$-integral:
\begin{widetext}
	\begin{align}
		\Gamma_T &= \sum_{A\in \{+,\times\}}\int_0^{2\pi}\int_0^\pi F^A(\hat{\Omega}) {F^A}'(\hat{\Omega}) \Delta h(\hat{\Omega}) \Delta h'(\hat{\Omega}) \sin\theta d\theta\,d\varphi
			= \int_0^\pi \oint_{C_1} f(z) dz\,d\theta = 2\pi\mathrm{i} \int_0^\pi \text{Res}[f(z),0] d\theta, \notag \\
		\vphantom{a}\notag\\
		f(z) &= \frac{\left(F^A F'^A\Delta h\Delta h'\right)(\theta,z)}{\mathrm{i}z}\sin\theta,
			\quad \Delta h = \frac{1-e^{\mathrm{i}L\omega[1+\gamma]}}{1+\gamma}, \quad z = e^{\mathrm{i}\varphi}.
	\end{align}
\end{widetext}

The poles from the denominators $\gamma = -1$ and $\gamma' = -1$ are cancelled by the nominators. Thus, the only pole left is the one at zero. To find the Residue we write the Laurent series of $f$ around zero and read out the $a_{-1}$-term:
\begin{align}
f(z) = \sum_{n\in\mathbb{Z}}a_n z^n, \quad Res[f(z),0] = a_{-1}.
\end{align}

We chose our reference frame such that the pulsar $a$ is located at $\hat{x} = (1,0,0)$ and the second pulsar $a'$ at $\hat{x}' = (\cos\phi,\sin\phi,0)$. Then complexified pattern functions form a "generalized" polynomial in z:\\
\begin{widetext}
	\begin{align}
		\left(F^AF'^A\right)&(\theta,\varphi) = \frac{\alpha^2-\beta^2}{2}\frac{\alpha'^2-\beta'^2}{2}
			+ \alpha\beta \alpha'\beta' \notag \\
		 =& \frac{1}{4}\left(-\frac{1}{4}\cos^2\theta\left(2 - 3\cos(2\phi) + \cos(2[\phi-2\varphi])\vphantom{\sqrt{2}}\right)
			 +\cos^4\theta\cos^2(\phi-\varphi)\cos^2\varphi + \sin^2(\phi-\varphi)\sin^2\varphi \right) \notag \\
		 =& \frac{1}{32} \left(\frac{e^{2\mathrm{i}\phi}\left(\cos^2\theta-1\right)^2}{2z^4} + \frac{\left(1+e^{2\mathrm{i}\phi}\right)\left(\cos^4\theta-1\right)}{z^2}
			 +\left(\left(1 + 6\cos^2\theta + \cos^4\theta \right)\cos(2\phi) + 2\sin^4\theta \vphantom{\sqrt{2}}\right)\right. \notag \\
		 &\qquad\left. +\ e^{-2\mathrm{i}\phi}\left(1 + e^{2\mathrm{i}\phi}\right)\left(\cos^4\theta - 1\right)z^2
			 + \frac{1}{2}e^{-2\mathrm{i}\phi}\left(\cos^2\theta - 1\right)^2 z^4
			 \vphantom{\frac{e^{2\mathrm{i}\phi}\left(\cos^2\theta-1\right)^2}{2z^4}}\right),
	\end{align}
	where $\alpha$ and $\beta$ are defined according to \cite{PTA1}.\\
	
	We can write down the Laurent series of the interference term using the geometric series:
	\begin{align}
		&\Delta h(\theta,\varphi) = \frac{1-e^{\mathrm{i}L\omega(1+\cos\varphi\sin\theta)}}{1+\cos\varphi\sin\theta}
			= \frac{2z}{\sin\theta}  \left( 1 - e^{\mathrm{i}L\omega } e^{\mathrm{i}\frac{L\omega}{2z}\sin\theta\left(1+z^2\right)} \right)\sum_{n=0}^\infty (-z)^n \left(z+\frac{2}{\sin\theta}\right)^n,
	\end{align}
	and thus
	\begin{align}
		&\left(\Delta h\Delta h'\right)(\theta,z) = \frac{4e^{-\mathrm{i}\phi}z^2}{\sin^2\theta} \\
		&\quad\cdot \left( 1 - e^{\mathrm{i}L\omega} e^{\frac{\mathrm{i}\omega}{2z}\sin\theta\, L\left(1+z^2\right)}
			- e^{-\mathrm{i}L'\omega} e^{-\frac{\mathrm{i}\omega}{2z}\sin\theta\, L'\left(e^{\mathrm{i}\phi}+e^{-\mathrm{i}\phi} z^2\right)}
			+ e^{\mathrm{i}(L-L')\omega} e^{\frac{\mathrm{i}\omega}{2z}\sin\theta\left(L-L'e^{\mathrm{i}\phi}
			+ \left[L-L'e^{-\mathrm{i}\phi}\right]z^2 \right)}\right) \notag \\
		&\quad\cdot \sum_{n=0}^\infty(-z)^n\left(z+\frac{2}{\sin\theta}\right)^n 
			\sum_{n=0}^\infty\left(-e^{-\mathrm{i}\phi}z\right)^n\left(\frac{2}{\sin\theta} + e^{-\mathrm{i}\phi}z\right)^n.
			\notag
	\end{align}\\
\end{widetext}

The powers of $z$ in the two geometric series are all positive but the exponential terms provide negative powers when we write them as series and the terms in the polynomial from the pattern functions will shift the indices of the series coefficients and thus a different $a_n$ will become the residual of that term (multiplied by the polynomial coefficient of said term).\\
Thus, we define the polynomial part $P$ and the geometric series part $G$
\begin{widetext}
	\begin{align}
		P(z) \coloneq& \frac{e^{2\mathrm{i}\phi}}{2}\left(\cos^2\theta-1\right)^2 z^{-3}
		+ \left(1+e^{2\mathrm{i}\phi}\right)\left(\cos^4\theta-1\right) z^{-1}
			+ \left( \left( 1 + 6\cos^2\theta + \cos^4\theta \right)\cos(2\phi) + 2\sin^4\theta \vphantom{\sqrt{2}}\right) z \notag \\
		 &+ \left(1+e^{-2\mathrm{i}\phi}\right)\left(\cos^4\theta-1\right) z^3 + \frac{e^{-2\mathrm{i}\phi}}{2}\left(\cos^2\theta-1\right)^2 z^5, \\
		\vphantom{a}\notag\\
		G(z) \coloneq& \sum_{n=0}^\infty (-z)^n \left(z + \frac{2}{\sin\theta}\right)^n 
			\sum_{n=0}^\infty \left(-e^{-\mathrm{i}\phi}z\right)^n \left(e^{-\mathrm{i}\phi}z + \frac{2}{\sin\theta}\right)^n,
	\end{align}
	to write $f$ in a simplified way, which allows us to calculate the Laurent series more efficiently:
	\begin{align}
		f(z) = -\frac{\mathrm{i}e^{-\mathrm{i}\phi}}{8\sin\theta} P(z)
			&\left( 1 - e^{\mathrm{i}L\omega} e^{\frac{\mathrm{i}\omega}{2z}\sin\theta\, L\left(1+z^2\right)}
			- e^{-\mathrm{i}L'\omega} e^{-\frac{\mathrm{i}\omega}{2z}\sin\theta\, L'\left(e^{\mathrm{i}\phi}+e^{-\mathrm{i}\phi} z^2\right)} \right.\notag\\
		&\quad\left. +\ e^{\mathrm{i}(L-L')\omega} e^{\frac{\mathrm{i}\omega}{2z}\sin\theta\left(L-L'e^{\mathrm{i}\phi}
			+ \left[L-L'e^{-\mathrm{i}\phi}\right]z^2 \right)}\right) G(z).
	\end{align}
\end{widetext}
We notice, that: $P(z) \eqcolon z\sum_{n=-4}^4 b_n z^n$; \quad $b_{-n} = \bar{b}_n$.

\subsection{Multiplying Series}
To get the Laurent series we use the Cauchy product on the two geometric series and the exponential terms, starting with $G(z)$ since this term also appears alone (multiplied with one):\\
\begin{equation}
\text{Cauchy product:} \quad \sum_{n=0}^\infty a_n \sum_{n=0}^\infty b_n = \sum_{n=0}^\infty \sum_{k=0}^n a_k b_{n-k},
\end{equation}

\begin{widetext}
	\begin{align}
		G(z) &= \sum _{n=0}^\infty \sum _{k=0}^n \left(-e^{-\mathrm{i}\phi}z\right)^k \left(\frac{2}{\sin\theta} + e^{-\mathrm{i}\phi}z\right)^k 
			(-z)^{n-k} \left(z + \frac{2}{\sin\theta}\right)^{n-k} \notag \\
		&=\sum _{n=0}^\infty \sum _{k=0}^n (-1)^n e^{-\mathrm{i}k\phi} \left(\frac{2}{\sin\theta}
			+ e^{-\mathrm{i}\phi}z\right)^k \left(\frac{2}{\sin\theta} + z\right)^{n-k} z^n.
	\end{align}

	Now we write the exponential terms as a Series and multiply it to the geometric series:
	\begin{align}
		E(z) \coloneq& - e^{\mathrm{i}L\omega} e^{\frac{\mathrm{i}\omega}{2z}\sin\theta\, L\left(1+z^2\right)}
			- e^{-\mathrm{i}L'\omega} e^{-\frac{\mathrm{i}\omega}{2z}\sin\theta\, L'\left(e^{\mathrm{i}\phi} + e^{-\mathrm{i}\phi}z^2\right)}
			+ e^{\mathrm{i}(L-L')\omega} e^{\frac{\mathrm{i}\omega}{2z}\sin\theta\left(L-L'e^{\mathrm{i}\phi}
			+ \left(L-L'e^{-\mathrm{i}\phi}\right)z^2 \right)} \notag \\
		=&\sum_{n=0}^\infty \frac{1}{n!} \left(\frac{\mathrm{i}\omega}{2z}\sin\theta\right)^n
			\left\lbrace -e^{\mathrm{i}L\omega}L^n\left(1+z^2\right)^n - e^{-\mathrm{i}L'\omega}(-L')^n\left(e^{\mathrm{i}\phi}
			+ e^{-\mathrm{i}\phi}z^2\right)^n \right. \notag \\
		 &\left.\qquad\qquad\qquad\qquad\quad +\ e^{\mathrm{i}(L-L')\omega}\left( L-L'e^{\mathrm{i}\phi}
			 + \left[L-L'e^{-\mathrm{i}\phi}\right]z^2 \right)^n\right\rbrace, \\
		\vphantom{a}\notag\\
		E(z)G(z) =& \sum _{n=0}^{\infty } \sum _{j=0}^n 
			\left(\sum_{k=0}^j (-1)^j e^{-\mathrm{i}k\phi} \left( \frac{2}{\sin\theta} + e^{-\mathrm{i}\phi}z \right)^k
			\left(\frac{2}{\sin\theta} + z\right)^{j-k} z^j\right) \notag \\
		 &\cdot \left( \frac{1}{(n-j)!} \left(\frac{\mathrm{i}\omega}{2z}\sin\theta\right)^{n-j}
			\left\lbrace -e^{\mathrm{i}L_a\omega}L^{n-j}\left(1+z^2\right)^{n-j} - e^{-\mathrm{i}L'\omega}(-L')^{n-j}\left(e^{\mathrm{i}\phi}
			+ e^{-\mathrm{i}\phi}z^2\right)^{n-j} \right.\right. \notag \\
		 &\qquad\left.\left. +\ e^{\mathrm{i}(L-L')\omega}\left( L-L'e^{\mathrm{i}\phi}
			 + \left[L-L'e^{-\mathrm{i}\phi}\right]z^2 \right)^{n-j}\right\rbrace 
			 \vphantom{\left(\frac{\mathrm{i}}{2}\right)^{j}}\right).
	\end{align}		
\end{widetext}

\subsection{Expanding Binomials}
We expand the binomials and collect all terms with the same power in $z$ using the binomial formula:\\
\begin{equation}
(a + b)^n = \sum_{k=0}^n \binom{n}{k} a^k b^{n-k},
\end{equation}
\begin{widetext}
	\begin{align}
		E(z)G(z) =& \sum _{n=0}^{\infty } \sum _{j=0}^n \left( \sum_{k=0}^j (-1)^j \sum _{m=0}^k \sum _{l=0}^{j-k} e^{-\mathrm{i}(k+m)\phi}
			\binom{k}{m} \binom{j-k}{l} \left(\frac{2}{\sin\theta}\right)^{j-l-m} z^{-n+2j+m+l} \right) \notag \\
		 &\cdot \left( \frac{1}{(n-j)!} \left(\frac{\mathrm{i}\omega}{2z}\sin\theta\right)^{n-j} \sum_{i=0}^{n-j} \binom{n-j}{i}
			 \left\lbrace\vphantom{\left[e^{-\mathrm{i}\phi}\right]^i} -L^{n-j} e^{\mathrm{i}L\omega}
			 - (-L')^{n-j} e^{-\mathrm{i}L'\omega} e^{\mathrm{i}(n-j-2i)\phi} \right.\right. \notag \\
		 &\qquad\left.\left. +\ e^{\mathrm{i}(L-L')\omega} \left[L - L'e^{\mathrm{i}\phi}\right]^{n-j-i}
			 \left[L - L'e^{-\mathrm{i}\phi}\right]^i \right\rbrace z^{2i} \vphantom{\left(\frac{\mathrm{i}}{2}\right)^{j}}\right) \notag \\
		=& \sum_{n=0}^\infty \sum_{j=0}^n \sum_{k=0}^j \sum_{m=0}^k \sum_{l=0}^{j-k} \binom{k}{m} \binom{j-k}{l} \left(\frac{2}{\sin\theta}\right)^{j-l-m}
			\sum_{i=0}^{n-j} \frac{(-1)^je^{-\mathrm{i}(k+m)\phi}}{(n-j-i)!i!}
		   \left(\frac{\mathrm{i}\omega}{2}\sin\theta\right)^{n-j} \notag \\
		 &\left\lbrace -L^{n-j} e^{\mathrm{i}L\omega} - (-L')^{n-j} e^{-\mathrm{i}L'\omega} e^{\mathrm{i}(n-j-2i)\phi} \right. \notag \\
			 &\qquad\left. +\ e^{\mathrm{i}(L-L')\omega} \left[L - L'e^{\mathrm{i}\phi}\right]^{n-j-i}
			 \left[L - L'e^{-\mathrm{i}\phi}\right]^i \right\rbrace z^{-n+2 j+m+l+2 i} \eqcolon EG(z).
	\end{align}
\end{widetext}

With $G(z)$ and $E\,G(z)$ written as a power series in $z$ and $P(z)$ being a polynomial in $z$ we now read out the $a_{-1}$-term of:
\begin{equation}
f(z) = -\frac{\mathrm{i}e^{-\mathrm{i}\phi}}{8\sin\theta} P(z) \left( G(z) + E\,G(z) \vphantom{\sqrt{2}}\right)
\end{equation}

\subsection{Residues}
Since $G(z)$ only has positive powers of $z$, the only contribution to the residue of that term comes from the negative powers of $P(z)$. The $E\,G(z)$ term however requires a bit more care. Thus we split our problem into two parts:
\begin{align}
	\text{Res}&[f(z),0] = a_{-1} = \text{Res}\left[\sum_{n=-\infty}^\infty a_n z^n,0\right]
	= \text{Res}\left[\sum_{n=-\infty}^\infty \left(b_n + c_n\right) z^n,0\right] = b_{-1} + c_{-1} \notag \\
	=& \text{Res}\left[\sum_{n=-\infty}^\infty b_n z^n,0\right] + \text{Res}\left[\sum_{n=-\infty}^\infty c_n z^n,0\right]
	= \text{Res}\left[ -\frac{\mathrm{i}e^{-\mathrm{i}\phi}}{8\sin\theta} P(z) \left( G(z) + E(z)G(z) \vphantom{\sqrt{2}}\right) \right] \notag \\
	=& \underbrace{\text{Res}\left[ -\frac{\mathrm{i}e^{-\mathrm{i}\phi}}{8\sin\theta} P(z) G(z) \right]}_{\text{Res}_1}
		+ \underbrace{\text{Res}\left[ -\frac{\mathrm{i}e^{-\mathrm{i}\phi}}{8\sin\theta} P(z) EG(z) \right]}_{\text{Res}_2}.
\end{align}

\begin{widetext}
	By writing the first terms of the first part up to zeroth order:
	\begin{align}
		-\frac{\mathrm{i}e^{-\mathrm{i}\phi}}{8\sin\theta} P(z) G(z) =& -\frac{\sin^3\theta(\mathrm{i}\cos\phi - \sin\phi)}{16z^3}
			+\frac{\mathrm{i}\left[1+e^{\mathrm{i}\phi}\right] \sin^2\theta}{8z^2}
			-\frac{\mathrm{i}\sin\theta\left(2+\cos\phi\sin^2\theta\right)}{8 z} \notag \\
			&+ \frac{1}{8}\left(\left[1+e^{\mathrm{i}\phi}\right] \sin^2\theta\right) (\mathrm{i}\cos\phi + \sin\phi) + \mathcal{O}\left(z^1\right),
	\end{align}
\end{widetext}
we see that its residue is given by:
\begin{equation}\label{eq.: Res1}
\text{Res}_1 = -\frac{\mathrm{i}}{8} \sin\theta \left( 2 + \cos\phi\sin^2\theta \right).
\end{equation}

\subsection{Residue 2}
Lets write out the second part.
\begin{widetext}
	\begin{align}\label{eq.: Res2}
		&\text{Res}_2 = \text{Res}\left[ -\frac{\mathrm{i}e^{-\mathrm{i}\phi}}{8\sin\theta}
			\left( \frac{e^{2\mathrm{i}\phi}}{2}\left(\cos^2\theta-1\right)^2 z^{-3} + \left(1+e^{2\mathrm{i}\phi}\right)\left(\cos^4\theta-1\right) z^{-1}
			\notag \right.\right. \\
		 &\quad\left.\left. + \left( \left( 1 + 6\cos^2\theta + \cos^4\theta \right)\cos(2\phi) + 2\sin^4\theta \vphantom{\sqrt{2}}\right) z
			 + \left(1+e^{-2\mathrm{i}\phi}\right)\left(\cos^4\theta-1\right) z^3 + \frac{e^{-2\mathrm{i}\phi}}{2}\left(\cos^2\theta-1\right)^2 z^5 \right)
			 \right. \notag \\
		 &\quad\left. \cdot \sum_{n=0}^\infty \sum_{j=0}^n \sum_{k=0}^j \sum_{m=0}^k \sum_{l=0}^{j-k} \binom{k}{m} \binom{j-k}{l} \left(\frac{2}{\sin\theta}\right)^{j-l-m}
			\sum_{i=0}^{n-j} \frac{(-1)^je^{-\mathrm{i}(k+m)\phi}}{(n-j-i)!i!}
			   \left(\frac{\mathrm{i}\omega}{2}\sin\theta\right)^{n-j} \right. \notag \\
		 &\quad\left.\left\lbrace -L^{n-j} e^{\mathrm{i}L\omega} - (-L')^{n-j} e^{-\mathrm{i}L'\omega} e^{\mathrm{i}(n-j-2i)\phi} + e^{\mathrm{i}(L-L')\omega} 
			 \left[L - L'e^{\mathrm{i}\phi}\right]^{n-j-i}
			 \left[L - L'e^{-\mathrm{i}\phi}\right]^i \right\rbrace z^{-n+2 j+m+l+2 i} \right]. \notag \\
		\vphantom{a}
	\end{align}
\end{widetext}

Let $N'$ be the power of a polynomial term:
\begin{equation}
P(z) = \sum_{N'} A_{N'} z^{N'},
\end{equation}
then a summand contributes to the residue, if:
\begin{align}
&z^{-n+2j+l+m+2i+N'} \overset{!}{=} z^{-1} \quad \Rightarrow \quad n = 2j+l+m+2i+N, \notag \\
&\text{with} \quad N = N'+1\in\{-2,0,2,4,6\}. \label{eq.: selection cond.}
\end{align}
But we need to make sure, that the combination of $(j,l,m,i)$ appears in the nested sums, thus we demand the conditions:
\begin{align}
n &\geqslant 0  &\Rightarrow &\quad &2j+l+m+2i+N \geqslant 0, \\
j &\leqslant n  &\overset{|-j}{\Rightarrow} &\quad &j+l+m+2i+N \geqslant 0, \\
i &\leqslant n-j  &\overset{|-i}{\Rightarrow} &\quad &\underline{j+l+m+i+N} \geqslant 0. \label{ineq.: selection cond.}
\end{align}
Condition~\eqref{ineq.: selection cond.} is the most stringent one. If it is satisfied the other two are as well.

\subsubsection{Case 1: $N \geqslant 0$ i.e. $z^{-1}, z, z^3, z^5$}
Condition~\eqref{ineq.: selection cond.} is always satisfied, as long as the indexes $j, l, m, i$ are chosen within their range since they all start at zero.\\
We can now write the sum of all contributing summands by replacing $n$ with $2j+l+m+2i+N$ as seen from~\eqref{eq.: selection cond.}. We only need to take care of the sum over $j$ and $i$ since their summation range depends on $n$:\\

For every $j \geqslant 0$ we choose there exists an $n \in \{0,..,\infty\}$ such that $j$ appears in the sum and we will have to pick that term form a different $n$ according to~\eqref{eq.: selection cond.}.\\
The same argument applies to $i$ for every valid choice of $j, l$ and $m$ and thus both sum from $0$ to $\infty$:
\begin{widetext}
	\begin{align}
		&S_{N-1} \coloneq \sum_{j=0}^\infty \sum_{k=0}^j \sum_{m=0}^k \sum_{l=0}^{j-k} \binom{k}{m} \binom{j-k}{l} \left(\frac{2}{\sin\theta}\right)^{j-l-m}
			\sum_{i=0}^\infty \frac{(-1)^j e^{-\mathrm{i}(k+m)\phi}}{(i+j+l+m+N)!i!} \left(\frac{\mathrm{i}\omega}{2}\sin\theta\right)^{2i+j+l+m+N} \notag \\
		&\cdot \left\lbrace-L^{2i+j+l+m+N} e^{\mathrm{i}L\omega} - (-L')^{2i+j+l+m+N} e^{-\mathrm{i}L'\omega} e^{\mathrm{i}(j+l+m+N)\phi}
			+ e^{\mathrm{i}(L-L')\omega} \left[L-L'e^{-\mathrm{i}\phi}\right]^i \left[L-L'e^{\mathrm{i}\phi}\right]^{i+j+l+m+N}\right\rbrace \notag \\
		&= \sum_{j=0}^\infty \sum_{k=0}^j \sum_{m=0}^k \sum_{l=0}^{j-k} (-1)^j e^{-\mathrm{i}(k+m)\phi} \binom{k}{m} \binom{j-k}{l} 
			\left(\frac{2}{\sin\theta}\right)^{j-l-m} \notag \\
		 &\quad\left\lbrace
			  -\ e^{\mathrm{i}L\omega} \sum_{i=0}^\infty \frac{1}{(i+j+l+m+N)!i!} \left(\frac{\mathrm{i}L\omega}{2}\sin\theta\right)^{2i+j+l+m+N} \right. \notag \\
			  &\qquad - e^{-\mathrm{i}L'\omega} e^{\mathrm{i}(j+l+m+N)\phi} \sum_{i=0}^\infty \frac{(-1)^{2 i+j+l+m+N}}{(i+j+l+m+N)!i!} 
			  \left(\frac{\mathrm{i}L'\omega}{2}\sin\theta\right)^{2 i+j+l+m+N} \notag \\
			  &\qquad\left. +\ e^{\mathrm{i}(L-L')\omega} \sum_{i=0}^\infty
			  \frac{\left[L-L'e^{-\mathrm{i}\phi}\right]^i \left[L-L'e^{\mathrm{i}\phi}\right]^{i+j+l+m+N}}{(i+j+l+m+N)!i!} 
			  \left(\frac{\mathrm{i}\omega}{2}\sin\theta\right)^{2i+j+l+m+N}
		 \right\rbrace.
	\end{align}
\end{widetext}
The series over the index $i$ can be written as Bessel functions:
\begin{equation}
J_n(z) \coloneq \sum_{i=0}^\infty \frac{(-1)^i}{i!(n+i)!} \left(\frac{z}{2}\right)^{2i+n}.
\end{equation}\\

For the third term we use:
\begin{align}
a &= L - L'e^{-\mathrm{i}\phi}, &\quad x &= \mathrm{i}\omega\sin\theta, &\quad
n &= j+l+m+N, &\quad s &= \text{sign}(\bar{a}) = \frac{\bar{a}}{|a|};
\end{align}
\begin{widetext}
	\begin{align}
		a^i\bar{a}^{i+n} = (a\bar{a})^i \bar{a}^n = |a|^{2i} \frac{\bar{a}^n}{|a|^n} |a|^n = s^n |a|^{2i+n},& \\
		\sum_{i=0}^\infty \frac{\left[L-L'e^{-\mathrm{i}\phi}\right]^i
			\left[L-L'e^{\mathrm{i}\phi}\right]^{i+j+l+m+N}}{(i+j+l+m+N)!i!}
			\left(\frac{\mathrm{i}\omega}{2}\sin\theta\right)^{2i+j+l+m+N}
			&= \sum_{i=0}^\infty \frac{a^i \bar{a}^{i+n}}{(i+n)!i!} \left(\frac{x}{2}\right)^{2i+n} = s^n I_n(|a|x), \notag
	\end{align}
\end{widetext}
where $I_n$ are the modified or hyperbolic Bessel functions of the first kind:
\begin{equation}
I_n(x) \coloneq \mathrm{i}^{-n}J_n(\mathrm{i}x) = \sum_{i=0}^\infty \frac{1}{i!(n+i)!} \left(\frac{x}{2}\right)^{2i+n},
\end{equation}
and thus:
\begin{align}
	&I_n(|a|x) = \mathrm{i}^{-n} J_n(-|a|\omega\sin\theta) = \mathrm{i}^{-n} \sum_{i=0}^\infty \frac{(-1)^{i+2i+n}}{i!(n+i)!} 
		\left(\frac{|a|\omega\sin\theta}{2}\right)^{2i+n} = \mathrm{i}^n J_n(|a|\omega\sin\theta).
\end{align}

The absolute value of $a$ and the sign $s$ can be expressed as:
\begin{equation}
|a|^2 = L^2 + L'^2 - 2LL'\cos\phi, \quad s = \frac{L - L'e^{\mathrm{i}\phi}}{\sqrt{L^2 + L'^2 - 2LL'\cos\phi}}.
\end{equation}

Since this combination of Bessel functions appear always in this fashion we can simplify our expressions by defining the function:
\begin{align}
	g^\pm_n(\theta) \coloneq&\ e^{\mathrm{i}(L-L')\omega}
		\left(\frac{L-e^{\pm\mathrm{i}\phi}L'}{\sqrt{L^2+L'^2-2LL'\cos\phi}}\right)^n
		J_n\left(\sqrt{L^2+L'^2-2LL'\cos\phi}\ \omega\sin\theta \right) \notag\\
	&- e^{\mathrm{i}L\omega} J_n(L_a\omega\sin\theta) - e^{\mathrm{i}(\pm n\phi -L'\omega)}
		J_n(-L'\omega\sin\theta).
\end{align}

Finally we get the contribution of the of $E\,G(z)$ for the polynomial term with power $N-1$ to the residue:
\begin{widetext}
	\begin{equation}
		S_{N-1} = \sum_{j=0}^\infty \sum_{k=0}^j \sum_{m=0}^k \sum_{l=0}^{j-k} (-1)^j e^{-\mathrm{i}(k+m)\phi} \binom{k}{m} \binom{j-k}{l}
			\left(\frac{2}{\sin\theta}\right)^{j-l-m} \mathrm{i}^{j+l+m+N} g^+_{j+l+m+N}(\theta), \quad N \geqslant 0.
	\end{equation}

	With this definition of $S_{N'} = S_{N-1}$ we can write the second residue as:
	\begin{align}\label{eq.: Res2(SN)}
		&\text{Res}_2 = \text{Res}\left[\vphantom{\sum_{i=0}^{n-j}} -\frac{\mathrm{i}e^{-\mathrm{i}\phi}}{8\sin\theta}
			\left( \frac{e^{2\mathrm{i}\phi}}{2}\left(\cos^2\theta-1\right)^2 z^{-3} + \left(1+e^{2\mathrm{i}\phi}\right)\left(\cos^4\theta-1\right) z^{-1}
			\notag \right.\right. \\
		&\quad\left.\left. + \left( \left( 1 + 6\cos^2\theta + \cos^4\theta \right)\cos(2\phi) + 2\sin^4\theta \vphantom{\sqrt{2}}\right) z
			+ \left(1+e^{-2\mathrm{i}\phi}\right)\left(\cos^4\theta-1\right) z^3 + \frac{e^{-2\mathrm{i}\phi}}{2}\left(\cos^2\theta-1\right)^2 z^5 \right)
			\right. \notag \\
		&\quad\left. \cdot \sum_{n=0}^\infty \sum_{j=0}^n \sum_{k=0}^j \sum_{m=0}^k \sum_{l=0}^{j-k} \binom{k}{m} \binom{j-k}{l} \left(\frac{2}{\sin\theta}\right)^{j-l-m}
			\sum_{i=0}^{n-j} \frac{(-1)^je^{-\mathrm{i}(k+m)\phi}}{(n-j-i)!i!}
			\left(\frac{\mathrm{i}\omega}{2}\sin\theta\right)^{n-j} \left\lbrace \dots \right\rbrace z^{-n+2 j+m+l+2 i} \right] \notag \\
		\vphantom{a}\notag\\
		&= -\frac{\mathrm{i}e^{-\mathrm{i}\phi}}{8\sin\theta}
			\left( \frac{e^{2\mathrm{i}\phi}}{2}\left(\cos^2\theta-1\right)^2 S_{-3} + \left(1+e^{2\mathrm{i}\phi}\right)\left(\cos^4\theta-1\right) S_{-1} \right. \\
		&\quad\left. + \left( \left( 1 + 6\cos^2\theta + \cos^4\theta \right)\cos(2\phi) + 2\sin^4\theta \vphantom{\sqrt{2}}\right) S_1
			+ \left(1+e^{-2\mathrm{i}\phi}\right)\left(\cos^4\theta-1\right) S_3 + \frac{e^{-2\mathrm{i}\phi}}{2}\left(\cos^2\theta-1\right)^2 S_5 \right). \notag
	\end{align}
\end{widetext}

\subsubsection{Case 2: $N = -2$ i.e. $z^{-3}$}
We still have to deal with $S_{-3}$. Since $N$ is negative in this case~\eqref{ineq.: selection cond.} is not automatically satisfied for any choice of $(j,l,m,i)$ within their summation ranges.\\
Our strategy to systematically go through all valid choices is to pick the first index large enough $j = 2$, then go through the remaining cases $j = 0, 1$ by choosing the second index large enough to satisfy~\eqref{ineq.: selection cond.} for all choices of the remaining indices and so on:
\begin{widetext}
	\begin{align}
		&j \geqslant 2 \quad \Rightarrow \quad j+m+l+i-2 \geqslant 0 \quad \forall\ k\in\{0,\dots,j\},\ m\in\{0,\dots,k\},\ l\in\{0,\dots,j-k\},\ i\in\{0,\dots,\infty\}
			\notag \\
		\vphantom{a}\notag\\
		&j = 1,\ m \geqslant 1: \quad \Rightarrow \quad k = m = 1,\ l = 0 \quad \Rightarrow \quad j+m+l+i-2 \geqslant 0 \quad \forall\ i\in\{0,\dots,\infty\} \notag \\
		&\qquad\quad m = 0,\ l \geqslant 1: \quad \Rightarrow \quad k = m = 0,\ l = 1 \quad \Rightarrow \quad j+m+l+i-2 \geqslant 0
			\quad \forall\ i\in\{0,\dots,\infty\} \notag \\
		&\qquad\qquad\qquad\ l = 0,\ i \geqslant 1 \quad \Rightarrow \quad k\in\{0,1\} \quad \Rightarrow \quad j+m+l+i-2 \geqslant 0 \notag \\
		\vphantom{a}\notag\\
		&j = 0 \quad \Rightarrow \quad k = 0 \quad \Rightarrow \quad m = l = 0 \quad \Rightarrow \quad j+m+l+i-2 \geqslant 0 \quad \forall\ i\in\{2,\dots,\infty\}.
	\end{align}

	And thus, we get:
	\begin{align}
		S_{-3} =& \sum_{j=2}^\infty \sum_{k=0}^j \sum_{m=0}^k \sum_{l=0}^{j-k} (-1)^j e^{-\mathrm{i}(k+m)\phi} \binom{k}{m} \binom{j-k}{l}
			\left(\frac{2}{\sin\theta}\right)^{j-l-m} \sum_{i=0}^\infty \frac{1}{(i+j+l+m-2)!i!} \left(\frac{\mathrm{i}\omega}{2}\sin\theta\right)^{2i+j+l+m-2} \notag \\
		&\qquad \cdot \left\lbrace-L^{2i+j+l+m-2} e^{\mathrm{i}L\omega} - (-L')^{2i+j+l+m-2} e^{-\mathrm{i}L'\omega} e^{\mathrm{i}(j+l+m-2)\phi} \right. \notag \\
		&\qquad\qquad\left. + e^{\mathrm{i}(L-L')\omega} \left[L-L'e^{-\mathrm{i}\phi}\right]^i \left[L-L'e^{\mathrm{i}\phi}\right]^{i+j+l+m-2}\right\rbrace
			\notag \\
		&- e^{-2\mathrm{i}\phi} \sum_{i=0}^\infty \frac{1}{i!i!} \left(\frac{\mathrm{i}\omega}{2}\sin\theta\right)^{2 i}
			\left\lbrace -L^{2i}e^{\mathrm{i}L\omega} - (-L')^{2i}e^{-\mathrm{i}L'\omega} + e^{\mathrm{i}(L-L')\omega}
			\left[L-L'e^{-\mathrm{i}\phi}\right]^i \left[L-L'e^{\mathrm{i}\phi}\right]^i\right\rbrace \notag \\
		&- \sum _{i=0}^\infty \frac{1}{i!i!} \left(\frac{\mathrm{i}\omega}{2}\sin\theta\right)^{2 i} 
			\left\lbrace -L^{2i}e^{\mathrm{i}L\omega} - (-L')^{2i}e^{-\mathrm{i}L'\omega} + e^{\mathrm{i}(L-L')\omega }
			\left[L-L'e^{-\mathrm{i}\phi}\right]^i \left[L-L'e^{\mathrm{i}\phi}\right]^i\right\rbrace \notag \\
		&- \frac{2}{\sin\theta} \sum_{k=0}^1 e^{-\mathrm{i}k\phi} \sum_{i=1}^\infty \frac{1}{(i-1)!i!} \left(\frac{\mathrm{i}\omega}{2}\sin\theta\right)^{2i-1} \notag \\
		&\qquad \cdot \left\lbrace -L^{2i-1}e^{\mathrm{i}L\omega} - (-L')^{2i-1}e^{-\mathrm{i}L'\omega}e^{-\mathrm{i}\phi} + e^{\mathrm{i}(L-L')\omega}
			\left[L-L'e^{-\mathrm{i}\phi}\right]^i \left[L-L'e^{\mathrm{i}\phi}\right]^{i-1}\right\rbrace \notag \\
		&+ \sum_{i=2}^\infty \frac{1}{(i-2)!i!} \left(\frac{\mathrm{i}\omega}{2}\sin\theta\right)^{2i-2}
			\left\lbrace -L^{2i-2}e^{\mathrm{i}L\omega} - L'^{2i-2}e^{-\mathrm{i}L'\omega}e^{-2\mathrm{i}\phi} + e^{\mathrm{i}(L-L')\omega}
			\left[L-L'e^{-\mathrm{i}\phi}\right]^i \left[L-L'e^{\mathrm{i}\phi}\right]^{i-2}\right\rbrace.
	\end{align}
\end{widetext}

We can shift the indexing for the cases, where $i$ does not start at zero to get Bessel series and write them as $g_n^\pm$ as well:
\begin{equation}
	\sum_{i=n}^\infty a_i = \sum_{i=0}^\infty a_{i+n} \ ;
\end{equation}
\begin{widetext}
	\begin{align}
		S_{-3} =& \sum_{j=2}^\infty \sum_{k=0}^j \sum_{m=0}^k \sum_{l=0}^{j-k} (-1)^j e^{-\mathrm{i}(k+m)\phi} \binom{k}{m} \binom{j-k}{l}
			\left(\frac{2}{\sin\theta}\right)^{j-l-m} i^{j+l+m-2} g^+_{j+l+m-2}(\theta) \notag \\
		&- \left(1+e^{-2\mathrm{i}\phi}\right) g^+_0(\theta) - \frac{2\mathrm{i}}{\sin\theta} \left(1+e^{-\mathrm{i}\phi}\right) g^-_1(\theta) - g^-_2(\theta).
	\end{align}
\end{widetext}

\subsection{$\varphi$ - Integral}
Now we have all the ingredients to assemble our the integral over $\varphi$. When we plug the $S_n$ into~\eqref{eq.: Res2(SN)}, we notice that they all have a common nested sum starting from $j = 2$ and all but $S_{-3}$ start from zero. Thus we can simplify by collecting the terms in that way and then add the two residues~\eqref{eq.: Res1} and~\eqref{eq.: Res2} to get the $\varphi$-integral:
\begin{widetext}
	\begin{align}\label{eq.: phi-int}
		&\int_0^{2\pi} F^A(\hat{\Omega}) F'^A(\hat{\Omega}) \Delta h(\hat{\Omega}) \Delta h'(\hat{\Omega}) \sin\theta\, d\varphi
			= \oint_{C_1} f(z) dz = 2\pi\mathrm{i}\,\text{Res}[f(z),0] = 2\pi\mathrm{i}\left( \text{Res}_1 + \text{Res}_2 \right) \notag \\
		&\quad = 2\pi\mathrm{i}\left( -\frac{\mathrm{i}}{8} \sin\theta \left( 2 + \cos\phi\sin^2\theta \right)
			-\frac{\mathrm{i}e^{-\mathrm{i}\phi}}{8\sin\theta} \sum_{N'} A_{N'} S_{N'} \right) \notag \\
		&\quad = \frac{\pi}{4}\sin\theta \left( 2 + \cos\phi\sin^2\theta \right) \notag \\
		 &\qquad + \frac{\pi e^{-\mathrm{i}\phi}}{4\sin\theta}
			\left(\sum_{j=2}^\infty \sum_{k=0}^j \sum_{m=0}^k \sum_{l=0}^{j-k} (-1)^j e^{-\mathrm{i}(k+m)\phi} \binom{k}{m} \binom{j-k}{l}
			\left(\frac{2}{\sin\theta}\right)^{j-l-m} \mathrm{i}^{j+m+l} \right. \notag \\
		 &\qquad\qquad\qquad\quad \cdot \left\lbrace -\frac{\sin^4\theta}{2} \left(e^{2\mathrm{i}\phi}g^+_{j+l+m-2}(\theta)
			 + e^{-2\mathrm{i}\phi} g^+_{j+l+m+6}(\theta)\right) \right. \notag \\
		 &\qquad\qquad\qquad\quad\quad\ \left. + \left[1+e^{2\mathrm{i}\phi}\right] \left(\cos^4\theta-1\right) \left(g^+_{j+l+m}(\theta )
			 + e^{-2\mathrm{i}\phi}g^+_{j+l+m+4}(\theta)\right) - A\,g^+_{j+l+m+2}(\theta) \vphantom{\frac{\sin^4}{2}}\right\rbrace \notag \\
		 &\qquad\qquad\qquad + \sum_{j=0}^1 \sum_{k=0}^j \sum_{m=0}^k \sum_{l=0}^{j-k} (-1)^j e^{-\mathrm{i}(k+m)\phi} \binom{k}{m} \binom{j-k}{l}
			 \left(\frac{2}{\sin\theta}\right)^{j-l-m} \mathrm{i}^{j+m+l} \notag \\
		 &\qquad\qquad\qquad\quad \cdot \left\lbrace -\frac{\sin^4\theta}{2}e^{-2\mathrm{i}\phi}g^+_{j+l+m+6}(\theta) \right. \notag \\
		 &\qquad\qquad\qquad\quad\quad\ \left.+ \left[1+e^{2\mathrm{i}\phi}\right] \left(\cos^4\theta-1\right) \left(g^+_{j+l+m}(\theta)
			 + e^{-2\mathrm{i}\phi}g^+_{j+l+m+4}(\theta)\right) - A\,g^+_{j+l+m+2}(\theta) \vphantom{\frac{\sin^4}{2}}\right\rbrace \notag \\
		 &\qquad\qquad\qquad\left. - \frac{e^{2\mathrm{i}\phi}}{2}\sin^4\theta \left( \left[1+e^{-2\mathrm{i}\phi}\right]g^+_0(\theta)
			 + \frac{2\mathrm{i}}{\sin\theta}\left[1+e^{-\mathrm{i}\phi}\right]g^-_1(\theta) + g^-_2(\theta) \right) \vphantom{\left(\frac{1}{2}\right)^j}\right),
	\end{align}
\end{widetext}
with the polynomial coefficient to the first power
\begin{equation}
A \coloneq A_1 = \left( 1 + 6\cos^2\theta + \cos^4\theta \right)\cos(2\phi) + 2\sin^4\theta.
\end{equation}

\section{$\theta$ - Integration}
The $g^\pm_n(\theta)$ are a combination of Bessel functions which depend on $\sin\theta$ and are thus a power series in $\sin\theta$. Looking at the $\theta$ dependence of the coefficients we see that the residue consists of powers of $\sin\theta$ and $\cos\theta$. Thus, it can be straightforwardly integrated.\\
The main task is to collect terms of the same form, calculate the integrals of those general forms and then simplifying the mess as good as we can.\\
Since integration is a linear operation and the series including the integral is absolutely convergent, as we show in appendix~\ref{sec: Conv}, we can pull the integral into the sum:
\begin{widetext}
	\begin{align}
		\Gamma_T =& 2\pi\mathrm{i} \int_0^\pi \text{Res}[f(z),0] d\theta = 2\pi\mathrm{i}\left( \int_0^\pi \text{Res}_1\, d\theta + \int_0^\pi \text{Res}_2\, d\theta \right) \notag \\
		=& \frac{\pi}{4} \left\{\vphantom{\sum_j^k} \int_0^{\pi } \sin\theta \left(2 + \cos\phi\sin^2\theta\right) \, d\theta \right. \notag \\
		 &\quad + e^{-\mathrm{i}\phi} \left(\sum _{j=2}^\infty \sum _{k=0}^j \sum_{m=0}^k \sum_{l=0}^{j-k} (-1)^j e^{-\mathrm{i}(k+m)\phi}
			\binom{k}{m} \binom{j-k}{l} \mathrm{i}^{j+m+l} 2^{j-m-l} \right. \notag \\
		 &\qquad\qquad\quad \cdot \int_0^{\pi} \left(\frac{1}{\sin\theta}\right)^{j-l-m+1}
				\left\lbrace -\frac{\sin^4\theta}{2} \left( e^{2\mathrm{i}\phi} g^+_{j+l+m-2}(\theta) + e^{-2\mathrm{i}\phi} g^+_{j+l+m+6}(\theta) \right)
				\right. \notag \\
		 &\qquad\qquad\qquad\left. + \left[1+e^{2\mathrm{i}\phi}\right] \left(\cos^4\theta-1\right) \left( g^+_{j+l+m}(\theta)
			+ e^{-2\mathrm{i}\phi} g^+_{j+l+m+4}(\theta\right) - A_1\, g^+_{j+l+m+2}(\theta) \vphantom{\frac{\sin^4}{2}}\right\rbrace \, d\theta \notag \\
		 &\qquad\qquad + \sum_{j=0}^1 \sum_{k=0}^j \sum_{m=0}^k \sum_{l=0}^{j-k} (-1)^j e^{-\mathrm{i}(k+m)\phi} \binom{k}{m} \binom{j-k}{l}
			\mathrm{i}^{j+m+l} 2^{j-m-l} \notag \\
		 &\qquad\qquad\quad \cdot \int_0^{\pi} \left(\frac{1}{\sin\theta}\right)^{j-l-m+1} \left\lbrace -\frac{\sin^4\theta}{2} e^{-2\mathrm{i}\phi}
			 g^+_{j+l+m+6}(\theta) \right. \notag \\
		 &\qquad\qquad\qquad\left.\left. + \left[1+e^{2\mathrm{i}\phi}\right] \left(\cos^4\theta-1\right) \left(g^+_{j+l+m}(\theta)
			 + e^{-2\mathrm{i}\phi} g^+_{j+l+m+4}(\theta)\right) - A_1\, g^+_{j+l+m+2}(\theta) \vphantom{\frac{\sin^4}{2}}\right\rbrace \, d\theta
			 \vphantom{\int_0^\infty\sum_0^k}\right) \notag \\
		 &\quad\left. -\frac{e^{\mathrm{i}\phi}}{2} \int_0^{\pi} \sin^3\theta \left( \left[1+e^{-2\mathrm{i}\phi}\right] g^+_0(\theta)
			 + \frac{2\mathrm{i}}{\sin\theta}\left[1+e^{-\mathrm{i}\phi}\right] g^-_1(\theta) + g^-_2(\theta) \right) \, d\theta \vphantom{\sum_j^k}\right\}.
	\end{align}
\end{widetext}

The integral of the first residue can directly be calculated to be:
\begin{equation}
\int_0^\pi \text{Res}_1\, d\theta = \int_0^{\pi } \sin\theta \left(2 + \cos\phi\sin^2\theta\right)\, d\theta = \frac{4}{3} (3 + \cos\phi).
\end{equation}

\subsection{$\theta$ - integral of the series-term}
Expanding the integral of the series term
\begin{widetext}
	\begin{align}
		&\int_0^{\pi} \left(\frac{1}{\sin\theta}\right)^{j-l-m+1} \left\lbrace \dots \right\rbrace \, d\theta
			= -\frac{1}{2} \int_0^{\pi} \left(\frac{1}{\sin\theta}\right)^{j-l-m-3} \left( e^{2\mathrm{i}\phi} g^+_{j+l+m-2}(\theta)
			+ e^{-2\mathrm{i}\phi} g^+_{j+l+m+6}(\theta)\right) \,d\theta \notag \\
		&\quad + \left[1+e^{2\mathrm{i}\phi}\right] \int_0^{\pi} \left(\frac{1}{\sin\theta}\right)^{j-l-m+1} \left(\cos^4\theta-1\right)
			\left(g^+_{j+l+m}(\theta) + e^{-2\mathrm{i}\phi} g^+_{j+l+m+4}(\theta) \right) \, d\theta \notag \\
		&\quad - \int_0^{\pi} \left(\frac{1}{\sin\theta}\right)^{j-l-m+1} \left( \left(1 + 6\cos^2\theta + \cos^4\theta \right) \cos(2\phi) + 2\sin^4\theta \right)
		g^+_{j+l+m+2}(\theta) \, d\theta
	\end{align}
	we see that it can be expressed as a sum of the following integrals:
	\begin{align}
		&\int_0^{\pi} \left(\frac{1}{\sin\theta}\right)^{j-l-m-3} g^+_{j+l+m+N}(\theta ) \, d\theta, 
			&\quad &\int_0^{\pi} \left(\frac{1}{\sin\theta}\right)^{j-l-m+1} g^+_{j+l+m+N}(\theta) \, d\theta, \notag \\
		&\int_0^{\pi} \left(\frac{1}{\sin\theta}\right)^{j-l-m+1} \cos^2(\theta) g^+_{j+l+m+N}(\theta) \, d\theta,
			&\quad &\int_0^{\pi} \left(\frac{1}{\sin\theta}\right)^{j-l-m+1} \cos^4(\theta) g^+_{j+l+m+N}(\theta) \, d\theta.
	\end{align}

	So, we try to solve the more general problem:
	\begin{align}
		&\int_0^{\pi} \left(\frac{1}{\sin\theta}\right)^{j-b-s} \cos^t\theta g^\pm_{j+b+N}(\theta) \, d\theta 
			= e^{\mathrm{i}(L-L')\omega} \left(\frac{L-e^{\pm\mathrm{i}\phi}L'}{\sqrt{L^2+L'^2 - 2LL'\cos\phi}}\right)^{j+b+N} \notag \\
		&\qquad \cdot \int_0^{\pi} \sin^{-j+b+s}\theta \cos^t\theta J_{j+b+N}\left( \sqrt{L^2+L'^2 - 2LL'\cos\phi}\,\omega \sin\theta \right) \, d\theta \notag \\
		&\quad - e^{\mathrm{i}L\omega} \int_0^{\pi} \sin^{-j+b+s}\theta \cos^t\theta J_{j+b+N}(L\omega\sin\theta) \, d\theta \notag \\
		&\quad - e^{\mathrm{i}(\pm(j+b+N)\phi - L'\omega)} \int_0^{\pi} \sin^{-j+b+s}\theta \cos^t\theta J_{j+b+N}(-L'\omega\sin\theta) \, d\theta,
	\end{align}
\end{widetext}
where we introduced the new index $b = m + l$ to simplify the expressions. The indices may take the following values:
\begin{align}
	j \geqslant 2, \quad \text{for } s =& 3: &\quad &t = 0, &\quad &N\in\{-2,6\} \notag \\
	\text{and for } s =& -1: &\quad &t\in\{0,2,4\}, &\quad &N\in\{0,2,4\}.
\end{align}

When we write the Bessel functions out as a series and then swap the integral with the sum, we get a series of integrals over powers in $\cos\theta$ and $\sin\theta$. We introduce $x$ as a placeholder for the pulsar distance terms in front of the $\sin\theta$. The factor 2 is just chosen out of convenience:
\begin{widetext}
	\begin{align}
		\int_0^{\pi} \sin^{-j+b+s}\theta \cos^t\theta J_{j+b+N}(2x\sin\theta) \, d\theta = \sum_{i=0}^\infty \frac{(-1)^i}{i!(j+b+N+i)!} x^{2i+j+b+N}
			\int_0^\pi \cos^t\theta\sin^{2i+2b+N+s}\theta \, d\theta.
	\end{align}
\end{widetext}

In the appendix \ref{sec: Integrations} we calculate the integral on the right hand side and get:\\
\begin{align}
\int_0^{\pi } \cos ^m(\theta ) \sin ^n(\theta ) \, d\theta
   =\frac{\left(1+(-1)^m\right) \Gamma \left(\frac{1+m}{2}\right) \Gamma
   \left(\frac{1+n}{2}\right)}{2 \Gamma \left(\frac{1}{2} (2+m+n)\right)},
\end{align}
which is valid for $\Re(m)$, and $\Re(n) > -1$.\\

This requirement is always met except for the term where $N = 0$ and $s = -1$. In this case $i = b = 0$ appear in the nested sum which leads us to the integral:
\begin{align}
	\int_0^\pi \frac{1}{\sin\theta} d\theta &= -\left[ \ln\left(\cos\frac{\theta}{2}\right) + \ln\left(\sin\frac{\theta}{2}\right) \right]_{\theta\to 0}^{\theta\to\pi} 
	= -\lim_{x\to 0}\ln x - \cancelto{0}{\ln 1} + \cancelto{0}{\ln 1} + \lim_{x\to 0}\ln x.
\end{align}
The poles of the integrand at $0$ and $\pi$ are bad enough to make an integration over them impossible. To be able to integrate this, we need to get rid of these poles. It does not come as a surprise that a physical quantity like this overlap reduction function also includes another term which cancels these poles. In other words: we accidentally split an integral over a smooth function into two integrals with poles. Thus we need to keep the combination $\left(\frac{1}{\sin\theta}\right)^{j-l-m-s} \left(\cos^4\theta-1\right)
   g^\pm_{j+l+m+N}(\theta)$ together for the integration:
\begin{equation}
\int_0^\pi \left(\cos^4\theta-1\right) \sin^n\theta \, d\theta =-\frac{\sqrt{\pi} (n+5) \Gamma\left(\frac{n+3}{2}\right)}{2\,\Gamma\left(\frac{n}{2}+3\right)}.
\end{equation}

Now, that we calculated the fundamental integrals we can start to put everything back together:
\begin{widetext}
	\begin{align}
		\int_0^{\pi} &\sin^{-j+b-1}\theta \left(\cos^4\theta-1\right) J_{j+b}(2 x \sin\theta)\, d\theta
			= -\sqrt{\pi} \sum _{i=0}^\infty \frac{(-1)^i}{i!(j+b+i)!} \frac{(b+i+2)\Gamma(b+i+1)}{\Gamma\left(b+i+\frac{5}{2}\right)}
			x^{2i+j+b} \notag \\
		&= -\sqrt{\pi}x^{j+b} \sum_{i=0}^\infty \frac{\Gamma(b+i+1) \Gamma(b+i+3)}{\Gamma(j+b+i+1) \Gamma(b+i+2)
			\Gamma\left(b+i+\frac{5}{2}\right)} \frac{(-x^2)^i}{i!} \notag \\
		&= -\sqrt{\pi} x^{j+b} \Gamma(b+1) \Gamma(b+3) \, _2\tilde{F}_3\left(b+1,b+3;j+b+1,b+2,b+\frac{5}{2};-x^2\right).
	\end{align}

	To simplify the series, we define:
	\begin{align}
		f_{j,b}(\phi) \coloneq& \int_0^{\pi} \sin^{-j+b-1}\theta \left(\cos^4\theta-1\right)g^+_{j+b}(\theta)\, d\theta \\
		=& e^{\mathrm{i}(L-L')\omega} \left(\frac{L-e^{\mathrm{i}\phi}L'}{\sqrt{L^2+L'^2-2LL'\cos\phi}}\right)^{j+b}
			\int_0^\pi \frac{\cos^4\theta-1}{\sin^{j-b+1}\theta}
			J_{j+b}\left(\sqrt{L^2+L'^2-2LL'\cos\phi}\omega\sin\theta\right)\, d\theta \notag \\
		&- e^{\mathrm{i}L\omega} \int_0^\pi \frac{\cos^4\theta-1}{\sin^{j-b+1}\theta} J_{j+b}(L\omega\sin\theta)\, d\theta
			- e^{\mathrm{i}((j+b)\phi-L'\omega)} \int_0^\pi \frac{\cos^4\theta-1}{\sin^{j-b+1}\theta}
			J_{j+b}(-L'\omega\sin\theta)\, d\theta \notag \\
		=& \sqrt{\pi}2^{-j-b}\Gamma(b+1)\Gamma(b+3) \left[ e^{\mathrm{i}L\omega}(L\omega)^{j+b}\, 
			_2\tilde{F}_3\left(b+1,b+3;j+b+1,b+2,b+\frac{5}{2};-\left(\frac{L\omega}{2}\right)^2\right) \right. \notag \\
		&+ e^{\mathrm{i}((j+b)\phi-L'\omega)}(-L'\omega)^{j+b} \,
			_2\tilde{F}_3\left(b+1,b+3;j+b+1,b+2,b+\frac{5}{2};-\left(\frac{L'\omega}{2}\right)^2\right) \notag \\
		&\left. - e^{\mathrm{i}(L-L')\omega} \left((L-e^{i\phi}L')\omega\right)^{j+b}\,
			_2\tilde{F}_3\left(b+1,b+3;j+b+1,b+2,b+\frac{5}{2};-(L^2+L'^2-2LL'\cos\phi)\frac{\omega^2}{4}\right)
			\vphantom{\left(\frac{L}{2}\right)^2}\right]. \notag
	\end{align}
\end{widetext}

Since $t$ is even we can conclude
\begin{equation}
	t\in\{0,2,4\} \quad \Rightarrow \quad \int_0^\pi \cos^t\theta\sin^n\theta\, d\theta
	= \frac{\Gamma\left(\frac{t+1}{2}\right) \Gamma\left(\frac{n+1}{2}\right)}{\Gamma\left(\frac{1}{2}(t+n+2)\right)},
\end{equation}
for all other cases and thus:
\begin{widetext}
	\begin{align}
		\int_0^\pi &\frac{\cos^t\theta}{\sin^{j-b-s}\theta} J_{j+b+N}(2x\sin\theta)\, d\theta
		= x^{j+b+N} \Gamma\left(\frac{t+1}{2}\right) \sum_{i=0}^\infty
			\frac{\Gamma\left(b+\frac{s+N+1}{2}+i\right)}{(j+b+N+i)!\Gamma\left(b+\frac{s+t+N}{2}+1+i\right)}
			\frac{(-x^2)^i}{i!} \notag \\
		=& x^{j+b+N} \Gamma\left(\frac{t+1}{2}\right) \Gamma\left(b+\frac{s+N+1}{2}\right)\, 
			_1\tilde{F}_2\left(b+\frac{s+N+1}{2};j+b+N+1,b+\frac{s+t+N}{2}+1;-x^2\right).
	\end{align}

	We define:
	\begin{align}\label{eq.: def h}
		h^\pm&_{j,b,s,t,N}(\phi) \coloneq \int_0^\pi \sin^{-j+b+s}\theta \cos^t\theta g_{j+b+N}(\theta)\, d\theta \\
		=& e^{\mathrm{i}(L-L')\omega} \left(\frac{L-e^{\pm\mathrm{i}\phi}L'}{\sqrt{L^2+L'^2-2LL'\cos\phi}}\right)^{j+b+N}
			\int_0^\pi \frac{\cos^t\theta}{\sin^{j-b-s}\theta} J_{j+b+N}\left(\sqrt{L^2+L'^2-2LL'\cos\phi}\ \omega \sin\theta\right)\, 
			d\theta \notag \\
		&- e^{\mathrm{i}L\omega} \int_0^\pi \frac{\cos^t\theta}{\sin^{j-b-s}\theta} J_{j+b+N}(L\omega\sin\theta)\, d\theta
			- e^{\mathrm{i}(\pm(j+b+N)\phi-L'\omega)} \int_0^\pi \frac{\cos^t\theta}{\sin^{j-b-s}\theta} 
			J_{j+b+N}(-L'\omega\sin\theta)\, d\theta \notag \\
		=& \Gamma\left(\frac{t+1}{2}\right) \Gamma\left(k+\frac{s+N+1}{2}\right) \left[
			e^{\mathrm{i}(L-L')\omega} \left(\frac{1}{2}(L-e^{\pm\mathrm{i}\phi}L')\omega\right)^{j+b+N} \right. \notag \\
		&\qquad \times\, _1\tilde{F}_2\left(b+\frac{s+N+1}{2};j+b+N+1,b+\frac{s+t+N}{2}+1;-(L^2+L'^2-2LL'\cos\phi)\frac{\omega^2}{4}
			\right) \notag \\
		&- e^{\mathrm{i}L\omega} \left(\frac{L\omega}{2}\right)^{j+b+N}\, 
			_1\tilde{F}_2\left(b+\frac{s+N+1}{2};j+b+N+1,b+\frac{s+t+N}{2}+1;-\left(\frac{L\omega}{2}\right)^2\right) \notag \\
		&\left. - e^{\mathrm{i}(\pm(j+b+N)\phi-L'\omega)} \left(-\frac{L'\omega}{2}\right)^{j+b+N}\, 
			_1\tilde{F}_2\left(b+\frac{s+N+1}{2};j+b+N+1,b+\frac{s+t+N}{2}+1;-\left(\frac{L'\omega}{2}\right)^2\right) \right]. \notag
	\end{align}
\end{widetext}

Now we can insert our definitions into the series term:
\begin{widetext}
	\begin{align}
		\sum _{j=2}^\infty& \sum_{k=0}^j \sum_{m=0}^k \sum_{l=0}^{j-k} (-1)^j e^{-\mathrm{i}(k+m)\phi} \binom{k}{m} \binom{j-k}{l}
			\mathrm{i}^{j+m+l} 2^{j-m-l} \int_0^\pi \left(\frac{1}{\sin\theta}\right)^{j-m-l+1} \notag \\
		 &\left\lbrace -\frac{\sin^4\theta}{2} \left(e^{2\mathrm{i}\phi} g^+_{j+m+l-2}(\theta)
			+ e^{-2\mathrm{i}\phi} g^+_{j+m+l+6}(\theta)\right) + \left(1+e^{2\mathrm{i}\phi}\right) (\cos^4\theta-1)
			\left(g^+_{j+m+l}(\theta) + e^{-2\mathrm{i}\phi} g^+_{j+m+l+4}(\theta)\right) \right. \notag \\
		 &\quad\left. - A g^+_{j+m+l+2}(\theta) \vphantom{\frac{\sin^4\theta}{2}}\right\rbrace \, d\theta \notag \\
		\vphantom{a}\notag\\
		=& \sum _{j=2}^\infty \sum_{k=0}^j \sum_{m=0}^k \sum_{l=0}^{j-k} (-2\mathrm{i})^j \left(\frac{\mathrm{i}}{2}\right)^{m+l}
			\binom{k}{m} \binom{j-k}{l} e^{-\mathrm{i}(k+m)\phi} \left\lbrace
			-\frac{e^{2\mathrm{i}\phi}}{2} \int_0^\pi \sin^{-j+m+l+3}\theta g^+_{j+m+l-2}(\theta)\, d\theta \right. \notag \\
		 &- \frac{e^{-2\mathrm{i}\phi}}{2} \int_0^\pi \sin^{-j+m+l+3}\theta g^+_{j+m+l+6}(\theta)\, d\theta
			 + \left(1+e^{2\mathrm{i}\phi}\right) \int_0^\pi \sin^{-j+m+l-1}\theta (\cos^4\theta-1) g^+_{j+m+l}(\theta)\, d\theta
			 \notag \\
		 &+ \left(e^{-2\mathrm{i}\phi}+1\right) \left( \int_0^\pi \sin^{-j+m+l-1}\theta \cos^4\theta g^+_{j+m+l+4}(\theta)\, d\theta
			- \int_0^\pi \sin^{-j+m+l-1}\theta g^+_{j+m+l+4}(\theta)\, d\theta \right) \notag \\
		 &- \cos(2\phi)\left( \int_0^\pi \sin^{-j+m+l-1}\theta g^+_{j+m+l+2}(\theta)\, d\theta
			 + 6\int_0^\pi \sin^{-j+m+l-1}\theta\cos^2\theta g^+_{j+m+l+2}(\theta)\, d\theta \right. \notag \\
		 &\qquad\qquad\quad\left.\left. + \int_0^\pi \sin^{-j+m+l-1}\theta\cos^4\theta g^+_{j+l+m+2}(\theta)\, d\theta \right)
			 - 2\int_0^\pi \sin^{-j+m+l+3}\theta g^+_{j+m+l+2}(\theta)\, d\theta \right\rbrace \notag \\
		\vphantom{a}\notag\\
		=& \sum_{j=2}^\infty (-2\mathrm{i})^j \sum_{k=0}^j \sum_{m=0}^k \sum_{l=0}^{j-k} \left(\frac{\mathrm{i}}{2}\right)^{m+l}
			\binom{k}{m} \binom{j-k}{l} e^{-\mathrm{i}(k+m)\phi} \left\lbrace -\frac{e^{2\mathrm{i}\phi}}{2} h^+_{j,m+l,3,0,-2}(\phi)
			- \frac{e^{-2\mathrm{i}\phi}}{2} h^+_{j,m+l,3,0,6}(\phi) \right. \notag \\
		 &+ \left(1+e^{2\mathrm{i}\phi}\right) f_{j,m+l}(\phi) + \left(e^{-2\mathrm{i}\phi}+1\right) \left( h^+_{j,m+l,-1,4,4}(\phi)
			 -h^+_{j,m+l,-1,0,4}(\phi) \right) \notag \\
		 &\left. - \cos(2\phi)\left( h^+_{j,m+l,-1,0,2}(\phi) + 6h^+_{j,m+l,-1,2,2}(\phi)
			 + h^+_{j,m+l,-1,4,2}(\phi) \right) - 2h^+_{j,m+l,3,0,2}(\phi) \vphantom{\frac{e^{2\mathrm{i}\phi}}{2}}\right\rbrace.
	\end{align}
\end{widetext}

\subsection{$\theta$ - integral of the sum-term and the remaining terms}
We can use the results of the series term on the sum as well:
\begin{widetext}
	\begin{align}
		&\sum_{j=0}^1 \sum_{k=0}^j \sum_{m=0}^k \sum_{l=0}^{j-k} (-1)^j e^{-\mathrm{i}(k+m)\phi} \binom{k}{m} \binom{j-k}{l}
			\mathrm{i}^{j+m+l} 2^{j-m-l} \int_0^\pi \left(\frac{1}{\sin\theta}\right)^{j-l-m+1} \notag \\
		 &\cdot \left\lbrace -\frac{\sin^4\theta}{2} e^{-2\mathrm{i}\phi} g^+_{j+l+m+6}(\theta)
			 + \left(1+e^{2\mathrm{i}\phi}\right) (\cos^4\theta-1) \left( g^+_{j+l+m}(\theta)
			 + e^{-2\mathrm{i}\phi} g^+_{j+l+m+4}\theta \right) - A g^+_{j+l+m+2}(\theta) \right\rbrace \, d\theta \notag \\
		&= \sum_{j=0}^1 (-2\mathrm{i})^j \sum_{k=0}^j \sum_{m=0}^k \sum_{l=0}^{j-k} \left(\frac{\mathrm{i}}{2}\right)^{m+l}
			\binom{k}{m} \binom{j-k}{l} e^{-\mathrm{i}(k+m)\phi} \left\lbrace -\frac{e^{-2\mathrm{i}\phi}}{2} h^+_{j,m+l,3,0,6}(\phi)
			+ \left(1+e^{2\mathrm{i}\phi}\right) f_{j,m+l}(\phi) \right. \notag \\
		 &\qquad + \left(1+e^{-2\mathrm{i}\phi}\right) \left(h^+_{j,m+l,-1,4,4}(\phi) - h^+_{j,m+l,-1,0,4}(\phi)\right) \notag \\
		 &\qquad\left. - \cos(2\phi) \left( h^+_{j,m+l,-1,0,2}(\phi) + 6h^+_{j,m+l,-1,2,2}(\phi) + h^+_{j,m+l,-1,4,2}(\phi) \right)
			 - 2h_{j,m+l,3,0,2}(\phi) \vphantom{\frac{e^{-2\mathrm{i}\phi}}{2}}\right\rbrace.
	\end{align}

	We can just apply definition~\eqref{eq.: def h} on the remaining terms and get:
	\begin{align}
		&-\frac{e^{2\mathrm{i}\phi}}{2} \int_0^\pi \sin^3\theta \left( \left(1+e^{-2\mathrm{i}\phi}\right) g^+_0(\theta)
			+\frac{2\mathrm{i}}{\sin\theta} \left(1+e^{-\mathrm{i}\phi}\right) g^-_1(\theta) + g^-_2(\theta) \right)\, d\theta \notag \\
		= &-\frac{e^{2\mathrm{i}\phi}}{2} \left( \left(1+e^{-2\mathrm{i}\phi}\right) h^+_{0,0,3,0,0}(\phi)
			+ 2\mathrm{i}\left(1+e^{-\mathrm{i}\phi}\right) h^-_{0,0,2,0,1}(\phi) + h^-_{0,0,3,0,2}(\phi) \right).
	\end{align}
\end{widetext}

\section{Optimization and convergence of the overlap reduction function}
Finally, we can put everything together to write down the overlap reduction function for the tensor mode to first order in $h$ and zeroth order in $\omega T_a$.
\begin{widetext}
	\begin{align}
		\Gamma_T =& \frac{\pi}{4} \left[ \int_0^\pi \sin\theta \left(2 + \cos\phi\sin^2\theta\right)\, d\theta
			+ e^{-\mathrm{i}\phi} \left( \sum_{j=2}^\infty \sum_{k=0}^j \sum_{m=0}^k \sum_{l=0}^{j-k} (-1)^j e^{-\mathrm{i}(k+m)\phi}
			\binom{k}{m} \binom{j-k}{l} \mathrm{i}^{j+m+l} 2^{j-m-l} \right.\right.\notag \\
		 &\cdot \int_0^\pi \left(\frac{1}{\sin\theta}\right)^{j-l-m+1} \left\lbrace -\frac{\sin^4\theta}{2} \left(
			e^{2\mathrm{i}\phi} g^+_{j+l+m-2}(\theta) + e^{-2\mathrm{i}\phi} g^+_{j+l+m+6}(\theta) \right) \right.\notag \\
		 &\qquad\left. + \left(1+e^{2\mathrm{i}\phi}\right) (\cos^4\theta-1) \left(g^+_{j+l+m}(\theta)
			+ e^{-2\mathrm{i}\phi} g^+_{j+l+m+4}(\theta)\right) - A\, g^+_{j+l+m+2}(\theta)
			\vphantom{\frac{\sin^4\theta}{2}}\right\rbrace\, d\theta \notag \\
		 &+ \sum_{j=0}^1 \sum_{k=0}^j \sum_{m=0}^k \sum_{l=0}^{j-k} (-1)^j e^{-\mathrm{i}(k+m)\phi} \binom{k}{m} \binom{j-k}{l}
			\mathrm{i}^{j+m+l} 2^{j-m-l} \int_0^\pi \left(\frac{1}{\sin\theta}\right)^{j-l-m+1} \notag \\
		 &\left\lbrace -\frac{\sin^4\theta}{2}  e^{-2\mathrm{i}\phi} g^+_{j+l+m+6}(\theta) + \left(1+e^{2\mathrm{i}\phi}\right)
			(\cos^4\theta-1) \left( g^+_{j+l+m}(\theta) + e^{-2\mathrm{i}\phi} g^+_{j+l+m+4}(\theta) \right)
			- A g^+_{j+l+m+2}(\theta) \right\rbrace \, d\theta \notag \\
		 &\left.\left. - \frac{e^{2\mathrm{i}\phi}}{2} \int_0^\pi \sin^3\theta \left( \left(1+e^{-2\mathrm{i}\phi}\right) g^+_0(\theta)
			+ \frac{2\mathrm{i}}{\sin\theta} \left(1+e^{-\mathrm{i}\phi}\right) g^-_1(\theta) + g^-_2(\theta) \right) \, d\theta
			\vphantom{\sum_{k=0}^j\binom{k}{m}}\right) \right] \notag \\
		\vphantom{a}\notag\\
		=& \frac{\pi}{3} (3+\cos\phi) - \frac{\pi}{8} e^{\mathrm{i}\phi} \left[ \left(1+e^{-2\mathrm{i}\phi}\right)
			h^+_{0,0,3,0,0}(\phi) + 2\mathrm{i}\left(1+e^{-\mathrm{i}\phi}\right) h^-_{0,0,2,0,1}(\phi) + h^-_{0,0,3,0,2}(\phi) \right] \notag \\
		 &+ \frac{\pi}{4} e^{-\mathrm{i}\phi} \left(\sum_{j=0}^1 (-2\mathrm{i})^j \sum_{k=0}^j \sum_{m=0}^k \sum_{l=0}^{j-k}
			\left(\frac{\mathrm{i}}{2}\right)^{m+l} \binom{k}{m} \binom{j-k}{l} e^{-\mathrm{i}(k+m)\phi} \right. \notag \\
		 &\cdot \left\lbrace -\frac{e^{-2\mathrm{i}\phi}}{2} h^+_{j,m+l,3,0,6}(\phi) + \left(1+e^{2\mathrm{i}\phi}\right) f_{j,m+l}(\phi)
			+ \left(1+e^{-2\mathrm{i}\phi}\right) \left(h^+_{j,m+l,-1,4,4}(\phi) - h^+_{j,m+l,-1,0,4}(\phi)\right) \right.\notag\\
		 &\left.\qquad - \cos(2\phi) \left( h^+_{j,m+l,-1,0,2}(\phi) + 6h^+_{j,m+l,-1,2,2}(\phi) + h^+_{j,m+l,-1,4,2}(\phi)\right)
			 - 2h^+_{j,m+l,3,0,2}(\phi) \vphantom{\frac{1}{2}}\right\rbrace \notag\\
		 &+ \sum_{j=2}^\infty (-2\mathrm{i})^j \sum_{k=0}^j \sum_{m=0}^k \sum_{l=0}^{j-k} \left(\frac{\mathrm{i}}{2}\right)^{m+l}
			 \binom{k}{m} \binom{j-k}{l} e^{-\mathrm{i}(k+m)\phi} \notag \\
		 &\cdot \left\lbrace -\frac{1}{2} \left( e^{2\mathrm{i}\phi} h^+_{j,m+l,3,0,-2}(\phi)
			 + e^{-2\mathrm{i}\phi} h^+_{j,m+l,3,0,6}(\phi) \right) + \left(1+e^{2\mathrm{i}\phi}\right) f_{j,m+l}(\phi) \right.\notag\\
		 &\qquad + \left(e^{-2\mathrm{i}\phi}+1\right) \left(h^+_{j,m+l,-1,4,4}(\phi) - h^+_{j,m+l,-1,0,4}(\phi)\right) \notag\\
		 &\left.\left.\qquad - \cos(2\phi) \left(h^+_{j,m+l,-1,0,2}(\phi) + 6h^+_{j,m+l,-1,2,2}(\phi) + h^+_{j,m+l,-1,4,2}(\phi)\right)
			 - 2h^+_{j,m+l,3,0,2}(\phi) \vphantom{\frac{1}{2}}\right\rbrace \vphantom{\sum_j^1}\right).
	\end{align}
\end{widetext}

To make the numerical evaluation of the series more efficient, we want to make sure, that we add the largest terms first, so we can choose the cutoff as low as possible for a given required precision. To achieve this, we reorder the sum.\\
As we discuss in appendix~\ref{sec: Conv} the fundamental series is given by:
\begin{widetext}
	\begin{align}
		&\Gamma\left(\frac{t+1}{2}\right) \sum_{j=2}^\infty (-2\mathrm{i})^j \sum_{k=0}^j \sum_{m=0}^k \sum_{l=0}^{j-k}
			\left(\frac{\mathrm{i}}{2}\right)^{m+l} \binom{k}{m} \binom{j-k}{l} e^{-\mathrm{i}(k+m)\phi} \notag \\
		 &\cdot x^{j+m+l+N} \sum_{i=0}^\infty \frac{\Gamma\left(m+l+\frac{s+N+1}{2}+i\right)}
			 {(j+m+l+N+i)! \Gamma\left(m+l+\frac{s+t+N}{2}+1+i\right)} \frac{(-x^2)^i}{i!}.
	\end{align}
\end{widetext}
This of course does not hold for the $f$-term but since this one converges faster than the $h$-terms we try to make the slower converging series converge as fast as possible.\\

\begin{figure}[h!]
	\centering\includegraphics[width=0.7\linewidth]{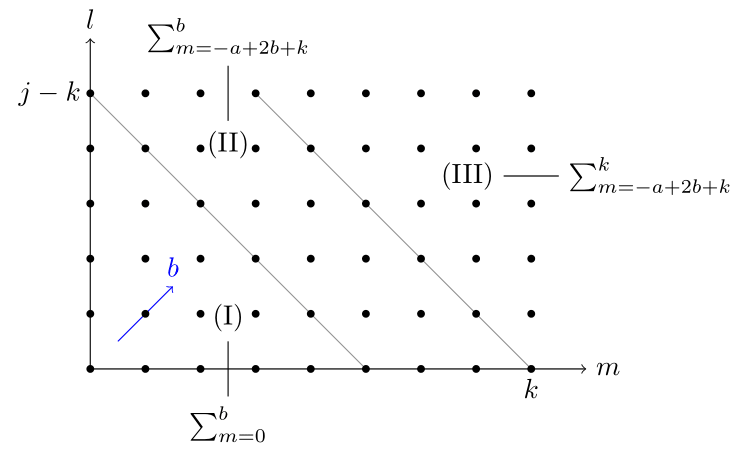}
	\caption{After the substitution we sum diagonally (indicated by the blue arrow) over the square shaped summation domain. This introduces different summation boundaries in the three different regions of the domain which are separated by the gray lines.}\label{fig: SumPts}
\end{figure}

To reorder the summation, we make the following substitution:
\begin{equation}
	j,\ k,\ m,\ l \quad \to \quad a = j+m+l,\ b = m+l,\ k,\ m
\end{equation}
\begin{align}
	&\Rightarrow \quad j = a-b,\ l = b-m \quad \wedge \quad \sum_{j=2}^\infty \sum_{k=0}^j \sum_{m=0}^k \sum_{l=0}^{j-k} \quad 
	\to \quad \sum_{a=2}^\infty \sum_{b=0}^{\left\lfloor\frac{a}{2}\right\rfloor} \sum_{k=0}^{a-b}
	\sum_{m=\max\{0,2 b+k-a\}}^{\min\{k,b\}}.
\end{align}
If we look at the fraction in the series over $i$ we see that what suppresses the terms for large indices is the factorial which apart from $i$ itself grows with $a$. This means, that for large $a$ the entire series is becoming small. The second new index $b$ appears in the nominator and denominator as argument of the gamma function. Since the gamma functions argument in the denominator is always bigger by at least $\frac{1}{2}$, a big value for $b$ would also suppress the series, but less than $a$. Thus, to add big terms first, we first sum over $a$, then over $b$ and then over $k$ and $m$:\\
\begin{equation}\label{eq: l-Limit}
	b-m = l \leqslant j-k = a-b-k \quad \Rightarrow \quad m \geqslant -a+2b+k.
\end{equation}
Due to the reordering we sum over $m$ only until $m = b$, region (I) depicted in Figure~\ref{fig: SumPts}. We also need to be careful, that we do not sum over $m$'s which we have already summed over at the previous values of $b$. The upper limit of the previous $l$-sum turns into this lower limit for the $m$-sum~\eqref{eq: l-Limit}, region (II). Since we now sum until $b$ instead of $k$ we still need to make certain that $m$ does not become bigger than $k$ otherwise we sum over our summation domain, region (III).

Our fundamental series becomes:
\begin{widetext}
	\begin{align}
		&\Gamma\left(\frac{t+1}{2}\right) \sum_{a=2}^\infty \sum_{b=0}^{\left\lfloor\frac{a}{2}\right\rfloor} \sum_{k=0}^{a-b}
			\sum_{m=\max\{0,2 b+k-a\}}^{\min\{k,b\}} (-2\mathrm{i})^{a-b} \left(\frac{\mathrm{i}}{2}\right)^b \binom{k}{m}
			\binom{a-b-k}{b-m} e^{-\mathrm{i}(k+m)\phi} \notag \\
		&\cdot x^{a+N} \sum_{i=0}^\infty \frac{\Gamma\left(b+\frac{s+N+1}{2}+i\right)}
			{(a+N+i)! \Gamma\left(b+\frac{s+t+N}{2}+1+i\right)} \frac{(-x^2)^i}{i!}.
	\end{align}
\end{widetext}

We appropriately redefine $h$ and $f$:
\begin{widetext}
	\begin{align}
		h^\pm&_{a,b,s,t,N}(\phi) \coloneq \Gamma\left(\frac{t+1}{2}\right) \Gamma\left(b+\frac{s+N+1}{2}\right)
			\left[ e^{\mathrm{i}(L-L')\omega} \left(\frac{1}{2}\left(L - e^{\pm\mathrm{i}\phi}L'\right)\omega\right)^{a+N}
			\right. \\
		 &\cdot\, _1\tilde{F}_2\left(b+\frac{s+N+1}{2};a+N+1,b+\frac{s+t+N}{2}+1;-(L^2 + L'^2 - 2LL'\cos\phi)\frac{\omega^2}{4}\right)
			\notag \\
		 &- e^{\mathrm{i}L\omega} \left(\frac{L\omega}{2}\right)^{a+N}
			\, _1\tilde{F}_2\left(b+\frac{s+N+1}{2};a+N+1,b+\frac{s+t+N}{2}+1;-\left(\frac{L\omega}{2}\right)^2\right) \notag \\
		 &\left. - e^{\mathrm{i}(\pm(a+N)\phi - L'\omega)} \left(-\frac{L'\omega}{2}\right)^{a+N}
			\, _1\tilde{F}_2\left(b+\frac{s+N+1}{2};a+N+1,b+\frac{s+t+N}{2}+1;-\left(\frac{L'\omega }{2}\right)^2\right) \right],
			\notag \\
		\vphantom{a}\notag\\
		f&_{a,b}(\phi ) \coloneq \sqrt{\pi} 2^{-a} \Gamma(b+1) \Gamma(b+3) \left[ e^{\mathrm{i}L\omega} (L\omega)^a
			\, _2\tilde{F}_3\left(b+1,b+3;a+1,b+2,b+\frac{5}{2};-\left(\frac{L\omega}{2}\right)^2\right) \right. \\
		 &+ e^{\mathrm{i}(a\phi-L'\omega)} (-L'\omega)^a
			\, _2\tilde{F}_3\left(b+1,b+3;a+1,b+2,b+\frac{5}{2};-\left(\frac{L'\omega}{2}\right)^2\right) \notag \\
		 &\left. - e^{\mathrm{i}(L-L')\omega} \left(\left(L - e^{\mathrm{i}\phi}L'\right)\omega\right)^a
			\, _2\tilde{F}_3\left(b+1,b+3;a+1,b+2,b+\frac{5}{2};-(L^2 + L'^2 - 2LL'\cos\phi)\frac{\omega^2}{4}\right)
			\vphantom{\left(\frac{L\omega}{2}\right)^2}\right], \notag
	\end{align}
	so we can write the full series as:
	\begin{align}
		&\sum_{a=2}^\infty \sum_{b=0}^{\left\lfloor\frac{a}{2}\right\rfloor} \sum_{k=0}^{a-b} \sum_{m=\max\{0,2 b+k-a\}}^{\min\{k,b\}}
			(-2\mathrm{i})^{a-b} \left(\frac{\mathrm{i}}{2}\right)^b \binom{k}{m} \binom{a-b-k}{b-m} e^{-\mathrm{i}(k+m)\phi} \notag \\
		&\cdot \left\lbrace -\frac{1}{2}\left( e^{2\mathrm{i}\phi} h^+_{a,b,3,0,-2}(\phi)
			+ e^{-2\mathrm{i}\phi} h^+_{a,b,3,0,6}(\phi) \right) + \left(1+e^{2\mathrm{i}\phi}\right) f_{a,b}(\phi)
			+ \left(e^{-2\mathrm{i}\phi}+1\right)\left(h^+_{a,b,-1,4,4}(\phi) - h^+_{a,b,-1,0,4}(\phi)\right) \right. \notag \\
		&\qquad\left. - \cos(2\phi) \left( h^+_{a,b,-1,0,2}(\phi) + 6h^+_{a,b,-1,2,2}(\phi) + h^+_{a,b,-1,4,2}(\phi) \right)
			- 2h^+_{a,b,3,0,2}(\phi) \vphantom{\frac{1}{2}}\right\rbrace.
	\end{align}
\end{widetext}

We notice, that the $h$- and $f$-terms are not dependent on $k$ and $m$ anymore and can thus be pulled out front, which is numerically more efficient. Furthermore, we can now simplify the double sum over the exponential and the binomial terms and get rid of the sum over $m$, which is again numerically much cheaper. As we show in~\ref{sec: 2F1} the sum over $m$ can be written as:
\begin{widetext}
	\begin{align}
		&\sum_{m=\max\{0,2 b+k-a\}}^{\min\{k,b\}} \binom{k}{m} \binom{a-b-k}{b-m} e^{-\mathrm{i}m\phi} \notag \\
		&\qquad = \begin{cases}
			\binom{a-b-k}{b}\, _2F_1\left(-b,-k;1+a-2b-k;e^{-\mathrm{i}\phi}\right) & a \geq 2b + k \\
			e^{\mathrm{i}(a-2b-k)\phi} \binom{k}{-a+2b+k}\, _2F_1\left(-a+2b,-a+b+k;1-a+2b+k;e^{-\mathrm{i}\phi}\right) & a < 2b + k
			\end{cases}.
	\end{align}
	
	Then finally we can write down the series in a reordered and simplified form which is more efficient for numerical evaluation:
	\begin{align}
	&\sum_{j=2}^\infty (-2\mathrm{i})^j \sum_{k=0}^j \sum_{m=0}^k \sum_{l=0}^{j-k} \left(\frac{\mathrm{i}}{2}\right)^{m+l}
	\left(\dots\vphantom{\frac{1}{2}}\right) \notag\\
	&= \sum_{a=2}^\infty (-2\mathrm{i})^a \sum_{b=0}^{\left\lfloor\frac{a}{2}\right\rfloor} \left(-\frac{1}{4}\right)^b
	\left\lbrace -\frac{1}{2}\left(e^{2\mathrm{i}\phi} h^+_{a,b,3,0,-2}(\phi) + e^{-2\mathrm{i}\phi} h^+_{a,b,3,0,6}(\phi)\right)
	+ \left(1+e^{2\mathrm{i}\phi}\right) f_{a,b}(\phi) \right.\notag\\
	&\qquad + \left(e^{-2\mathrm{i}\phi}+1\right) \left(h^+_{a,b,-1,4,4}(\phi) - h^+_{a,b,-1,0,4}(\phi)\right)
		- \cos(2\phi)\left(h^+_{a,b,-1,0,2}(\phi) + 6h^+_{a,b,-1,2,2}(\phi) + h^+_{a,b,-1,4,2}(\phi)\right) \notag\\
	&\left.\qquad- 2 h^+_{a,b,3,0,2}(\phi) \vphantom{\frac{1}{2}}\right\rbrace \sum _{k=0}^{a-b}
	\begin{cases}
		e^{-\mathrm{i}k\phi} \binom{a-b-k}{b}\, _2F_1\left(-b,-k;1+a-2b-k;e^{-\mathrm{i}\phi}\right) & a \geq 2b + k \\
		e^{\mathrm{i}(a-2(b+k))\phi} \binom{k}{a-2b}\, _2F_1\left(-a+2b,-a+b+k;1-a+2b+k;e^{-\mathrm{i}\phi}\right) & a < 2b + k
	\end{cases}.
	\end{align}
\end{widetext}

\clearpage

\appendix
\section{Integrations}\label{sec: Integrations}
To calculate the $\theta$-integral of $\Gamma_T$ we needed the following integral:
\begin{equation}
\int_0^\pi \cos^m\theta \sin^n\theta \, d\theta.
\end{equation}

We first integrate from zero to an arbitrary $\theta\in[0,\frac{\pi}{2}]$ using the substitution $t = \sin^\vartheta$, $dt = 2\sin\vartheta\cos\vartheta\,d\vartheta$:
\begin{widetext}
	\begin{align}
		&\int_0^\theta \cos^m\vartheta \sin^n\vartheta \, d\vartheta = \frac{s^{n-1}\sigma^{m-1}}{2} \int_0^\theta \left(\sin^2\vartheta\right)^\frac{n-1}{2}
			\left(1-\sin^2\vartheta\right)^\frac{m-1}{2} 2\sin\vartheta\cos\vartheta\,d\vartheta \notag \\
		&\qquad = \frac{s^{n-1}\sigma^{m-1}}{2} \int_0^{\sin^2\theta} t^{\frac{n+1}{2}-1} \left(1-t\right)^{\frac{m+1}{2}-1} dt
			= \frac{s^{n-1}\sigma^{m-1}}{2} B_{\sin^2\theta}\left( \frac{n+1}{2}, \frac{m+1}{2} \right),
	\end{align}
	where we defined $s \coloneq \text{sign}(\sin\vartheta)$ and $\sigma \coloneq \text{sign}(\cos\vartheta)$, which we could pull out of the integral since they are constant on the integration domain for $\theta < \frac{\pi}{2}$ and $B_x$ denotes the incomplete beta function.\\

	The beta function and its incomplete version are defined as:
	\begin{equation}
	B(x,y) \coloneq \int_0^1 t^{x-1} \left(1-t\right)^{y-1} dt, \quad \text{for }\ \Re(x),\, \Re(y) > 0, \qquad B_x(a,b) \coloneq \int_0^x t^{a-1} \left(1-t\right)^{b-1} dt.
	\end{equation}

	To integrate over $[0,\pi]$ we split the integral into the two parts since $\sigma$ changes its sign at $\frac{\pi}{2}$:
	\begin{align}
		&\int_0^\pi \cos^m\theta \sin^n\theta \, d\theta = \int_0^\frac{\pi}{2} \cos^m\theta \sin^n\theta \, d\theta + \int_\frac{\pi}{2}^\pi \cos^m\theta \sin^n\theta \, d\theta \notag \\
		&\qquad = \frac{1}{2} \int_0^1 t^{\frac{n+1}{2}-1} \left(1-t\right)^{\frac{m+1}{2}-1} dt
			+ \frac{(-1)^{m-1}}{2} \int_1^0 t^{\frac{n+1}{2}-1} \left(1-t\right)^{\frac{m+1}{2}-1} dt \notag \\
		&\qquad = \frac{1+(-1)^m}{2} B_1\left( \frac{n+1}{2}, \frac{m+1}{2} \right) = \frac{1+(-1)^m}{2} \frac{\Gamma\left(\frac{m+1}{2}\right) \Gamma\left(\frac{n+1}{2}\right)}{\Gamma\left(\frac{1}{2}(m+n+2)\right)}, \quad \text{for }\ \Re(m),\, \Re(n) > -1,
	\end{align}
\end{widetext}
where we used the relation between the beta and the gamma function: $B(x,y) = \frac{\Gamma(x)\Gamma(y)}{\Gamma(x+y)}$.\\

For the special case $N = 0$ and $s = -1$ we need to calculate the integral:
\begin{equation}
	\int_0^\pi (\cos^4\theta-1)\sin^n\theta\, d\theta.
\end{equation}

We try to make it look like an incomplete beta function as in the previous case, by writing everything in terms of cosine and then substituting $\cos^2\vartheta$. This means that we need to make sure that we have $-2\sin\vartheta\cos\vartheta$ in the integral before the substitution:
\begin{widetext}
	\begin{align}
		&\int_0^\theta (\cos^4\vartheta-1) \sin^n\vartheta\, d\vartheta
			= -\int_0^\theta (1-\cos^2\vartheta)(1+\cos^2\vartheta) \sin^n\vartheta\, d\vartheta \notag \\
		&= -\int_0^\theta (1-\cos^2\vartheta) \left[(1 - \cos^2\vartheta) + 2\cos\vartheta\right] \sin^n\vartheta\, d\vartheta \notag \\
		&= \frac{1}{2}\int_0^\theta \left[ (1-\cos^2\vartheta)^2\sigma(\cos^2\vartheta)^{-\frac{1}{2}} 
			+ 2(1-\cos^2\vartheta)\sigma(\cos^2\vartheta)^\frac{1}{2} \right]s^{n-1}(1-\cos^2\vartheta)^\frac{n-1}{2}
			(-2\sin\vartheta\cos\vartheta)\, d\vartheta \notag \\
		&= \frac{\sigma s^{n-1}}{2}\int_0^{\cos^2\theta} t^{-\frac{1}{2}} (1-t)^\frac{n+3}{2} + 2t^\frac{1}{2}(1-t)^\frac{n+1}{2} dt
			\notag \\
		&= \sigma s^{n-1}\left\{ \frac{1}{2}\int_0^{\cos^2\theta} t^{\frac{1}{2}-1} (1-t)^{\frac{n+5}{2}-1} dt
			+ \int_0^{\cos^2\theta} t^{\frac{3}{2}-1} (1-t)^{\frac{n+3}{2}-1} dt \right\} \notag \\
		&= \sigma s^{n-1}\left\{ \frac{1}{2}B_{\cos^2\theta}\left(\frac{1}{2},\frac{n+5}{2}\right)
			+ B_{\cos^2\theta}\left(\frac{3}{2},\frac{n+3}{2}\right) \right\} = \dots
	\end{align}

	Now we use the identity $B_x(a,b) = B_x(a,b+1) + B_x(a+1,b)$ on $\frac{1}{2}B_{\cos^2\theta}\left(\frac{1}{2},\frac{n+3}{2}+1\right) + B_{\cos^2\theta}\left(\frac{1}{2}+1,\frac{n+3}{2}\right)$ to pull the $\frac{1}{2}$ out front:
	\begin{equation}
		\dots = \frac{s^{n-1}\sigma}{2}\left\{ B_{\cos^2\theta}\left(\frac{1}{2},\frac{n+3}{2}\right)
			+ B_{\cos^2\theta}\left(\frac{3}{2},\frac{n+3}{2}\right) \right\} \eqcolon G_n(\theta).
	\end{equation}
\end{widetext}

Using another identity of the beta functions $B(a+1,b) = B(a,b)\frac{a}{a+b}$ we can compute the integral for our special case:
\begin{align}
	&\int_0^\pi (\cos^4\theta-1) \sin^n\theta\, d\theta = G_n(\pi) - G_n(0) 
	= -B\left(\frac{1}{2},\frac{n+3}{2}\right) - B(\frac{3}{2},\frac{n+3}{2})
	= -\frac{n+5}{n+4}B\left(\frac{1}{2},\frac{n+3}{2}\right) \notag \\
	&= -\frac{n+5}{n+4}\frac{\Gamma\left(\frac{1}{2}\right) \Gamma\left(\frac{n+3}{2}\right)}{\Gamma\left(\frac{n+4}{2}\right)} = -\frac{\sqrt{\pi}(n+5)\Gamma\left(\frac{n+3}{2}\right)} 
		{2\Gamma\left(\frac{n}{2}+3\right)}.
\end{align}

\section{Absolute convergence of the Series}\label{sec: Conv}
We show that the series converges absolutely by proving, that the absolute value of the series is finite. Or equivalently, if one would stop the summation at a finite cutoff number $N_c$, then the rest-term goes to zero if the cutoff tends to infinity:
\begin{equation}
	\sum_{n=n_0}^\infty \vert a_n \vert < \infty \quad \Leftrightarrow \quad \lim_{N_c\to\infty}\sum_{n=N_c}^\infty \vert a_n \vert.
\end{equation}

The series in the overlap reduction function consists of a sum of series. We are thus interested in the convergence of a general type of such a series. We will treat the terms with the function $h$ first and deal with the special case $f$ after.

\subsection{h-terms}
\begin{widetext}
	\begin{align}
		\sum_{j=2}^\infty& (-2\mathrm{i})^j \sum_{k=0}^j \sum_{m=0}^k \sum_{l=0}^{j-k} \left(\frac{\mathrm{i}}{2}\right)^{m+l}
			\binom{k}{m} \binom{j-k}{l} e^{-\mathrm{i}(k+m)\phi} h^+_{j,m+l,s,t,N}(\phi) \notag \\
		=& \sum_{j=2}^\infty (-2\mathrm{i})^j \sum_{k=0}^j \sum_{m=0}^k \sum_{l=0}^{j-k} \left(\frac{\mathrm{i}}{2}\right)^{m+l}
			\binom{k}{m} \binom{j-k}{l} e^{-\mathrm{i}(k+m)\phi} \notag \\
		&\cdot \Gamma\left(\frac{t+1}{2}\right) \Gamma\left(m+l+\frac{s+N+1}{2}\right) \left\lbrace
			e^{\mathrm{i}(L-L')\omega} \left(\frac{1}{2}\left(L-e^{\pm\mathrm{i}\phi}L'\right)\omega\right)^{j+m+l+N} \right. \notag \\
		&\cdot\, _1\tilde{F}_2\left(m+l+\frac{s+N+1}{2};j+m+l+N+1,m+l+\frac{s+t+N}{2}+1;
			-\left(L^2+L'^2-2LL'\cos(\phi)\right)\frac{\omega^2}{4} \right) \notag \\
		&- e^{\mathrm{i}L\omega} \left(\frac{L\omega}{2}\right)^{j+m+l+N}\, 
			_1\tilde{F}_2\left(m+l+\frac{s+N+1}{2};j+m+l+N+1,m+l+\frac{s+t+N}{2}+1;-\left(\frac{L\omega}{2}\right)^2\right) \notag \\
		&- e^{\mathrm{i}(\pm(j+m+l+N)\phi-L'\omega)} \left(-\frac{1}{2}(L'\omega)\right)^{j+m+l+N} \notag \\
		&\left.\cdot\, _1\tilde{F}_2\left(m+l+\frac{s+N+1}{2};j+m+l+N+1,m+l+\frac{s+t+N}{2}+1;-\left(\frac{L'\omega}{2}\right)^2\right).
			\right\rbrace
	\end{align}

	This series can again be split into three of the same type. To investigate their convergence, we write the hypergeometric function as a series:
	\begin{align}
		\sum_{j=2}^\infty& (-2\mathrm{i})^j \sum_{k=0}^j \sum_{m=0}^k \sum_{l=0}^{j-k} \left(\frac{\mathrm{i}}{2}\right)^{m+l}
			\binom{k}{m} \binom{j-k}{l} e^{-\mathrm{i}(k+m)\phi} x^{j+m+l+N} \notag \\
		&\cdot \Gamma\left(\frac{t+1}{2}\right) \Gamma\left(m+l+\frac{s+N+1}{2}\right)\, 
			_1\tilde{F}_2\left(m+l+\frac{s+N+1}{2};j+m+l+N+1,m+l+\frac{s+t+N}{2}+1;-x^2\right) \notag \\
		=& \sum_{j=2}^\infty (-2\mathrm{i})^j \sum_{k=0}^j \sum_{m=0}^k \sum_{l=0}^{j-k} \left(\frac{\mathrm{i}}{2}\right)^{m+l}
			\binom{k}{m} \binom{j-k}{l} e^{-\mathrm{i}(k+m)\phi} x^{j+m+l+N} \notag \\
		&\cdot \Gamma\left(\frac{t+1}{2}\right) \sum_{i=0}^\infty
			\frac{\Gamma\left(m+l+\frac{s+N+1}{2}+i\right)}{(j+m+l+N+i)! \Gamma\left(m+l+\frac{s+t+N}{2}+1+i\right)} \frac{(-x^2)^i}{i!}.
	\end{align}
\end{widetext}

We try to find an upper bound for the series for which we can show that it is finite. So, for
\begin{equation}
	x\in\mathbb{R}, \ s\in\{-1,3\}, \ t\in\{0,2,4\}, \ N\in\{2,4,6\}
\end{equation}
consider:
\begin{widetext}
	\begin{align}
		\sum_{j=2}^\infty& \left| (-2\mathrm{i})^j \sum_{k=0}^j \sum_{m=0}^k \sum_{l=0}^{j-k} \left(\frac{\mathrm{i}}{2}\right)^{m+l} 
			\binom{k}{m} \binom{j-k}{l} e^{-\mathrm{i}(k+m)\phi} x^{j+m+l+N} \right. \notag \\
		 &\left.\cdot \Gamma\left(\frac{t+1}{2}\right) \sum_{i=0}^\infty
			\frac{\Gamma \left(m+l+\frac{s+N+1}{2}+i\right)}{(j+m+l+N+i)! \Gamma\left(m+l+\frac{s+t+N}{2}+1+i\right)}
			\frac{(-x^2)^i}{i!} \right| \notag \\
		\leq&\, \Gamma\left(\frac{t+1}{2}\right) \sum_{j=2}^\infty 2^j \left|x\right|^{j+N} \sum_{k=0}^j \sum_{m=0}^k \sum_{l=0}^{j-k}
			2^{-m-l} \binom{k}{m} \binom{j-k}{l} \left|x\right|^{m+l} \notag \\
		 &\cdot \left| \sum_{i=0}^\infty \frac{\Gamma\left(m+l+\frac{s+N+1}{2}+i\right)}{\Gamma(j+m+l+N+i+1)
			 \Gamma\left(m+l+\frac{s+t+N}{2}+1+i\right)} \frac{(-x^2)^i}{i!} \right|,
	\end{align}
\end{widetext}
where we used the triangle inequality.\\

Our next goal is to reformulate the fraction in the series over $i$, such that we can divide out the nominator. To achieve this we use Stirling's approximation and the following identity:
\begin{equation}\label{eq: GHalfInt}
	\Gamma\left(n+\frac{1}{2}\right) = \frac{\sqrt{\pi}(2n)!}{4^n n!}.
\end{equation}

To simplify the calculation and better illustrate, what we are doing we substitute:
\begin{equation}
	n \coloneq m+l+\frac{1+N+s}{2}+i-1 \in \mathbb{N}\cup\{-1\}, \ \text{ since } \ s\in \{-1,3\}.
\end{equation}

So, the argument of the second gamma function in the denominator can be written as:
\begin{align}
	&m+l+\frac{s+t+N}{2}+1+i = n+\frac{s+t}{2}+2 = n+K+\frac{1}{2}, \quad
	K \coloneq \frac{s+t+1}{2}+1\in \mathbb{N}.
\end{align}

Using the identity $\Gamma(z+1) = z\Gamma(z)$ and the falling factorial $(x)_n = x(x-1)\ldots(x-n+1)$ we can write said gamma function as:
\begin{align}
	\Gamma&\left(m+l+\frac{s+t+N}{2}+1+i\right) = \Gamma\left(n+K+\frac{1}{2}\right)
	= \left(n+K-\frac{1}{2}\right)_K \Gamma\left(n+\frac{1}{2}\right).
\end{align}

Using~\eqref{eq: GHalfInt} we can rewrite the fraction to:
\begin{align}
	&\frac{\Gamma(n+1)}{\Gamma(j+m+l+N+1+i) \left(n+K+\frac{1}{2}\right)_K
		\Gamma\left(n+\frac{1}{2}\right)}
	= \frac{\Gamma(n+1)}{\Gamma(j+m+l+N+1+i) \left(n+K+\frac{1}{2}\right)_K
		\sqrt{\pi}\frac{(2n)!}{4^n n!}} \notag \\
	&= \frac{4^n \Gamma(n+1)^2}{\sqrt{\pi}\, \Gamma(j+m+l+N+1+i)
		\left(n+K+\frac{1}{2}\right)_K \Gamma(2n+1)}
	= \frac{4^n n^2 \Gamma(n)^2}{\sqrt{\pi}\, \Gamma(j+m+l+N+1+i) \left(n+K+\frac{1}{2}\right)_K
		2n \Gamma(2n)} \notag \\
	&=\frac{4^n n B(n,n)}{2 \sqrt{\pi}\, \Gamma(j+m+l+N+1+i) \left(n+K+\frac{1}{2}\right)_K},
\end{align}
where we used the beta function on two identical arguments, which asymptotically tends to Stirling's formula for large real values:
\begin{align}
	B(x,y) &\coloneq \frac{\Gamma(x)\Gamma(y)}{\Gamma(x+y)} \sim \frac{\sqrt{2\pi}
		\left(x^{x-\frac{1}{2}} y^{y-\frac{1}{2}}\right)}{(x+y)^{x+y-\frac{1}{2}}}, \quad x, y\to\infty. 
\end{align}

We approximate the fraction, since we are interested in the limit of rest-term after the cutoff:
\begin{align}
	&\sim \frac{\sqrt{2\pi} 4^n n
		\frac{\left(n^{n-\frac{1}{2}}\right)^2}{(2 n)^{2 n-\frac{1}{2}}}}
		{2 \sqrt{\pi } \Gamma (j+m+l+N+1+i) \left(n+K+\frac{1}{2}\right)_K}
	= \frac{\sqrt{n}}{\Gamma(j+m+l+N+1+i) \left(n+K+\frac{1}{2}\right)_K}.
\end{align}

The nominator $\sqrt{n}$ can now be divided by the falling factorial using the following inequality:
\begin{equation}
	\frac{\sqrt{n}}{n+\frac{1}{2}} < \frac{\sqrt{n}}{n} = \frac{1}{\sqrt{n}},
\end{equation}
\begin{widetext}
	\begin{align}
		\frac{\sqrt{n}}{\Gamma(\dots) \left(n+K+\frac{1}{2}\right)_K} < \frac{1}{\Gamma(j+m+l+N+1+i) \sqrt{n}\left(n+K+\frac{1}{2}\right)_{K-1}}.
	\end{align}

	Thus, the series is smaller than:
	\begin{align}
		\Gamma\left(\frac{t+1}{2}\right) \sum_{j=2}^\infty \ldots <&\, \Gamma\left(\frac{t+1}{2}\right) 
			\sum_{j=2}^\infty 2^j \left|x\right|^{j+N} \sum_{k=0}^j \sum_{m=0}^k \sum_{l=0}^{j-k}
			2^{-m-l} \binom{k}{m} \binom{j-k}{l} \left|x\right|^{m+l} \notag \\
		&\cdot \left| \sum_{i=0}^\infty \frac{1}{\Gamma(j+m+l+N+1+i) \sqrt{m+l+\frac{s+N-1}{2}+i}
			\left(n+K+\frac{1}{2}\right)_{K-1}} \frac{(-x^2)^i}{i!} \right| \notag \\
		<& \Gamma\left(\frac{t+1}{2}\right) e^{-x^2} \sum_{j=2}^\infty
			\frac{\left(2^j\left| x\right| ^{j+N}\right) }{\Gamma (j+N+1)}\sum _{k=0}^j \sum _{m=0}^k \sum _{l=0}^{j-k}
			2^{-m-l} \binom{k}{m} \binom{j-k}{l} \left| x\right| ^{m+l}.
	\end{align}
\end{widetext}

In the last inequality we dropped the square root and the falling factorial as well as $m,l$ and $i$ in the argument of the gamma function in the denominator. This is quite a rough estimate (large upper bound), but still finite and thus sufficient for our purposes.\\
One could get a smaller upper bound by leaving the falling factorial and get
\begin{align}
	\frac{\, _1F_1\left(m+l+\frac{N+3}{2}; m+l+\frac{s+t+N}{2}+2; -x^2\right)}
		{\left(m+l+\frac{s+t+N}{2}+1\right)_{\frac{s+t+1}{2}}},
\end{align}
instead of $e^{-x^2}$. For simplicity we continue with the former version.\\

We now use the binomial theorem on
\begin{equation}
	\sum_{l=0}^{j-k} \binom{j-k}{l} \left| \frac{x}{2}\right|^l 1^{j-k-l}
		= \left( 1 + \left|\frac{x}{2}\right| \right)^{j-k},
\end{equation}
to get rid of the sums over $m$ and $l$:
\begin{widetext}
	\begin{align}
		&\Gamma\left(\frac{t+1}{2}\right) e^{-x^2} \sum_{j=2}^\infty \frac{2^j\left|x\right|^{j+N}}{\Gamma(j+N+1)}
			\sum_{k=0}^j \sum_{m=0}^k \sum_{l=0}^{j-k} \binom{k}{m} \binom{j-k}{l} \left|\frac{x}{2}\right|^{m+l} \notag \\
		   &=\Gamma \left(\frac{t+1}{2}\right) e^{-x^2} \sum_{j=2}^\infty
			\frac{2^j \left|x\right|^{j+N}}{\Gamma(j+N+1)} \sum_{k=0}^j \left( 1 + \left|\frac{x}{2}\right| \right)^j
			= \Gamma\left(\frac{t+1}{2}\right) e^{-x^2} \left|x\right|^N \sum_{j=2}^\infty
			\frac{(j+1) \left(\left|2x\right| + x^2\right)^j}{\Gamma(j+N+1)},
	\end{align}
	since the summand of the $k$-sum is independent of $k$.\\

	The series over $j$ gives us the finite result:
	\begin{align}
		\sum_{j=2}^\infty \frac{(j+1) \left(\left|2x\right| + x^2\right)^j}{\Gamma(1+j+N)}
			= \frac{(N+2) \left(\left|2x\right| + x^2\right)^2}{\Gamma(N+3)}
			+ e^{\left|2x\right| + x^2}
			\frac{\left(1-N+\left|2x\right| + x^2\right)}{\left(\left|2x\right| + x^2\right)^N}
			\left(1-\frac{(N+2) \Gamma\left(N+2,\left|2x\right| + x^2\right)}{\Gamma(N+3)}\right),
	\end{align}
\end{widetext}
where $\Gamma(a,z)$ is the lower incomplete gamma function:
\begin{equation}
	\Gamma(a,z) \coloneq \int_z^\infty t^{a-1} e^{-t} dt.
\end{equation}

And the rest term tends towards zero as the cutoff goes to infinity:
\begin{equation}
	\lim_{N_c\to\infty} \sum_{j=N_c}^\infty \frac{(j+1) \left(\left|2x\right| + x^2\right)^j}{\Gamma(j+N+1)} \to 0.
\end{equation}

\subsection{f-term}
We start out in the same way by identifying the fundamental series which we need to proof to be absolutely convergent:
\begin{widetext}
	\begin{align}
		\sum_{j=2}^\infty& (-2\mathrm{i})^j \sum_{k=0}^j \sum_{m=0}^k \sum_{l=0}^{j-k}
			\left(\frac{\mathrm{i}}{2}\right)^{m+l} \binom{k}{m} \binom{j-k}{l} e^{-\mathrm{i}(k+m)\phi} f_{j,m+l}(\phi) \\
		=& \sum_{j=2}^\infty (-2\mathrm{i})^j \sum_{k=0}^j \sum_{m=0}^k \sum_{l=0}^{j-k} \left(\frac{\mathrm{i}}{2}\right)^{m+l}
			\binom{k}{m} \binom{j-k}{l} e^{-\mathrm{i}(k+m)\phi} \sqrt{\pi} 2^{-j-m+l} \Gamma(m+l+1) \Gamma(m+l+3) \notag \\
		 &\cdot \left\lbrace e^{\mathrm{i}L\omega}(L\omega)^{j+m+l}
			\, _2\tilde{F}_3\left(m+l+1,m+l+3;j+m+l+1,m+l+2,m+l+\frac{5}{2};-\left(\frac{L\omega}{2}\right)^2\right) \right. \notag \\
		 &+ e^{\mathrm{i}((j+m+l)\phi - L'\omega)} (-L'\omega)^{j+m+l}
			\, _2\tilde{F}_3\left(m+l+1,m+l+3;j+m+l+1,m+l+2,m+l+\frac{5}{2};-\left(\frac{L'\omega}{2}\right)^2\right) \notag \\
		 &- e^{\mathrm{i}(L-L')\omega} \left(\left(L - e^{\mathrm{i}\phi}L'\right)\omega\right)^{j+m+l} \notag \\
		 &\left. \cdot \, _2\tilde{F}_3\left(m+l+1,m+l+3;j+m+l+1,m+l+2,m+l+\frac{5}{2};-(L^2+L'^2-2LL'\cos\phi)\frac{\omega^2}{4}\right) 	\vphantom{\left(\frac{L\omega}{2}\right)^2}\right\rbrace, \notag
	\end{align}
	\begin{align}
		\sum_{j=2}^\infty& (-2\mathrm{i})^j \sum_{k=0}^j \sum_{m=0}^k \sum_{l=0}^{j-k} \left(\frac{\mathrm{i}}{2}\right)^{m+l}
			\binom{k}{m} \binom{j-k}{l} e^{-\mathrm{i}(k+m)\phi} \sqrt{\pi} 2^{-j-m+l} \notag \\
		 &\Gamma(m+l+1) \Gamma(m+l+3) x^{j+m+l} \, _2\tilde{F}_3\left(m+l+1,m+l+3;j+m+l+1,m+l+2,m+l+\frac{5}{2};-x^2\right) \notag \\
		=& - \sqrt{\pi} \sum_{j=2}^\infty (-2\mathrm{i})^j \sum_{k=0}^j \sum_{m=0}^k \sum_{l=0}^{j-k}
			\left(\frac{\mathrm{i}}{2}\right)^{m+l} \binom{k}{m} \binom{j-k}{l} e^{-\mathrm{i}(k+m)\phi} 2^{-j-m+l} x^{j+m+l} \notag \\
		 &\cdot \sum_{i=0}^\infty \frac{\Gamma(m+l+i+1) \Gamma(m+l+i+3)}
			 {\Gamma(j+m+l+i+1) \Gamma(m+l+i+2) \Gamma\left(m+l+i+\frac{5}{2}\right)} \frac{(-x^2)^i}{i!}.
	\end{align}
	Let us find an upper bound for the absolute value series $\forall x\in\mathbb{R}$ using the same idea as above. We need to deal with two more terms in the fraction of the series over $i$, which happen to cancel to a fraction without nominator however:
	\begin{align}
		\sqrt{\pi}& \sum_{j=2}^\infty \left| (-2\mathrm{i})^j \sum_{k=0}^j \sum_{m=0}^k \sum_{l=0}^{j-k}
			\left(\frac{\mathrm{i}}{2}\right)^{m+l} \binom{k}{m} \binom{j-k}{l} e^{-\mathrm{i}(k+m)\phi} 2^{-j-m+l} x^{j+m+l}
			\right. \notag \\
		&\left. \cdot \sum_{i=0}^\infty \frac{1}{\Gamma(j+m+l+i+1)} \frac{\Gamma(m+l+i+1)}{\Gamma(m+l+i+2)}
			\frac{\Gamma(m+l+i+3)}{\Gamma\left(m+l+i+\frac{5}{2}\right)} \frac{(-x^2)^i}{i!} \right| \notag \\
		\leq& \sqrt{\pi} \sum_{j=2}^\infty 2^j \sum_{k=0}^j \sum_{m=0}^k \sum_{l=0}^{j-k} 2^{-m-l} \binom{k}{m} \binom{j-k}{l}
		2^{-j-m-l} |x|^{j+m+l} \notag \\
		&\cdot \left| \sum_{i=0}^\infty \frac{1}{\Gamma(j+m+l+i+1)} \frac{1}{m+l+i+1}
			\frac{\Gamma(m+l+i+3)}{\Gamma\left(m+l+i+2+\frac{1}{2}\right)} \frac{(-x^2)^i}{i!} \right|.
	\end{align}
\end{widetext}

We define $n \coloneq m+l+i+2$ and do the same procedure as with the $h$-terms:
\begin{align}
	&\frac{\Gamma(n+1)}{\Gamma\left(n+\frac{1}{2}\right)}
		= \frac{\Gamma(n+1)}{\frac{\sqrt{\pi}(2n)!}{4^n n!}}
		= \frac{4^n \Gamma(n+1)^2}{\sqrt{\pi} \Gamma(2n+1)}
		= \frac{4^n n^2 \Gamma(n)^2}{\sqrt{\pi} 2n \Gamma(2n)}
	= \frac{4^n n B(n,n)}{2\sqrt{\pi}}
		\sim \frac{\sqrt{2\pi} 4^n n \left(n^{n-\frac{1}{2}}\right)^2}{2\sqrt{\pi}(2n)^{2n-\frac{1}{2}}} = \sqrt{n}.
\end{align}

Using, that
\begin{equation}
	\frac{\sqrt{n}}{n-1} = \frac{1}{\sqrt{n}-\frac{1}{\sqrt{n}}} \sim \frac{1}{\sqrt{n}}, \quad n \gg 1
\end{equation}
we can approximate the series as:
\begin{widetext}
	\begin{align}
		\sqrt{\pi}& \sum_{j=2}^\infty 2^j \sum_{k=0}^j \sum_{m=0}^k \sum_{l=0}^{j-k} 2^{-m-l} \binom{k}{m} \binom{j-k}{l}
			2^{-j-m-l} |x|^{j+m+l} \left| \sum_{i=0}^\infty \frac{1}{\Gamma(j+m+l+i+1)} \frac{\sqrt{m+l+i+2}}{m+l+i+1}
			\frac{(-x^2)^i}{i!} \right| \notag \\
		\sim& \sqrt{\pi} \sum_{j=2}^\infty 2^j \sum_{k=0}^j \sum_{m=0}^k \sum_{l=0}^{j-k} 2^{-m-l} \binom{k}{m} \binom{j-k}{l}
			2^{-j-m-l} |x|^{j+m+l} \left| \sum_{i=0}^\infty \frac{1}{\Gamma(j+m+l+i+1)} \frac{1}{\sqrt{m+l+i+2}}
			\frac{(-x^2)^i}{i!} \right| \notag \\
		\leq& \sqrt{\pi} \sum_{j=2}^\infty |x|^j \left| \sum_{i=0}^\infty \frac{1}{\Gamma(j+i+1)} \frac{1}{\sqrt{i+2}}
			\frac{(-x^2)^i}{i!} \right|  \sum_{k=0}^j \sum_{m=0}^k \sum_{l=0}^{j-k} \binom{k}{m} \binom{j-k}{l}
			\left|\frac{x}{4}\right|^{m+l} \notag \\
		\leq& \sqrt{\pi} \left| \sum_{i=0}^\infty \frac{1}{\sqrt{i+2}} \frac{(-x^2)^i}{i!} \right|
			\sum_{j=2}^\infty \frac{|x|^j}{\Gamma(j+i+1)} \sum_{k=0}^j \left( 1 + \left|\frac{x}{4}\right| \right)^j
			\leq \sqrt{\pi} e^{-x^2} \sum_{j=2}^\infty \frac{j+1}{\Gamma(j+1)} \left( |x| + \frac{x^2}{4} \right)^j.
	\end{align}

	This series can be calculated to be:
	\begin{equation}
		\sum_{j=2}^\infty \frac{j+1}{j!} \left( |x| + \frac{x^2}{4} \right)^j
			= e^{|x| + \frac{x^2}{4}} \left(1 + |x| + \frac{x^2}{4}\right)
			- 1 - 2|x| - \frac{x^2}{2},
	\end{equation}
	and thus, is finite for finite $x$.

	And again, the limit of the rest term for can be bounded from above with:
	\begin{align}
		\lim_{N_c\to\infty}& \sum_{j=N_c}^\infty \left| (-2\mathrm{i})^j \sum_{k=0}^j \sum_{m=0}^k \sum_{l=0}^{j-k}
			\left(\frac{\mathrm{i}}{2}\right)^{m+l} \binom{k}{m} \binom{j-k}{l} e^{-\mathrm{i}(k+m)\phi} f_{j,m+l}(\phi) \right|
			\notag \\
		\leq& \sqrt{\pi} e^{-x^2} \lim_{N_c\to\infty} \sum_{j=N_c}^\infty \frac{j+1}{\Gamma(j+1)} \left( |x| + \frac{x^2}{4} \right)^j
			\to 0.
	\end{align}
\end{widetext}

\section{Hypergeometric function on negative integers}\label{sec: 2F1}
We can simplify the series in the overlap reduction function and get the four nested sums down to three by using the hypergeometric function with negative integers as the first two arguments:
\begin{align}\label{eq: 2F1}
	\, _2F_1&\left(-b,-k; 1+a-2b-k; e^{-\mathrm{i}\phi}\right)
	= \sum_{m=0}^k (-1)^m \binom{k}{m} \frac{(-b)^{(m)}}{(1+a-2b-k)^{(m)}}
		\left(e^{-\mathrm{i}\phi}\right)^m \\
	&= \sum_{m=0}^b (-1)^m \binom{b}{m} \frac{(-k)^{(m)}}{(1+a-2b-k)^{(m)}}
		\left(e^{-\mathrm{i}\phi}\right)^m, \notag
\end{align}
As always for hypergeometric functions the sequence of the first and the second set of arguments does not matter and thus, we get two identical but expressions which look quite different at first glance.\\

We use the following definitions and identities for the rising $x^{(n)}$ and falling $(x)_n$ factorials:
\begin{widetext}
	\begin{align}\label{eq: factorials}
		x^{(n)} &\coloneq \prod_{i=0}^{n-1} (x + i) = x(x+1)\cdot\ldots\cdot(x+n-1),
			&\quad (x)_n &\coloneq \prod_{i=0}^{n-1}(x - i) = x(x-1)\cdot\ldots\cdot(x-n+1), \notag \\
		(x)_n &= (-1)^n(-x)^{(n)}, &\quad (m)_n &= \frac{m!}{(m-n)!}.
	\end{align}
	
	Let us first look at the case:
	\begin{align}
		&\sum_{m=0}^{\min\{k,b\}} \binom{k}{m} \binom{a-b-k}{b-m} e^{-\mathrm{i}m\phi}
			= \sum_{m=0}^{\min\{k,b\}} \binom{k}{m} \frac{(a-b-k)!}{(a-2b-k+m)! (b-m)!} \left(e^{-\mathrm{i}\phi}\right)^m \notag \\
		&= \sum_{m=0}^{\min\{k,b\}} \binom{k}{m} \binom{a-b-k}{b-m} e^{-\mathrm{i}m\phi}
		= \sum_{m=0}^{\min\{k,b\}} \binom{k}{m} \frac{(b-m+1)\cdot\ldots\cdot b (a-b-k)!}
			{(a-2b-k)! (a-2b-k+1)\cdot\ldots\cdot(a-2b-k+m) b!} \left(e^{-\mathrm{i}\phi}\right)^m \notag \\
		&= \binom{a-b-k}{b} \sum_{m=0}^{\min\{k,b\}} \binom{k}{m} \frac{(b)_m}{(a-2b-k+1)^{(m)}} \left(e^{-\mathrm{i}}\phi\right)^m
			\notag \\
		&= \binom{a-b-k}{b} \sum_{m=0}^{\min\{k,b\}} \binom{k}{m} \frac{(-1)^m(-b)^{(m)}}{(a-2b-k+1)^{(m)}} 
			\left(e^{-\mathrm{i}}\phi\right)^m.
	\end{align}
\end{widetext}

If we sum to $k$ or rewrite the last expression and using:
\begin{equation}
	\binom{k}{m}(b)_m = \frac{k!}{(k-m)!m!} \frac{b!}{(b-m)!} = (k)_m\binom{b}{m},
\end{equation}
we see, that in both cases we get on of the two versions of~\eqref{eq: 2F1}. Thus, it does not matter whether we sum until $k$ or be $b$ and we can write the sum as a hypergeometric function:
\begin{align}
	\sum_{m=0}^{\min\{k,b\}}& \binom{k}{m} \binom{a-b-k}{b-m} e^{-\mathrm{i}m\phi}
	= \binom{a-b-k}{b}\, _2F_1\left(-b,-k; a-2b-k+1; e^{-\mathrm{i}\phi}\right).
\end{align}
This result is used for the case where $a \geqslant 2b + k$, since then $\max\{0,2b+k-a\} = 0$.\\

For the other case $a < 2b + k$ we do the analogue computation. However, to use the identity~\ref{eq: 2F1} we need the sum to start at zero and thus relabel it. This time we start with the sum until $k$ and then show that we get to the same result with $b$:
\begin{widetext}
	\begin{align}
		&\sum_{m=2b+k-a}^k \binom{k}{m} \binom{a-b-k}{b-m} e^{-\mathrm{i}m\phi} = \sum_{m=0}^{a-2b} \binom{k}{m+2b+k-a}
			\binom{a-b-k}{b-m-2b-k+a} e^{-\mathrm{i}(m+2b+k-a)\phi} \notag \\
		&\quad = e^{-\mathrm{i}(2b+k-a)\phi} \sum_{m=0}^{a-2b} \binom{a-b-k}{a-b-k-m} \binom{k}{m+2b+k-a} \left(e^{-\mathrm{i}\phi}\right)^m,
	\end{align}
	we use $\binom{n}{k} = \binom{n}{n-k}$ to get:
	\begin{align}
		=& e^{-\mathrm{i}(2b+k-a)\phi} \sum_{m=0}^{a-2b} \binom{a-b-k}{m} \binom{k}{a-2b-m} \left(e^{-\mathrm{i}\phi}\right)^m \notag \\
		=& e^{-\mathrm{i}(2b+k-a)\phi} \sum_{m=0}^{a-2b} \binom{a-b-k}{m} \frac{(a-2b)\cdot\ldots\cdot(a-2b-m+1) k!}
			{(k-a+2b)! (k-a+2b+1)\cdot\ldots\cdot(k-a+2b+m) (a-2b)!} \left(e^{-\mathrm{i}\phi}\right)^m \notag \\
		=& e^{-\mathrm{i}(2b+k-a)\phi} \binom{k}{a-2b} \sum_{m=0}^{a-2b} \frac{(a-b-k)!(a-2b)_m}{(a-b-k-m)! m! (k-a+2b+1)^{(m)}}
			\left(e^{-\mathrm{i}\phi}\right)^m.
	\end{align}

	Using~\ref{eq: factorials} we can verify that $\frac{(a-b-k)!}{(a-b-k-m)!} (a-2b)_m = (a-b-k)_m \frac{(a-2b)!}{(a-2b-m)!}$ and thus:
	\begin{align}
		&= e^{-\mathrm{i}(2b+k-a)\phi} \binom{k}{a-2b} \sum_{m=0}^{a-2b} \binom{a-2b}{m} \frac{(-1)^m (k+b-a)^{(m)}}{(k-a+2b+1)^{(m)}}
			\left(e^{-\mathrm{i}\phi}\right)^m \notag \\
		&= e^{-\mathrm{i}(2b+k-a)\phi} \binom{k}{a-2b}\, _2F_1\left(2b-a,k+b-a; 2b+k-a+1; e^{-\mathrm{i}\phi}\right).
	\end{align}

	For the $b$ case everything works the same except that we sum up to a different value:
	\begin{align}
		&\sum_{m=2b+k-a}^b \binom{k}{m} \binom{a-b-k}{b-m} e^{-\mathrm{i}m\phi} = \sum_{m=0}^{a-b-k} \binom{k}{m+2b+k-a}
			\binom{a-b-k}{a-b-k-m} e^{-\mathrm{i}m\phi} \notag \\
		&= e^{-\mathrm{i}(2b+k-a)\phi} \binom{k}{a-2b} \sum_{m=0}^{a-2b} \binom{a-2b}{m} \frac{(-1)^m (k+b-a)^{(m)}}{(k-a+2b+1)^{(m)}}
			\left(e^{-\mathrm{i}\phi}\right)^m.
	\end{align}

	Then, as before, we use:
	\begin{equation}
		\binom{a-2b}{m} (a-b-k)_m = \frac{(a-2b)!}{(a-2b-m)!m!} \frac{(a-b-k)!}{(a-b-k-m)!} 
			= (a-2b)_m \binom{a-b-k}{m},
	\end{equation}
	to rewrite the sum as:
	\begin{align}
		&= e^{-\mathrm{i}(2b+k-a)\phi} \binom{k}{a-2b} \sum_{m=0}^{a-b-k} \binom{a-b-k}{m} \frac{(-1)^m(a-2b)^{(m)}}{(k-a+2b+1)^{(m)}}
			\left(e^{-\mathrm{i}\phi}\right)^m \notag \\
		&= e^{-\mathrm{i}(2b+k-a)\phi} \binom{k}{a-2b}\, _2F_1\left(k+b-a,2b-a; 2b+k-a+1; e^{-\mathrm{i}\phi}\right).
	\end{align}
	And again, we get the same expression as for the $k$ case.
\end{widetext}

\bibliographystyle{apsrev4-1}
\bibliography{PTARef}